\documentclass[twocolumn]{aastex631}
\usepackage{threeparttable}
\usepackage{booktabs}
\usepackage{subfigure}
\usepackage{amsmath}
\usepackage{epstopdf}
\usepackage{color}
\usepackage{enumitem}
\usepackage{tabularx}
\usepackage{multirow}
\usepackage{makecell}
\usepackage{xcolor}
\usepackage{comment}
\setlist[enumerate]{label=\textbf{\arabic*.}}

\newcommand{\reffig}[1]{Figure \ref{#1}}

\newcommand{\refsection}[1]{Section \ref{#1}}
\newcommand{\refsubsection}[1]{Section \ref{#1}}
\newcommand{\refsubsubsection}[1]{Section \ref{#1}}
\newcommand{\refformula}[1]{Equation (\ref{#1})}

\begin{document}
	\title{Scalable Stellar Parameter Inference Using Python-based LASP: From CPU Optimization to GPU Acceleration}
	\author{Jun-Chao Liang}
	\affil{CAS Key Laboratory of Optical Astronomy, National Astronomical Observatories, Chinese Academy of Sciences, Beijing 100101, People’s Republic of China}
	\affil{School of Astronomy and Space Science, University of Chinese Academy of Sciences, Beijing 100049, People’s Republic of China}
	
	\author{Yin-Bi Li $^\star$}
	\affil{CAS Key Laboratory of Optical Astronomy, National Astronomical Observatories, Chinese Academy of Sciences, Beijing 100101, People’s Republic of China}
	\email{$^\star$ ybli@bao.ac.cn}
	
	\author{A-Li Luo $^\star$}
	\affil{CAS Key Laboratory of Optical Astronomy, National Astronomical Observatories, Chinese Academy of Sciences, Beijing 100101, People’s Republic of China}
	\affil{School of Astronomy and Space Science, University of Chinese Academy of Sciences, Beijing 100049, People’s Republic of China}
	\email{$^\star$ lal@nao.cas.cn}
	
	\author{Fang Zuo}
	\affil{CAS Key Laboratory of Optical Astronomy, National Astronomical Observatories, Chinese Academy of Sciences, Beijing 100101, People’s Republic of China}
	
	\author{Bing Du}
	\affil{CAS Key Laboratory of Optical Astronomy, National Astronomical Observatories, Chinese Academy of Sciences, Beijing 100101, People’s Republic of China}
	
	\author{Shuo Li}
	\affil{CAS Key Laboratory of Optical Astronomy, National Astronomical Observatories, Chinese Academy of Sciences, Beijing 100101, People’s Republic of China}
	\affil{School of Astronomy and Space Science, University of Chinese Academy of Sciences, Beijing 100049, People’s Republic of China}
	
	\author{Xiao-Xiao Ma}
	\affil{CAS Key Laboratory of Optical Astronomy, National Astronomical Observatories, Chinese Academy of Sciences, Beijing 100101, People’s Republic of China}
	\affil{School of Astronomy and Space Science, University of Chinese Academy of Sciences, Beijing 100049, People’s Republic of China}
	
	\author{Shu-Guo Ma}
	\affil{CAS Key Laboratory of Optical Astronomy, National Astronomical Observatories, Chinese Academy of Sciences, Beijing 100101, People’s Republic of China}
	
	\author{Hai-Ling Lu}
	\affil{CAS Key Laboratory of Optical Astronomy, National Astronomical Observatories, Chinese Academy of Sciences, Beijing 100101, People’s Republic of China}
	\affil{School of Astronomy and Space Science, University of Chinese Academy of Sciences, Beijing 100049, People’s Republic of China}
	
	\author{Ke-Fei Wu}
	\affil{CAS Key Laboratory of Optical Astronomy, National Astronomical Observatories, Chinese Academy of Sciences, Beijing 100101, People’s Republic of China}
	
	\author{Zhi-Hua Zhong}
	\affil{Center for Spatial Information Science, The University of Tokyo, Tokyo 153-8505, Japan}
	
	\author{Wen Hou}
	\affil{CAS Key Laboratory of Optical Astronomy, National Astronomical Observatories, Chinese Academy of Sciences, Beijing 100101, People’s Republic of China}
	
	\author{Xiao Kong}
	\affil{CAS Key Laboratory of Optical Astronomy, National Astronomical Observatories, Chinese Academy of Sciences, Beijing 100101, People’s Republic of China}
	
	\author{Shuo Ye}
	\affil{CAS Key Laboratory of Optical Astronomy, National Astronomical Observatories, Chinese Academy of Sciences, Beijing 100101, People’s Republic of China}
	\affil{School of Astronomy and Space Science, University of Chinese Academy of Sciences, Beijing 100049, People’s Republic of China}
	
	\author{Li-Li Wang}
	\affil{School of Computer and Information, Dezhou University, Dezhou 253023, People’s Republic of China}
	
	\author{Hugh R. A. Jones}
	\affiliation{School of Physics, Astronomy and Mathematics, University of Hertfordshire, UK}
	
	\begin{abstract}
		{To enhance the efficiency, scalability, and cross-survey applicability of stellar parameter inference in large spectroscopic datasets, we  present a modular, parallelized Python framework with automated error estimation, built on the LAMOST Atmospheric Parameter Pipeline (LASP) originally implemented in IDL. Rather than a direct code translation, this framework refactors LASP with two complementary modules:
		LASP-CurveFit, a new implementation of the LASP fitting procedure that runs on a CPU, preserving legacy logic while improving data I/O and multithreaded execution efficiency; and LASP-Adam-GPU, a GPU-accelerated method that introduces grouped optimization by constructing a joint residual function over multiple observed and model spectra, enabling high-throughput parameter inference across tens of millions of spectra. Applied to 10 million LAMOST spectra, the framework reduces runtime from 84 to 48 hr on the same CPU platform and to 7 hr on an NVIDIA A100 GPU, while producing results consistent with those from the original pipeline. The inferred errors agree well with the parameter variations from repeat observations of the same target (excluding radial velocities), while the official empirical errors used in LASP are more conservative. When applied to DESI DR1, our effective temperatures and surface gravities agree better with APOGEE than those from the DESI pipeline, particularly for cool giants, while the latter performs slightly better in radial velocity and metallicity. These results suggest that the framework delivers reliable accuracy, efficiency, and transferability, offering a practical approach to parameter inference in large spectroscopic surveys. The code and DESI-based catalog are available via \dataset[DOI: 10.12149/101679]{https://doi.org/10.12149/101679} and \dataset[DOI: 10.12149/101675]{https://doi.org/10.12149/101675}, respectively.}
	\end{abstract}
	
	
	\section{Introduction}
	In recent years, the rapid development of large-scale spectroscopic surveys---such as the Radial Velocity Experiment \citep{Steinmetz2006}, the Large Sky Area Multi-object Fiber Spectroscopic Telescope (LAMOST; \citeauthor{Zhao2012} \citeyear{Zhao2012}; \citeauthor{Cui2012} \citeyear{Cui2012}; \citeauthor{Luo2015} \citeyear{Luo2015}), the Galactic Archaeology with HERMES (GALAH; \citeauthor{DeSilva2015} \citeyear{DeSilva2015}), the Sloan Digital Sky Survey (SDSS; \citeauthor{Yanny2009} \citeyear{Yanny2009}; \citeauthor{Majewski2017} \citeyear{Majewski2017}; \citeauthor{Almeida2023} \citeyear{Almeida2023}), the Dark Energy Spectroscopic Instrument (DESI; \citeauthor{DESICollaboration2025} \citeyear{DESICollaboration2025}), Gaia Radial Velocity Spectrometer \citep{GaiaCollaboration2023,RecioBlanco2023}, the 4.2 m William Herschel Telescope Enhanced Area Velocity Explorer \citep{Jin2024}, the Multi-Object Optical and Near-infrared Spectrograph \citep{Cirasuolo2020}, the upcoming the 4 m Multi-Object Spectroscopic Telescope \citep{Jong2022}, and the Chinese Space Station Telescope (CSST; \citealt{Gong2019,CSSTCollaboration2025})---has provided, or is expected to provide, an unprecedented observational foundation for studies of stellar physics and Galactic structure.
	
	These datasets enable the measurement of key stellar parameters, such as radial velocity ($\mathrm{RV}$), effective temperature ($T_{\mathrm{eff}}$), surface gravity ($\log g$), metallicity ($[\mathrm{Fe}/\mathrm{H}]$), and dozens of elemental abundances. To efficiently and reliably infer these parameters, various automated pipelines have been developed, including the APOGEE Atmospheric Parameter and Chemical Abundance Pipeline (ASPCAP; \citealt{GarciaPerez2016}), the Cannon for GALAH DR1/DR2 \citep{Ness2015,Martell2017,Buder2018}, the Payne for GALAH DR4 \citep{Ting2019,Buder2025}, the General Stellar Parametriser from spectroscopy in Gaia DR3 \citep{RecioBlanco2023}, the RVS and SP pipelines in DESI \citep{Cooper2023,Koposov2024,Koposov2025}, and the LAMOST Atmospheric Parameter Pipeline (LASP; \citealt{Luo2015}). These pipelines have become essential tools for large-scale stellar parameter estimation, offering robust and automated solutions across diverse spectroscopic surveys.
	
	However, with increasing sample sizes and expanding parameter dimensionality, existing pipelines still leave room for improvement in terms of computational efficiency and joint-modeling capabilities. Most do not systematically exploit GPU acceleration, making it challenging to process tens of millions of spectra efficiently. Moreover, except for the Payne \citep{Ting2019} and its extension TransformerPayne \citep{Rozanski2025}, atmospheric parameters and elemental abundances are often modeled independently, limiting the ability to capture interparameter correlations in high-dimensional spaces. 
	
	These limitations are particularly evident in the modular design of LASP. In the current LAMOST low-resolution survey, $\mathrm{RV}$, $T_{\mathrm{eff}}$, $\log g$, and [Fe/H] for AFGK-type stars are inferred using LASP, which incorporates the University of Lyon Spectroscopic analysis Software (ULySS\footnote{\url{http://ulyss.univ-lyon1.fr/}}), implemented in IDL and optimized via the MPFit algorithm \citep{Markwardt2009} (hereafter LASP-MPFit). In contrast, projected rotational velocity ($v \sin i$) and $\alpha$-element abundance ($[\alpha/\mathrm{M}]$) are inferred using two separate Python-based modules developed by \citet{Zuo2024} and W. Hou (2025, personal communication), which rely on LASP-MPFit outputs as strong priors. This modular separation prevents fully coupled multiparameter inference, which not only propagates prior errors across sequential modules but also limits the framework’s capacity to enforce joint constraints and capture parameter correlations, thereby potentially compromising both its accuracy and extensibility. Additionally, the IDL-based implementation is increasingly difficult to maintain and limits flexibility for adaptation to other surveys. Such architectural and methodological fragmentation is not unique to LASP; similar inconsistencies frequently arise when integrating stellar parameters derived from different survey pipelines. Differences in algorithmic design and modeling assumptions across surveys can introduce hard-to-quantify deviations in the resulting stellar parameters, complicating their cross-survey integration. Therefore, developing an efficient and readily implementable inference framework for high-dimensional joint optimization, with the flexibility to adapt to different surveys, will facilitate consistent analysis and integrated modeling of multisurvey data.
	
	Motivated by these challenges, we develop a new Python-based architecture for LASP\footnote{\url{https://github.com/LiangJunC/PyLASP}} (hereafter PyLASP) that supports efficient inference, multitask joint modeling, and cross-survey compatibility. The framework incorporates two optimization strategies: LASP-CurveFit, a CPU-based implementation of the Levenberg-Marquardt algorithm \citep{Nocedal2006,Vugrin2007}, and LASP-Adam-GPU, a GPU-accelerated method based on the Adam optimizer \citep{Kingma2014}. LASP-CurveFit preserves the logic of LASP-MPFit while restructuring the codebase for improved I/O and multithreading performance using Python. LASP-Adam-GPU performs grouped optimization by minimizing a joint residual function over multiple spectra in each group, using the PyTorch library \citep{Paszke2019} to support scalable and parallel inference for large datasets and high-dimensional parameter spaces. Thanks to the flexibility of Python and its extensive scientific ecosystem, this framework is easily extensible and transferable across surveys such as DESI and CSST. Given that LASP-MPFit forms the foundation for other parameter modules in LAMOST, its IDL-based implementation has certain limitations in computational efficiency and scalability. We therefore prioritize the reimplementation of this module, which serves as the baseline method for performance and consistency evaluation in subsequent experiments. Other modules, such as $v \sin i$ and $[\alpha/\mathrm{M}]$, adopt different modeling strategies and are not yet included in the current framework. Their integration will be explored in future work as part of a unified framework for high-dimensional abundance inference.

	This paper is organized as follows. In Section~\ref{Methodology}, we introduce the overall structure of PyLASP. Section~\ref{Data introduction} describes the datasets used to evaluate the performance of LASP-CurveFit and LASP-Adam-GPU. In Section~\ref{Data Experiment}, we present results on inference efficiency, robustness, error modeling, and generalization to DESI DR$1$. Section~\ref{Conclusions} provides an overall summary of this work.
	
	\section{Methodology} \label{Methodology}
	To support large-scale stellar parameter inference, we develop LASP-CurveFit and LASP-Adam-GPU based on LASP-MPFit. LASP-MPFit is designed to infer $\mathrm{RV}$, $T_{\mathrm{eff}}$, $\log g$, and [Fe/H] for AFGK-type stars in the LAMOST survey \citep{Wu2011,Wu2014,Luo2015}, using version $3.2$ of the ELODIE library \citep{Wu2011a} and the ULySS package implemented in IDL \citep{Koleva2009a,Koleva2009,Wu2011a}. LASP-CurveFit is designed for CPU environments, with a restructured parameter inference procedure that improves I/O efficiency and multithreading performance. In contrast, LASP-Adam-GPU introduces further optimization across multiple modules---such as data loading, spectral generation, and wavelength resampling---and adopts a grouped optimization strategy that minimizes the rms residual across multiple spectra, thereby enhancing inference throughput and enabling future extensions to multielement abundance inference.
	
	\subsection{LASP-CurveFit} \label{LASP-CurveFit-method}
	 LASP-CurveFit is primarily designed for efficient stellar parameter inference from individual spectra and is scalable to large spectroscopic datasets through parallelization with the \texttt{joblib} library \citep{Varoquaux2024}. Spectral data are read from FITS files using the \texttt{Astropy} library \citep{AstropyCollaboration2022} and organized into dictionary-based structures to facilitate rapid access during parameter inference.
	 
	 The core components of LASP-CurveFit include modules for $\chi^2$ computation, wavelength resampling, resolution degradation, and correction of shape differences between model and observed spectra. These modules retain logical consistency with LASP-MPFit. The main modifications are as follows: (1) Since LAMOST spectra are in relative flux, the observed spectra are median-normalized before fitting to standardize the flux scale and improve numerical stability, and to maintain consistency with LASP-Adam-GPU. This also facilitates potential future comparison of $\chi^2$ values across spectra. (2) All computations are implemented using the \texttt{NumPy}  library \citep{Harris2020} and \texttt{SciPy} library \citep{Virtanen2020}. (3) The optimization step replaces the MPFit algorithm \citep{Markwardt2009} with \texttt{scipy.optimize.curve\_fit}, which provides covariance-based error estimates for stellar parameters \citep{Vugrin2007}, thereby eliminating the need for the empirical error correction previously required in LASP-MPFit.
	
	\subsection{LASP-Adam-GPU}\label{LASP-Python-GPU}
	LASP-Adam-GPU is designed for efficient and scalable stellar parameter inference from large spectroscopic datasets. The current implementation targets the inference of atmospheric parameters ($T_{\mathrm{eff}}$, $\log g$, and [Fe/H]) and RV, leveraging GPU-based parallel optimization to enable high-throughput processing. 
	
	The framework consists of nine sequential stages: (1) definition of input spectral data formats; (2) generation of model spectra for arbitrary atmospheric parameters; (3)-(5) spectral processing to ensure consistency with observed spectra in wavelength, resolution, and continuum shape; (6) identification of bad pixels; (7) spectral comparison between processed models and observations; (8) iterative optimization of parameters, with repeated execution of steps (2)-(7) until convergence criteria are met; and (9) error estimation for the final parameters. The entire workflow is implemented with GPU parallelization to maximize computational efficiency. Details of each stage are described below.
	\begin{enumerate} 
		\item \textbf{Storing batch data.} 
		Spectroscopic data are read using the \texttt{Astropy} library and serialized into PyTorch tensor files (\texttt{.pt}), each containing $20{,}000$ spectra. These files support efficient memory access during inference. In each optimization cycle, up to $\texttt{N} {<} 20{,}000$ spectra are loaded as a group and jointly processed on the GPU for parallel inference of $T_{\mathrm{eff}}$, $\log g$, [Fe/H], and $\mathrm{RV}$.
		\item \textbf{Generating \texttt{N} model spectra for each group.} 
		For any set of \texttt{N} stellar atmospheric parameters $\theta_{1}, \theta_{2}, \dots, \theta_{N}$ ($\theta_{i} = {T_{\mathrm{eff}}, \log g, \mathrm{[Fe/H]}}$), we use PyTorch to generate a group of \texttt{N} model spectra:
		{\small
			\begin{equation}
				F {=}
				\begin{pmatrix}
					F_{1}\\
					\vdots\\
					F_{N}
				\end{pmatrix}
				{=}
				\begin{pmatrix}
					f_1(\theta_1, \theta_2, \dots, \theta_{m_1})\\
					f_2(\theta_{m_1+1}, \dots, \theta_{m_2})\\
					f_3(\theta_{m_2+1}, \dots, \theta_{m_3})\\
					f_4(\theta_{m_3+1}, \dots, \theta_{m_4})\\
					f_5(\theta_{m_4+1}, \dots, \theta_{N})
				\end{pmatrix}
				{=}
				\begin{pmatrix}
					T_{1},\dots, T_{q}
				\end{pmatrix},
				\label{T}
		\end{equation}}
	
		where $F_i$ is the $i$th model spectrum, $T_j$ denotes the $j$th column of $F$, corresponding to the flux values at the $j$th wavelength pixel across all \texttt{N} spectra, and $q$ is the number of flux points in the model spectrum. The functions $f_1$ through $f_5$ are used to generate spectra over different $T_{\mathrm{eff}}$ regimes---$T_{\mathrm{eff}} {\leq} 4000$ K, $4000 {<} T_{\mathrm{eff}} {\leq} 4550$ K, $4550 {<} T_{\mathrm{eff}} {\leq} 7000$ K, $7000 {<} T_{\mathrm{eff}} {\leq} 9000$ K, and $T_{\mathrm{eff}} {>} 9000$ K, respectively---and satisfy
		\begin{equation*}
			\begin{pmatrix}
				f_1(\theta_1, \theta_2, \dots, \theta_{m_1})\\
				f_2(\theta_{m_1+1}, \dots, \theta_{m_2})\\
				f_3(\theta_{m_2+1}, \dots, \theta_{m_3})\\
				f_4(\theta_{m_3+1}, \dots, \theta_{m_4})\\
				f_5(\theta_{m_4+1}, \dots, \theta_{N})
			\end{pmatrix}
			=
			\begin{pmatrix}
				f_{1}(\theta_{1}) \\
				\vdots \\
				f_{1}(\theta_{m_{1}})\\
				f_{2}(\theta_{m_{1}+1})\\
				\vdots \\
				f_{5}(\theta_{N})
			\end{pmatrix},
		\end{equation*}
		where $f_1(\theta_1), \dots, f_5(\theta_N)$ are model spectra generated using the ELODIE spectral emulator in ULySS\footnote{\url{http://ulyss.univ-lyon1.fr/uly\_tgm\_eval.html}}.
		\item \textbf{Resampling to the LAMOST linear wavelength grid.}
		We adopt a flux-conserving interpolation method (for a discussion of flux conservation, see \citealt{Carnall2017}) to resample the model spectrum $F$ (Equation~\ref{T}), originally defined on a linearly spaced wavelength grid $\lambda_{11}, \lambda_{12}, \dots, \lambda_{1q}$ with a step size of $\Delta \lambda_1$, onto the LAMOST wavelength grid:
		\begin{equation*}
			\lambda_{21}^{\prime} = e^{\lambda_{21}}, \dots, \lambda_{2p}^{\prime} = e^{\lambda_{2p}},
		\end{equation*}
		where $\lambda_{2i}$ $(i=1, \dots, p)$ denotes the logarithmically spaced wavelength sequence of LAMOST (base $e$), with a step size of $\Delta \lambda_{2}$, and $p$ is the number of flux points in the observed spectrum. The entire resampling procedure consists of four steps:
		\begin{itemize}
			\item \textbf{Constructing the wavelength integration nodes.}
			First, construct the wavelength integration nodes for the model spectrum $F$ as
			\begin{equation*}
				\lambda_{11}-\frac{\Delta \lambda_{1}}{2}, \lambda_{11}+\frac{\Delta \lambda_{1}}{2}, \dots, \lambda_{11}+\frac{(2q-1)\Delta \lambda_{1}}{2},
			\end{equation*}
			and the corresponding wavelength integration nodes for LAMOST:
			\begin{equation*}
				\lambda_{21}^{\prime\prime} = e^{\lambda_{21}-\frac{\Delta \lambda_{2}}{2}}, \dots, \lambda_{2(p+1)}^{\prime\prime} = e^{\lambda_{21}+\frac{(2p-1)\Delta \lambda_{2}}{2}}.
			\end{equation*}
			
			\item \textbf{Computing the integrated flux of the model spectrum.}
			The model spectrum $F$ is cumulatively integrated column-wise over its wavelength integration nodes to obtain
			\begin{equation}
				0,\ T_1 \Delta \lambda_1,\ \sum_{i=1}^{2} T_i \Delta \lambda_1,\ \dots,\ \sum_{i=1}^{q} T_i \Delta \lambda_1.
				\label{integrated}
			\end{equation}
			\item \textbf{Computing the integrated flux at LAMOST integration nodes.}
			The integrated flux of $F$ (Equation~\ref{integrated}) is linearly interpolated onto the LAMOST wavelength integration nodes to yield
			\begin{equation}
				\mathcal{F}_{\lambda_{21}^{\prime\prime}},\ \mathcal{F}_{\lambda_{22}^{\prime\prime}},\ \dots,\ \mathcal{F}_{\lambda_{2(p+1)}^{\prime\prime}}.
				\label{integratedLAMOST1}
			\end{equation}
			\item \textbf{Computing the resampled flux on the LAMOST linear wavelength grid.}
			The fluxes at $\lambda_{21}^{\prime}$, $\lambda_{22}^{\prime}$, \dots, $\lambda_{2p}^{\prime}$ are obtained by differencing the integrated values from \refformula{integratedLAMOST1}. The \texttt{N} model spectra obtained through batch resampling are denoted as
			\begin{equation}
				F' 	=
				\begin{pmatrix}
					F_1^{\prime} \\
					F_2^{\prime} \\
					\vdots \\
					F_N^{\prime}
				\end{pmatrix}
				= 
				\begin{pmatrix}
					T_1^{\prime},\ T_2^{\prime},\ \ldots,\ T_p^{\prime}
				\end{pmatrix},
				\label{integratedLAMOST}
			\end{equation}
			where
			\begin{equation*}
				T_i^{\prime} = \frac{\mathcal{F}_{\lambda_{2(i+1)}^{\prime\prime}} - \mathcal{F}_{\lambda_{2i}^{\prime\prime}}}{\lambda_{2(i+1)}^{\prime\prime} - \lambda_{2i}^{\prime\prime}},
			\end{equation*}
			and $F_{i}^{\prime}$ denotes the resampled model spectrum corresponding to the original $F_{i}$ (Equation~\ref{T}).
		\end{itemize}
		\item \textbf{Convolving model spectra with a Gaussian broadening kernel to match the resolution of the observed spectra.}
		To match the resolution of the LAMOST spectra, we convolve the resampled model spectra $F^{\prime}$ (Equation~\ref{integratedLAMOST}) using \texttt{torch.nn.functional.conv1d} in batches:
		\begin{equation}
			F^{\prime\prime} = 
		\begin{pmatrix}
			F^{\prime\prime}_{1}\\
			F^{\prime\prime}_{2}\\
			\vdots\\
			F^{\prime\prime}_{N}
		\end{pmatrix}
		=
		\begin{pmatrix}
			F_{1}^{\prime} \otimes G(\mu_{1}, \sigma_{1})\\
			F_{2}^{\prime} \otimes G(\mu_{2}, \sigma_{2})\\
			\vdots\\
			F_{N}^{\prime} \otimes G(\mu_{N}, \sigma_{N})
		\end{pmatrix},
		\label{integratedLAMOSTCON1D}
		\end{equation}
		where $G(\mu_i, \sigma_i)$ is a Gaussian kernel of shape $1 \times (2 \cdot \lceil |\mu_i| + 5\sigma_i \rceil + 1)$, and $F^{\prime\prime}_{i}$ denotes the spectrum obtained by convolving $F^{\prime}_{i}$ with this kernel. The kernel parameters $\mu_i$ and $\sigma_i$ are specific to each spectrum and are introduced to accommodate varying resolution and RV offsets across the LAMOST spectra. In particular, $\mu_i$ is used to model the $\mathrm{RV}$ of the $i$th observed spectrum~\citep{VanDerMarel1993,Koleva2009,Wu2011a,Wu2011,Prugniel2011,Koleva2012,Luo2015,Sharma2016,Cappellari2016,Kumar2025}, and $\sigma_{i}$ encompasses both the instrumental broadening and the effects of rotation \citep{Luo2015}. In the subsequent optimization process, all kernel parameters (a total of $2\texttt{N}$) are jointly inferred with the stellar atmospheric parameters (a total of $3\texttt{N}$).
		\item \textbf{Correcting shape differences between model and LAMOST spectra.}
		To correct the shape differences between the model spectra $F^{\prime\prime}$ (Equation~\ref{integratedLAMOSTCON1D}) and the LAMOST spectra---which arise from flux calibration, Galactic extinction, and other effects that alter the spectral shape \citep{Luo2015}---we apply a multiplicative correction factor $P(x)b_{i}$ to each $F^{\prime\prime}_i$, yielding:
		\begin{equation}
		\begingroup
		\renewcommand{\arraystretch}{1.5}
		\begin{pmatrix}
			\dfrac{L_1}{w_1} \\
			\dfrac{L_2}{w_2} \\
			\vdots \\
			\dfrac{L_N}{w_N}
		\end{pmatrix}
		\endgroup
		=
		\begin{pmatrix}
			(P(x)b_{1})^{T} \odot F^{\prime\prime}_{1} \\
			(P(x)b_{2})^{T} \odot F^{\prime\prime}_{2} \\
			\vdots \\
			(P(x)b_{N})^{T} \odot F^{\prime\prime}_{N}
		\end{pmatrix},
		\label{continuum}
		\end{equation}
		where $\odot$ denotes the Hadamard product. $L_i$ is the $i$th LAMOST spectrum (a $1 \times p$ row vector), and $w_i$ is a weighting factor applied to $L_i$ to reduce the flux differences among the \texttt{N} LAMOST spectra (see \refsection{Data Preprocessing} for details). The matrix $P(x) \in \mathbb{R}^{p \times 51}$ contains the values of Legendre basis functions of degrees $0$ through $50$, evaluated at evenly spaced nodes,
		\[
		x_j = -1 + \frac{2(j - 1)}{p}, \quad j = 1, \dots, p,
		\]
		using \texttt{scipy.special.eval\_legendre}. The coefficient vector $b_i \in \mathbb{R}^{51 \times 1}$ is obtained by solving Equation~\ref{continuum} using weighted least squares~\citep{Lawson1995}.
		\item \textbf{Masking Outlier Pixels.}
		Based on the flux residuals between the observed and model spectra,
		\begin{equation}
			\epsilon_i = \dfrac{L_i}{w_i} - (P(x)b_{i})^{T} \odot F^{\prime\prime}_{i}=\dfrac{L_i}{w_i} -  F^{\prime\prime\prime}_{i},
			\label{wucha}
		\end{equation}
		we use a masking vector $A_i$ to mitigate the impact of anomalous pixels on parameter inference. Like LASP-MPFit, LASP-Adam-GPU supports two masking strategies: No Clean and Clean. The No Clean strategy masks only predefined problematic regions (e.g., the Na D absorption lines), while the Clean strategy incorporates the clipping algorithm\footnote{\url{http://ulyss.univ-lyon1.fr/uly\_fit.html}} that iteratively identifies and masks outlier pixels through a three-step filtering process. To enable large-scale spectroscopic inference, we reimplement the Clean strategy from LASP-MPFit using PyTorch. While preserving the original masking logic, this implementation is restructured to support efficient spectrum-wise parallel execution across \texttt{N} spectra, making it suitable for GPU-accelerated workflows. It consists of the following three steps:
		\begin{itemize}
			\item \textbf{Compute the primary outlier mask $B_i$.} For the $j$th pixel of the $i$th spectrum, if
			\begin{equation*}
				|\epsilon_{i,j}| - f_{\mathrm{shift}} \cdot G_{i,j} > k \cdot \sigma_{\epsilon_i},
			\end{equation*}
			then $B_{i,j} {=} 0$, otherwise $B_{i,j} {=} 1$. Here, $\sigma_{\epsilon_i}$ is the standard deviation of $\epsilon_i$ computed from the unmasked pixels, and $G_{i,j} = \max(|F^{\prime\prime\prime}_{i,j} - F^{\prime\prime\prime}_{i,j-1}|, |F^{\prime\prime\prime}_{i,j} - F^{\prime\prime\prime}_{i,j+1}|)$ is the flux gradient at pixel $j$, used to assess whether $\epsilon_{i,j}$ can be attributed to minor wavelength shifts. The factor $f_{\mathrm{shift}}$ controls the comparison scale between the residual and the flux gradient. Following ULySS conventions, $f_{\mathrm{shift}}$ is set to $0.5$ in the first iteration and reduced to $0.2$ in subsequent iterations to implement a progressive filtering strategy. The threshold $k$ starts from $3$ and is relaxed to $k \in \{4, 5, 7\}$ if the outlier fraction exceeds $3\%$. This step targets sharp outliers that cannot be explained by the model, such as sky residuals, observational defects, or cosmic rays.
			\item \textbf{Compute the adjacent outlier mask $C_i$.} For each pixel $j$ already marked as an outlier in $B_i$, its neighboring pixels $j' \in \{j-1, j+1\}$ are examined. If
			\begin{equation*}
				|\epsilon_{i,j'}| - f_{\mathrm{shift}} \cdot G_{i,j'} > 2 \cdot \sigma_{\epsilon_i},
			\end{equation*}
			then $C_{i,j'} {=} 0$, otherwise $C_{i,j'} {=} 1$.  This step aims to eliminate secondary outliers that are adjacent to primary outliers but exhibit slightly smaller residuals, helping to suppress small-scale structures such as residual sky features.
			\item \textbf{Compute the neighborhood outlier mask $D_i$.} For pixels $j$ identified as an outlier in $B_i$ or $C_i$, their neighbors $j' \in \{j-1, j+1\}$ are checked. If
			\[
			\begin{cases}
				\epsilon_{i,j} \cdot \epsilon_{i,j'} > 0, \\
				|\epsilon_{i,j'}| \leq |\epsilon_{i,j}|, \\
				|\epsilon_{i,j'}| - f_{\mathrm{shift}} \cdot G_{i,j'} > \sigma_{\epsilon_i}^{\prime},
			\end{cases}
			\]
			then $D_{i,j'} {=} 0$, otherwise $D_{i,j'} {=} 1$. Here, $\sigma_{\epsilon_i}^{\prime}$ is the standard deviation of $\epsilon_i$ computed from the unmasked pixels. This step is repeated until the mask converges, i.e., $D_i {=} D_{i+1}$, with a maximum of $20$ iterations. This step further eliminates boundary pixels near primary outliers, allowing effective removal of continuous spectral anomalies that cannot be fitted by the model.
		\end{itemize}
		
		After the above three steps, the final mask matrix for the \texttt{N} spectra is defined as
		\begin{equation}
			A=\begin{pmatrix}
				A_{1} \\
				A_{2} \\
				\vdots \\
				A_{N}
			\end{pmatrix}
			=
			\begin{pmatrix}
				B_{1} \odot C_{1} \odot D_{1} \\
				B_{2} \odot C_{2} \odot D_{2} \\
				\vdots \\
				B_{N} \odot C_{N} \odot D_{N}
			\end{pmatrix}.
			\label{MASKA}
		\end{equation}
		\item \textbf{Constructing the objective function.} 
		To quantify the discrepancy between the \texttt{N} LAMOST spectra and the corresponding model spectra $F^{\prime\prime}$, we define the objective function as
		\begin{equation}
			\mathcal{L} = \sqrt{ \frac{1}{p N} \sum_{i=1}^{N} (A_i \odot \epsilon_i) (A_i \odot \epsilon_i)^{T} },
			\label{loss}
		\end{equation}
		where $\epsilon_i$ and $A_i$ are defined by Equations~(\ref{wucha}) and (\ref{MASKA}), respectively, representing the flux residuals and the pixel masking vector for the $i$th spectrum. To mitigate potential numerical instabilities introduced by wavelength resampling and multiplicative correction near the spectrum edges, edge pixels are excluded from the objective function by default.
		\item \textbf{Minimizing the objective function.}
		We adopt the Adam optimizer~\citep{Kingma2014} to minimize the objective function defined in \refformula{loss}, using the initial parameter values provided by the correlation function interpolation (CFI) method~\citep{Du2012}, which is consistent with the initialization strategy used in LASP-MPFit. To avoid distortion of the optimization trajectory caused by differences in parameter scales, we follow the variable rescaling strategy proposed by \citet{Nocedal2006} and normalize all free parameters to the range $[-1, 1]$. The learning rate is set to $0.1$, with all other Adam parameters left at their default values. The optimization is considered converged when any of the following three criteria is satisfied: (1) a maximum of $5000$ iterations is reached; (2) the change in loss over $50$ consecutive steps is less than $10^{-5}$; or (3) the loss increases monotonically over $50$ consecutive steps. This configuration of the learning rate and convergence threshold is designed to balance convergence stability and computational efficiency, and is examined in detail in \refsection{Sensitivity}. Upon convergence, the resulting $5\texttt{N}$ parameters are taken as the minimizer of \refformula{loss}. To ensure consistent mask handling across \texttt{N} spectra, the Clean strategy updates $A_i$ (Equation~\ref{MASKA}) every $30$ steps based on residuals, allowing up to $11$ mask updates during the optimization.
		\item \textbf{Estimating parameter errors.}
		Once the minimizer of \refformula{loss} is obtained, we estimate the parameter errors by computing the Jacobian matrix $J_i$ of the residual vector $\epsilon_i$ (Equation~\ref{wucha}) at the solution point, using central finite differences. The covariance matrix $\Sigma_{i}$ is then approximated via linear error propagation~\citep{Vugrin2007,Drosg2009} as $\Sigma_{i} = \frac{\epsilon_i\epsilon_i^{T}}{p-5}\left( J_i^{T} J_i \right)^{-1}$, where $p - 5$ is the number of degrees of freedom. Parameter errors are given by the square roots of the diagonal elements of $\Sigma_{i}$, and computed in parallel for all \texttt{N} spectra using PyTorch.
	\end{enumerate}
	
	Beyond the current implementation, the modular architecture of LASP-Adam-GPU enables the possibility of future extension to the joint modeling of multidimensional elemental abundances. Such extensions could follow two main paths: (1) integrating an abundance module into the optimization loop via a multitask scheme to simultaneously infer multiple parameters---a strategy that has been widely adopted in deep learning tasks such as computer vision, natural language processing, and speech recognition \citep{Collobert2008,Huang2013,Kokkinos2017,Cipolla2018}---or (2) incorporating a high-dimensional spectral emulator, inspired by The Payne \citep{Ting2019}, trained on theoretical spectra to model all target parameters in a unified manner. Once developed, the new modules can be directly applied to different spectroscopic surveys, with resolution matching and related adjustments handled dynamically during inference---without retraining.
	
	\section{Data} \label{Data introduction}
	LASP-MPFit performs parameter inference for low-resolution AFGK-type stellar spectra from LAMOST using a spectral emulator constructed from the ELODIE spectral library~\citep{Luo2015}. To evaluate the consistency between the Python and IDL versions of LASP, we use a crossmatched dataset of common stars observed by both LAMOST DR$10$ and APOGEE DR$16$ as the test sample.
	
	\subsection{ELODIE library}
	The ELODIE library~\citep{Prugniel2004,Moultaka2004,Prugniel2007} is based on echelle spectra taken with the eponym spectrograph attached to the $1.93$ m telescope of Observatoire de Haute-Provence. It contains $1962$ spectra of $1388$ stars, covering the wavelength range $3900{-}6800$\,\AA{} at a resolving power of $R {\approx} 42{,}000$. Version 3.2 of the library spans a wide range of atmospheric parameters: $T_{\mathrm{eff}}$ from 3100\,K to 59{,}000\,K, $\log g$ from 0.0 to 5.0\,dex, and $\mathrm{[Fe/H]}$ from $-2.8$ to $+1.0$\,dex~\citep{Wu2011,Luo2015}.
	
	To improve interpolation performance near the edges and sparsely populated regions of parameter space, ELODIE incorporates a set of \lq semiempirical\rq \ spectra constructed by combining observed and synthetic spectra~\citep{Wu2011a}. This enhances the emulator's extrapolation capability. The spectral emulator used in LASP is built on the ELODIE spectra degraded to a resolution of $R {\approx} 10{,}000$ \citep{Wu2011}, and uses separate high-order polynomial regression models for three stellar temperature classes, following the ULySS convention \citep{Koleva2009}: hot ($T_{\rm eff} {\ge} 7000$\,K), warm ($4000 {<} T_{\rm eff} {\le} 9000$\,K), and cold ($T_{\rm eff} {\le} 4550$\,K). Each model maps three atmospheric parameters to $14{,}501$ flux points. The hot, warm, and cold models contain $19$, $26$, and $25$ nonzero polynomial coefficients, respectively.
	
	\begin{table*}[tp]
		\caption{Configuration ranges of efficiency-related parameters for LASP modules on different devices.}
		\label{laspdeviceconfig}
		\renewcommand{\arraystretch}{1.2}
		\begin{tabular*}{\textwidth}{@{\extracolsep{\fill}}ccccc@{}}
			\toprule
			\multirow{2}{*}{\centering Device \textsuperscript{a}} & LASP-MPFit & LASP-CurveFit & LASP-Adam-CPU & LASP-Adam-GPU \\
			\cmidrule{2-3} \cmidrule{4-5}
			& \multicolumn{2}{c}{\texttt{n\_jobs}\textsuperscript{b}} & \multicolumn{2}{c}{\texttt{N}\textsuperscript{c}} \\
			\midrule
			Ryzen 9 7945HX
			& {[1, 32]} 
			& {[1, 32]} 
			& $\begin{array}{c}
				\{d \mid d = 1 \vee (10 \mid d, \\
				d \leq 100) \vee (100 \mid d,
				d \leq 1000) \}
			\end{array}$
			& $\begin{array}{c}
				\{d \mid d = 1 \vee (d \mid 5000, 10 \mid d) \}
			\end{array}$ \\
			
			Xeon Silver 4214
			& --- \textsuperscript{d}
			& {[1, 48]} 
			& $\begin{array}{c}
				\{d \mid d = 1 \vee (10 \mid d, \\
				d \leq 100) \vee (100 \mid d,
				d \leq 1000) \}
			\end{array}$
			& --- \\
			
			RTX 3090
			& --- 
			& --- 
			& --- 
			& $\begin{array}{c}
				\{d \mid d = 1 \vee (d \mid 10000, 10 \mid d) \}
			\end{array}$ \\
			
			A100
			& --- 
			& --- 
			& --- 
			& $\begin{array}{c}
				\{d \mid d = 1 \vee (d \mid 10000, 10 \mid d) \}
			\end{array}$ \\
			\bottomrule
		\end{tabular*}
		
		\raggedright
		\noindent
		\textbf{Notes.}
		
		\vspace{0.2em}
		\noindent
		\parbox{\textwidth}{
			\textsuperscript{a} Device: Ryzen 9 7945HX with RTX 4060 is a 16-core/32-thread AMD laptop processor; Xeon Silver 4214 is a 24-core/48-thread Intel server processor; and RTX 3090 and A100 are NVIDIA GPUs for desktop and data center applications, respectively.\\
			\textsuperscript{b} \texttt{n\_jobs} represents the number of CPU multiprocessing processes. \\
			\textsuperscript{c} \texttt{N} represents the number of spectra in \refformula{loss}.\\
			\textsuperscript{d} \lq ---\rq\ indicates configurations not tested: LASP-MPFit requires licensed IDL software installed only on specific platforms, LASP-Adam-GPU was benchmarked exclusively on dedicated GPU accelerators (RTX 4060, RTX 3090 and A100), and CPU-optimized modules were evaluated on Xeon blade servers lacking high-performance GPUs.}\\
	\end{table*}
	\begin{figure*}[!htbp]
		\centering
		\subfigure{
			\begin{minipage}[t]{0.31\linewidth}
				\centering
				\includegraphics[width=2.2in, height=1.673in]{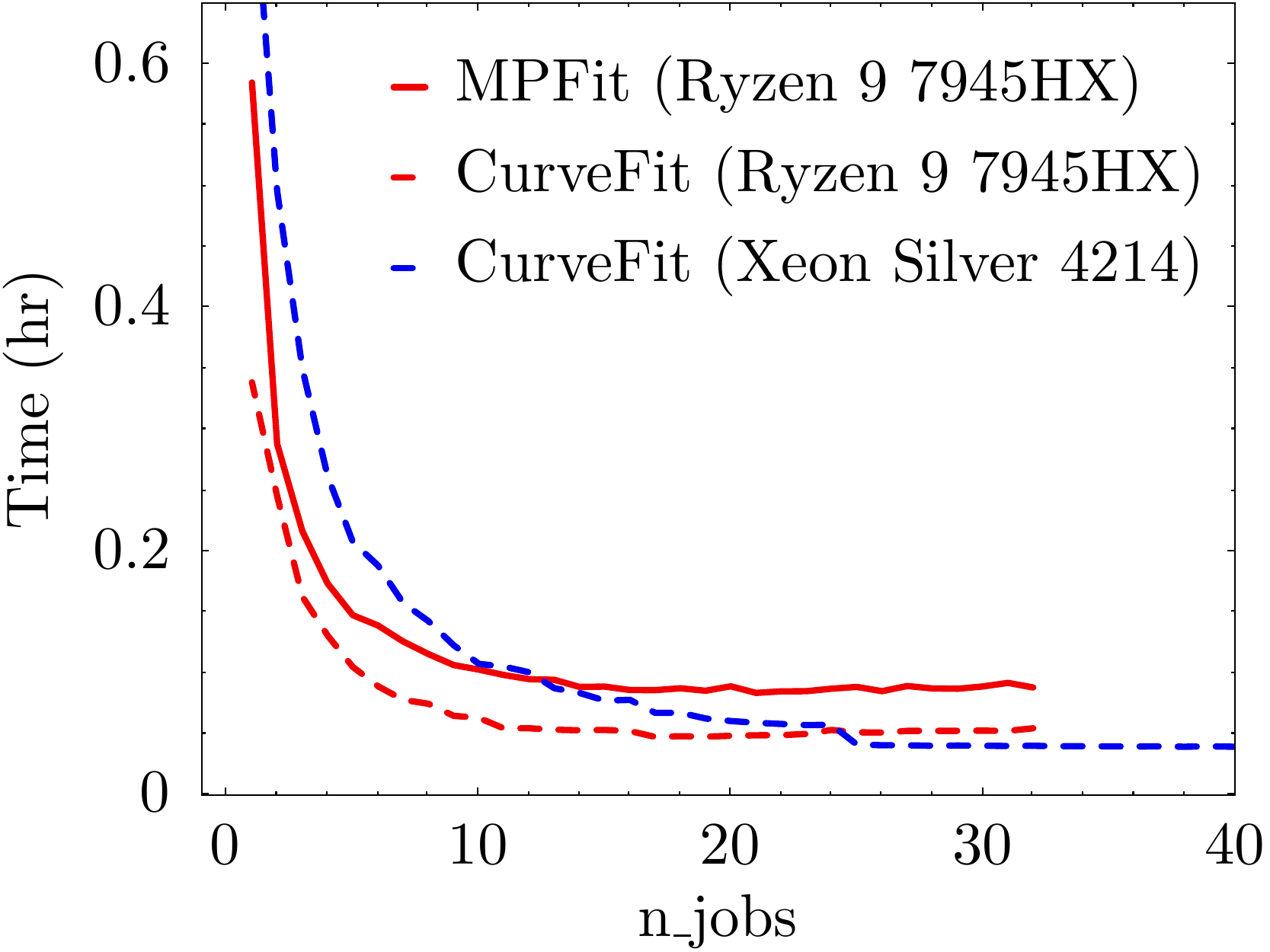}
				\centering
			\end{minipage}
		}
		\centering
		\subfigure{
			\begin{minipage}[t]{0.31\linewidth}
				\centering
				\includegraphics[width=2.2in, height=1.673in]{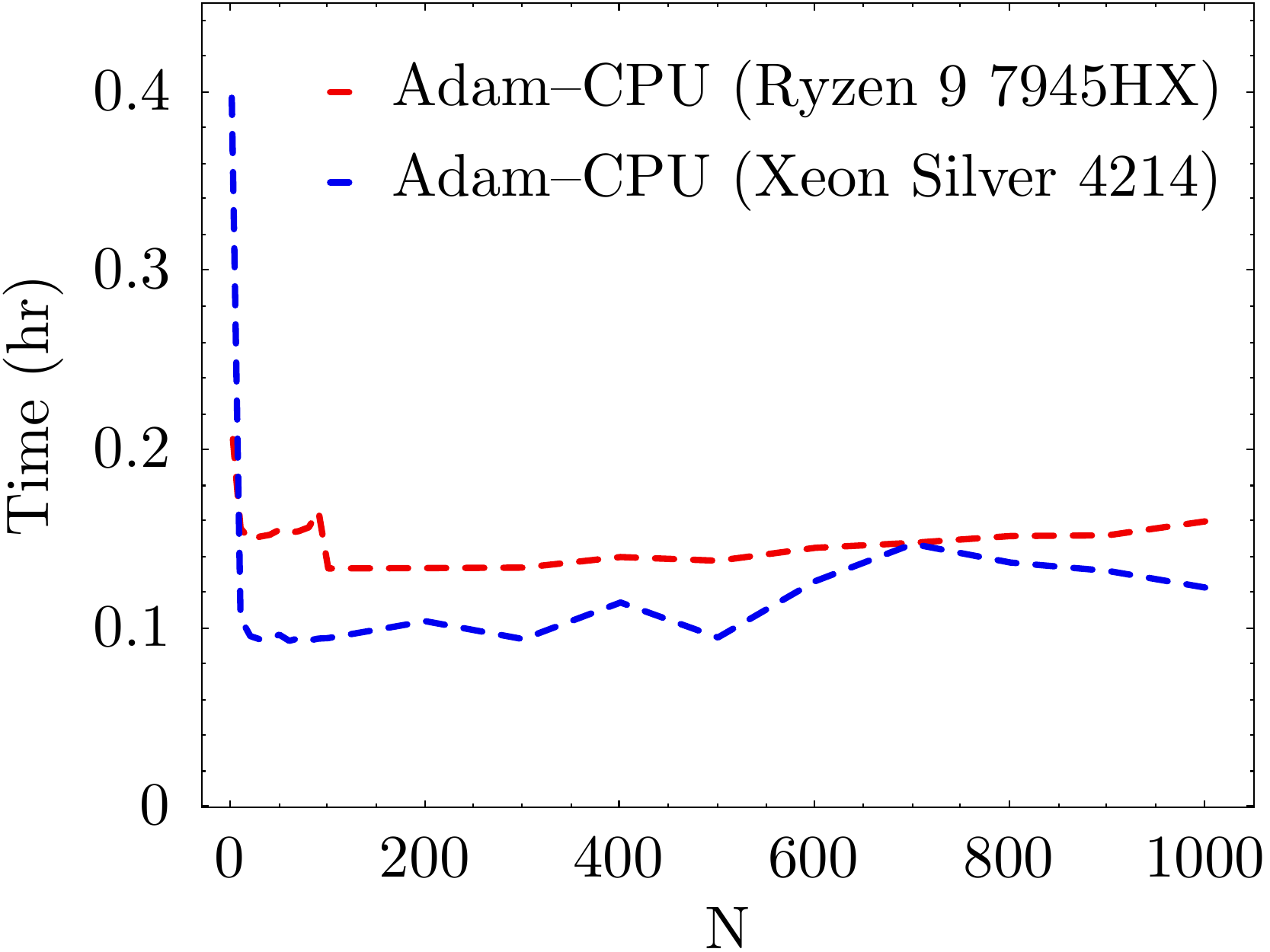}
				\centering
			\end{minipage}
		}
		\centering
		\subfigure{
			\begin{minipage}[t]{0.31\linewidth}
				\centering
				\includegraphics[width=2.2in, height=1.673in]{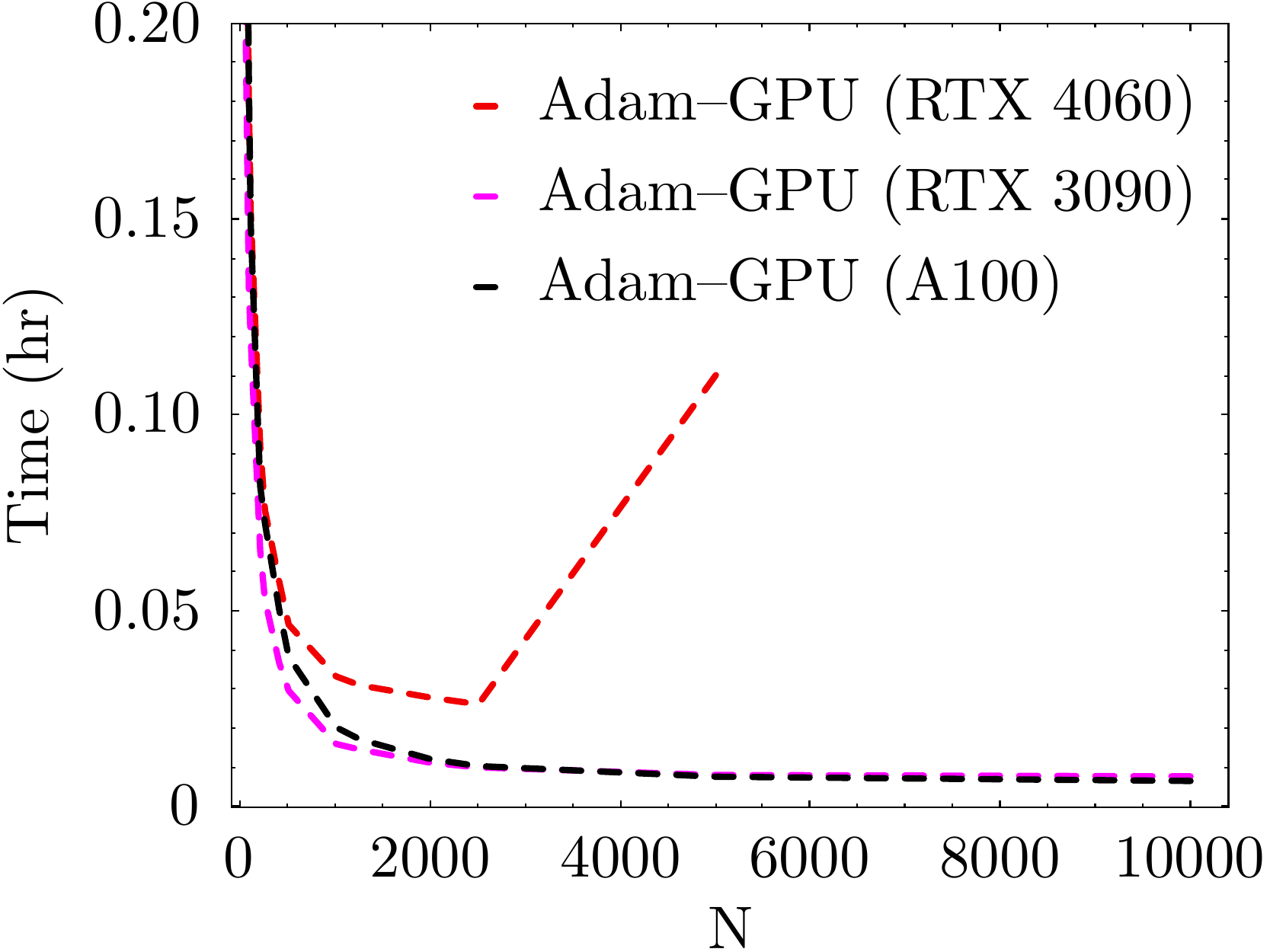}
				\centering
			\end{minipage}
		}
		\caption{Comparison of the computational efficiency of different LASP versions across hardware platforms. The left panel shows the runtime of LASP-MPFit and LASP-CurveFit for processing $10{,}000$ spectra under various efficiency parameter settings. The middle and right panels present the runtime of LASP-Adam on CPU and GPU platforms, respectively. The solid line indicates the IDL-based implementation, while dashed lines correspond to the Python-based version; line colors distinguish different hardware platforms. RTX 4060 refers to a mobile GPU integrated into a laptop with a Ryzen 9 7945HX processor. For clarity, the \lq LASP\rq\ prefix has been omitted from all legend labels in the figure.}
		\centering
		\label{IDL-Python-Time}
	\end{figure*}
	
	\subsection{Survey data and reference labels} \label{SurveyData}
	\subsubsection{The low-resolution spectra of LAMOST DR10} \label{LRS}
	LAMOST is a Schmidt telescope located at the Xinglong Observatory northeast of Beijing, China. It is capable of simultaneously obtaining $4000$ low-resolution ($R{\sim}1800$) spectra in a single exposure, covering the wavelength range of  $3700{-}9000$\,\AA{}, with blue and red arms that overlap in the $5700{-}5900$\,\AA{} region~\citep{Zhao2012,Cui2012,Luo2015}.
	
	In the $10$th data release (DR$10$ v$1.0$)\footnote{\url{https://www.lamost.org/dr10/v1.0/}}, a total of $11{,}817{,}430$ low-resolution spectra were released, of which $11{,}473{,}644$ were classified as stellar spectra. The remainder include $263{,}444$ galaxy spectra and $80{,}342$ quasar spectra. LASP-MPFit has been applied to $7{,}478{,}650$ AFGK-type stellar spectra in DR$10$ v$1.0$, yielding inferred values of $T_{\mathrm{eff}}$, $\log g$, [Fe/H], and $\mathrm{RV}$.
	
	\subsubsection{APOGEE Labels} \label{APOGEELabels}
	APOGEE is one of the programs in SDSS-III~\citep{Majewski2017} and SDSS-IV~\citep{Blanton2017}. It employs a high-resolution ($R {\sim} 22{,}500$) near-infrared spectrograph covering the wavelength range $1.51{-}1.70$\,$\mu$m, targeting primarily red giant stars in the Milky Way, as well as stars in the Large Magellanic Cloud, nearby dwarf galaxies, and a substantial number of cool dwarfs (FGKM types)~\citep{Smith2021}.
	
	In its $16$th data release (DR$16$), APOGEE provided $\mathrm{RV}$, atmospheric parameters, and up to $26$ elemental abundances for approximately $430{,}000$ stars~\citep{Joensson2020}. RV is determined via cross correlation, and ASPCAP derives stellar parameters with FERRE using precomputed, MARCS-based synthetic spectral grids that are compressed with principal component analysis (PCA); interpolation is performed in PCA coefficient space, with radial basis functions bridging gaps in these grids. The pipeline first determines the global parameters---$T_{\mathrm{eff}}$, $\log g$, overall metallicity [M/H], [$\alpha$/M], [C/M], and [N/M]---together with microturbulent velocity and $v \sin i$ for dwarfs and macroturbulent velocity for giants; then, holding these fixed, it infers individual elemental abundances from element-specific spectral windows. Additionally, APOGEE employs independent external methods to calibrate the spectroscopic measurements: $T_{\mathrm{eff}}$ is calibrated using the infrared flux method; $\log g$ for giants is calibrated using asteroseismic data from Kepler field stars, while dwarf stars use asteroseismic values for warmer stars and isochrone-derived calibrations for cooler stars; and elemental abundances are calibrated through zero-point shifts to ensure solar neighborhood stars with solar [M/H] have mean [X/M]=0. Owing to its high precision and stability, the APOGEE catalog has been widely used as a training reference for label transfer methods \citep{Li2021,Wang2022,Liang2022}. In this study, we adopt $T_{\mathrm{eff}}$, $\log g$, [Fe/H], and $\mathrm{RV}$ from APOGEE DR$16$ as external references to evaluate the consistency between the Python and IDL versions of LASP.
	
	\subsection{Data Preprocessing} \label{Data Preprocessing}
	We crossmatch the LAMOST DR$10$ AFGK-type stellar parameter catalog with APOGEE DR$16$, and select LAMOST spectra that have both APOGEE labels and CFI initial values, yielding a sample of $177{,}848$ spectra. The restriction to spectra with CFI initialization is intended to systematically assess the sensitivity of PyLASP to different starting conditions (see \refsubsubsection{Sensitivity}). Before applying LASP-Adam-GPU for parameter inference, the following preprocessing steps are performed:
	\begin{enumerate} 
		\item \textbf{Wavelength system conversion.}
		To ensure consistency in the wavelength reference system between LAMOST spectra and the ELODIE spectral library, we convert the LAMOST vacuum wavelengths to air wavelengths prior to parameter inference. Since LASP-MPFit performs spectral fitting in the wavelength range $4200{-}5700$\,\AA{} during the first stage\footnote{LASP uses a two-stage inference~\citep{Luo2015}: $T_{\mathrm{eff}}$, $\log g$, [Fe/H], and $\mathrm{RV}$ are first inferred from the original spectrum, then reinferred after continuum correction. The final values are adopted if both  inferences agree within a set threshold.}, the conversion is applied only within this range. Each spectrum is truncated to its $1327$ flux points within this range (the maximum number of pixels), and saved together with the corresponding wavelengths, the CFI initial values, the wavelength ranges before and after resampling (see step~3 in \refsubsection{LASP-Python-GPU}), and the polynomial coefficients used for generating the model spectra, in a single \texttt{.pt} file.
		\item \textbf{Setting spectral weighting factors.} 
		The fluxes of different LAMOST spectra exhibit differences in scale, which may affect the parameter-inference accuracy of LASP-Adam-GPU. To alleviate this, we assign each spectrum an empirical weight factor \(w_i\) (used in Equation~\ref{wucha}), defined as the median ($0.5$ quantile) of the flux values. This factor depends only on survey data quality, is fixed during optimization, and is not treated as a free parameter. We find that this setting performs well on LAMOST data in this work because it reduces intergroup numerical differences in LASP-Adam-GPU and decreases cases where unstable weight factors cause normalized flux values to become very large or near zero. In practice, once the weight factor is well behaved (i.e., not near-zero or excessively large), the specific choice between $0.5$ and $0.75$ quantiles plays a secondary role for the inferred parameters; we therefore treat \(q\) as a fixed value rather than a tunable hyperparameter. However, when applying LASP-Adam-GPU to DESI spectra, the $0.5$ quantile can cause the normalized flux of some spectra to become very large, whereas using the $0.75$ quantile yields better results. Therefore, we recommend testing different quantiles (e.g., $0.5$, $0.75$) when adapting the method to other spectroscopic surveys.
	\end{enumerate} 
	
	\begin{table*}
		\centering
		\caption{Recommendation and optimal efficiency parameters for LASP modules across devices.}
		\label{laspdeviceconfigBest}
		\renewcommand{\arraystretch}{1.2}
		\begin{tabular*}{\textwidth}{@{\extracolsep{\fill}}lcccccccc@{}}
			\toprule
			\multirow{3}{*}{Device} & \multicolumn{4}{c}{\texttt{n\_jobs}} & \multicolumn{4}{c}{\texttt{N}} \\
			\cmidrule{2-5} \cmidrule{6-9}
			& \multicolumn{2}{c}{LASP-MPFit} & \multicolumn{2}{c}{LASP-CurveFit} & \multicolumn{2}{c}{LASP-Adam-CPU} & \multicolumn{2}{c}{LASP-Adam-GPU} \\
			\cmidrule{2-3} \cmidrule{4-5} \cmidrule{6-7} \cmidrule{8-9}
			& RecRange\textsuperscript{a} & OptVal\textsuperscript{b} & RecRange\textsuperscript{a} & OptVal\textsuperscript{b} & RecRange\textsuperscript{a} & OptVal\textsuperscript{b} & RecRange\textsuperscript{a} & OptVal\textsuperscript{b} \\
			\midrule
			Ryzen 9 7945HX
			& [10, 32] & 21
			& [10, 32] & 17
			& [100, 700] & 100 
			& [1000, 2500] & 2500
			\\
			\addlinespace[3pt]        
			Xeon Silver 4214
			& ---  & --- 
			& [24, 48] & 47
			& [20, 500] & 60
			& ---  & ---  \\
			\addlinespace[3pt]
			RTX 3090
			& ---  & --- 
			& ---  & --- 
			& ---  & --- 
			& [2000, 10,000] & 10,000 \\
			\addlinespace[3pt]
			A100
			& ---  & --- 
			& ---  & --- 
			& ---  & --- 
			& [2000, 10,000] & 10,000 \\
			\bottomrule
		\end{tabular*}
		
		\vspace{0.5em}
		\raggedright
		\noindent
		\textbf{Notes.}
		
		\vspace{0.2em}
		\small
		\noindent
		\textsuperscript{a} RecRange (recommended range): the recommended range of the efficiency parameters; \lq---\rq \ indicates that the configuration is not applicable.\\
		\textsuperscript{b} OptVal (optimal value): the optimal value of the efficiency parameters determined in this study.
	\end{table*}
	\begin{table*}
		\centering
		\caption{Sample count statistics of common targets between the Python and IDL versions of LASP.}
		\label{nsample}
		\renewcommand{\arraystretch}{1.2}
		\begin{tabular*}{\textwidth}{@{\extracolsep{\fill}}lcccccccc@{}}
			\toprule
			\multirow{3}{*}{Stellar parameters} & \multicolumn{4}{c}{No \ Clean \ (vs \ LASP-MPFit)} & \multicolumn{4}{c}{Clean \ (vs \ LASP-MPFit)} \\
			\cmidrule{2-5} \cmidrule{6-9}
			& \multicolumn{2}{c}{LASP-CurveFit} & \multicolumn{2}{c}{LASP-Adam-GPU} & \multicolumn{2}{c}{LASP-CurveFit} & \multicolumn{2}{c}{LASP-Adam-GPU} \\
			\cmidrule{2-3} \cmidrule{4-5} \cmidrule{6-7} \cmidrule{8-9}
			& Matched\textsuperscript{a} & Outlier\textsuperscript{b} & Matched\textsuperscript{a} & Outlier\textsuperscript{b} & Matched\textsuperscript{a} & Outlier\textsuperscript{b} & Matched\textsuperscript{a} & Outlier\textsuperscript{b} \\
			\midrule
			$\mathrm{RV}$ & 171{,}713 & 32 & 172{,}246 & 78 & 171{,}618 & 234 & 172{,}238 & 60 \\
			\addlinespace[3pt]        
			$T_{\mathrm{eff}}$ & 171{,}713 & 63 & 172{,}246 & 205 & 171{,}618 & 27 & 172{,}238 & 113 \\
			\addlinespace[3pt]
			$\log g$ & 171{,}713 & 88 & 172{,}246 & 294 & 171{,}618 & 417 & 172{,}238 & 601 \\
			\addlinespace[3pt]
			$\mathrm{[Fe/H]}$ & 171{,}713 & 73 & 172{,}246 & 277 & 171{,}618 & 228 & 172{,}238 & 465 \\
			\bottomrule
		\end{tabular*}
		
		\vspace{0.5em}
		\raggedright
		\noindent
		\textbf{Notes.}
		
		\vspace{0.2em}
		\small
		\noindent
		\parbox{\textwidth}{
		\textsuperscript{a} Number of matched samples for which both PyLASP and LASP-MPFit successfully infer the parameter, before outlier removal. \\
		\textsuperscript{b} Number of outliers, defined as matched samples where the parameter difference between PyLASP and LASP-MPFit exceeds $5$ standard deviations.}
	\end{table*}
	
	\begin{figure*}[!htbp]
		\centering
		\subfigure{\includegraphics[width=0.25\textwidth, height=0.19\textwidth]{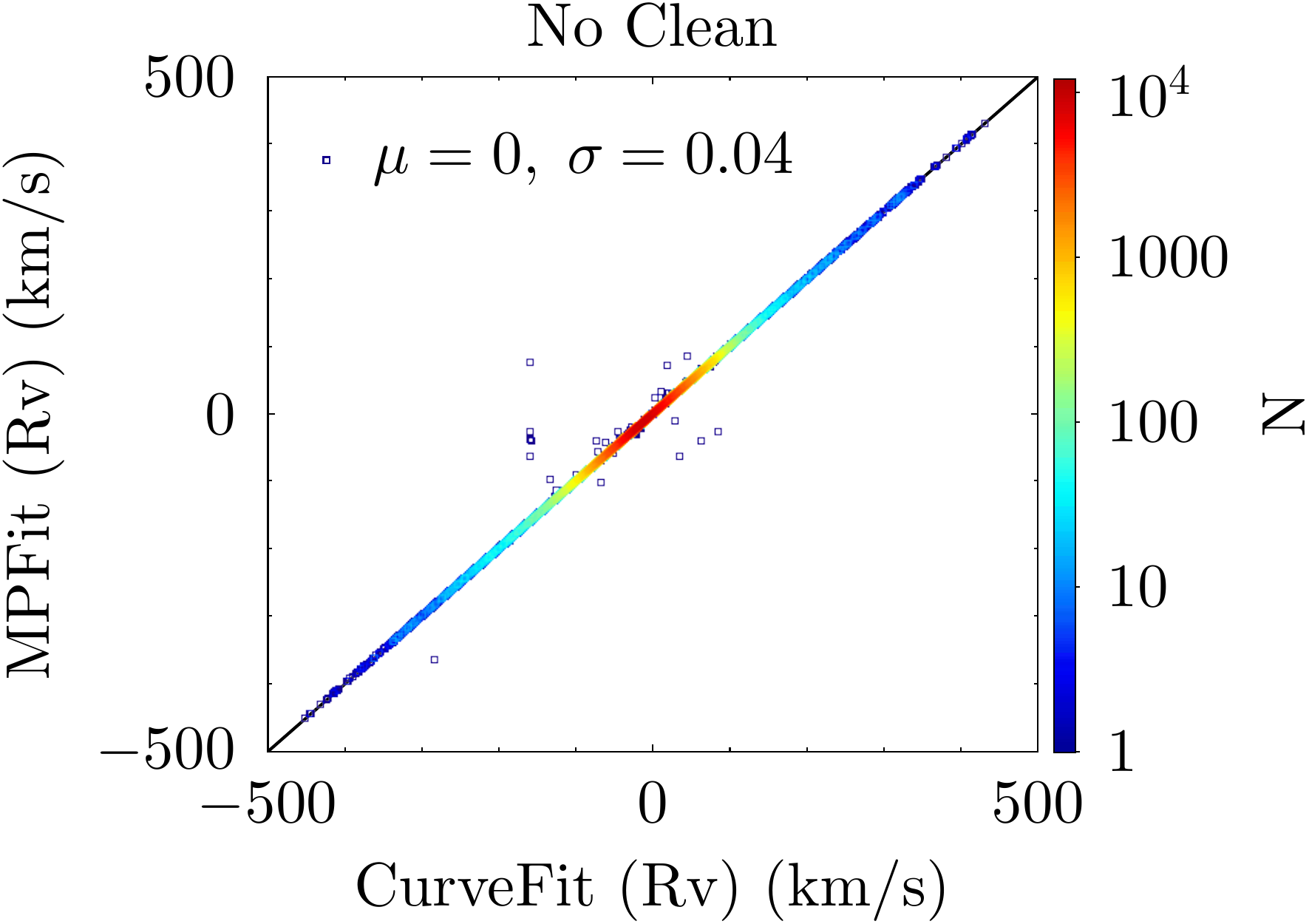}}
		\subfigure{\includegraphics[width=0.25\textwidth, height=0.19\textwidth]{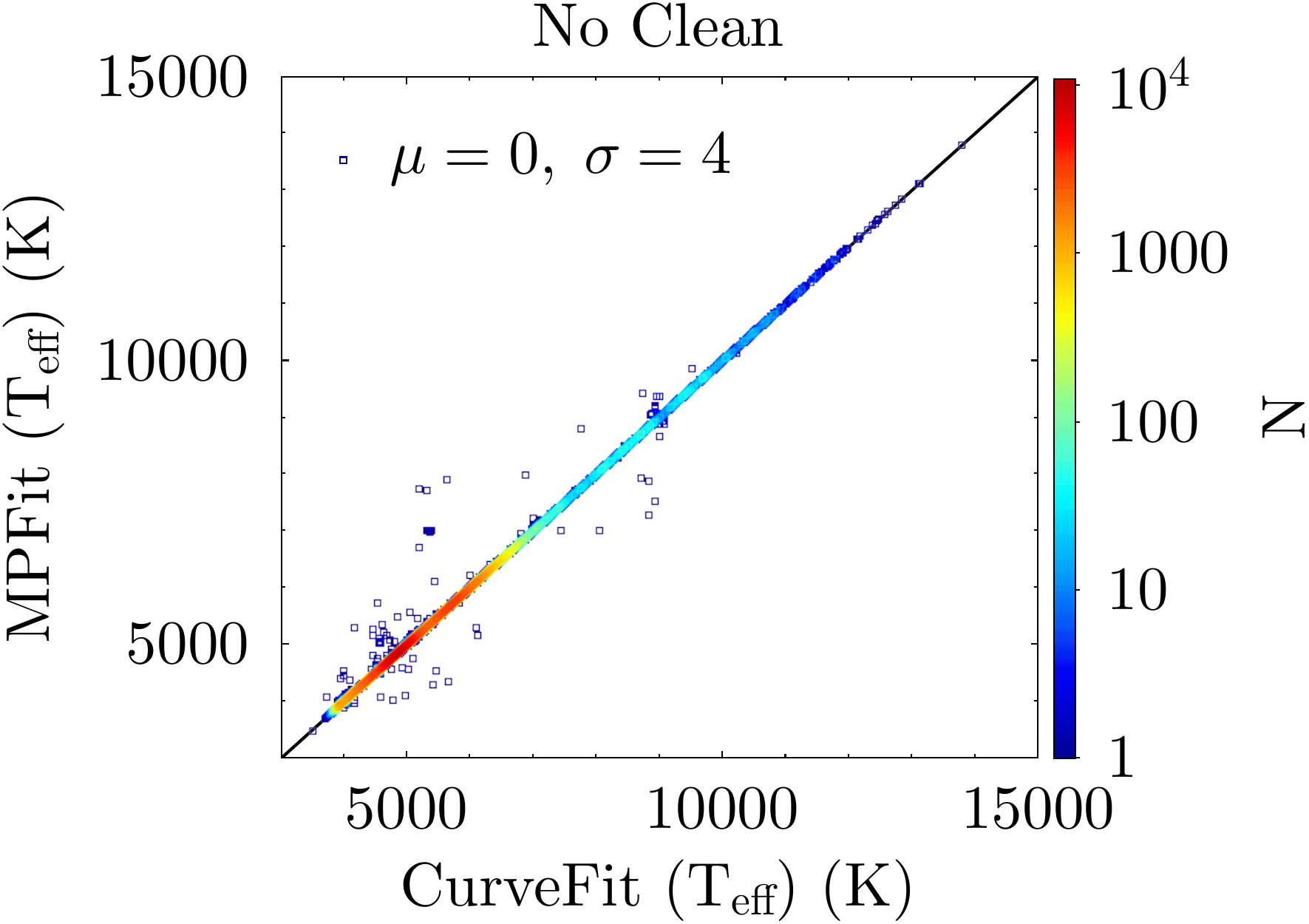}}
		\subfigure{\includegraphics[width=0.242\textwidth, height=0.19\textwidth]{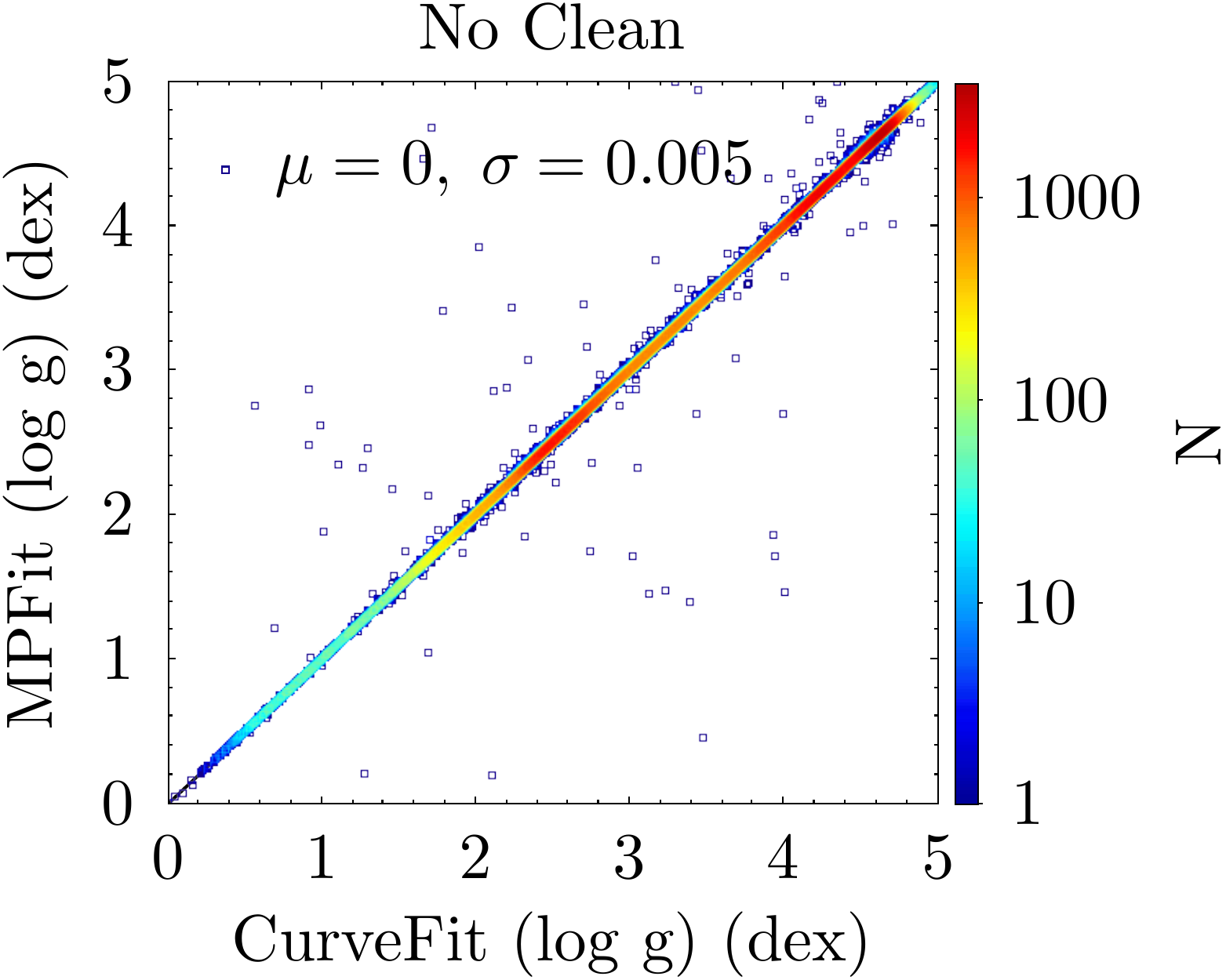}}
		\subfigure{\includegraphics[width=0.242\textwidth, height=0.19\textwidth]{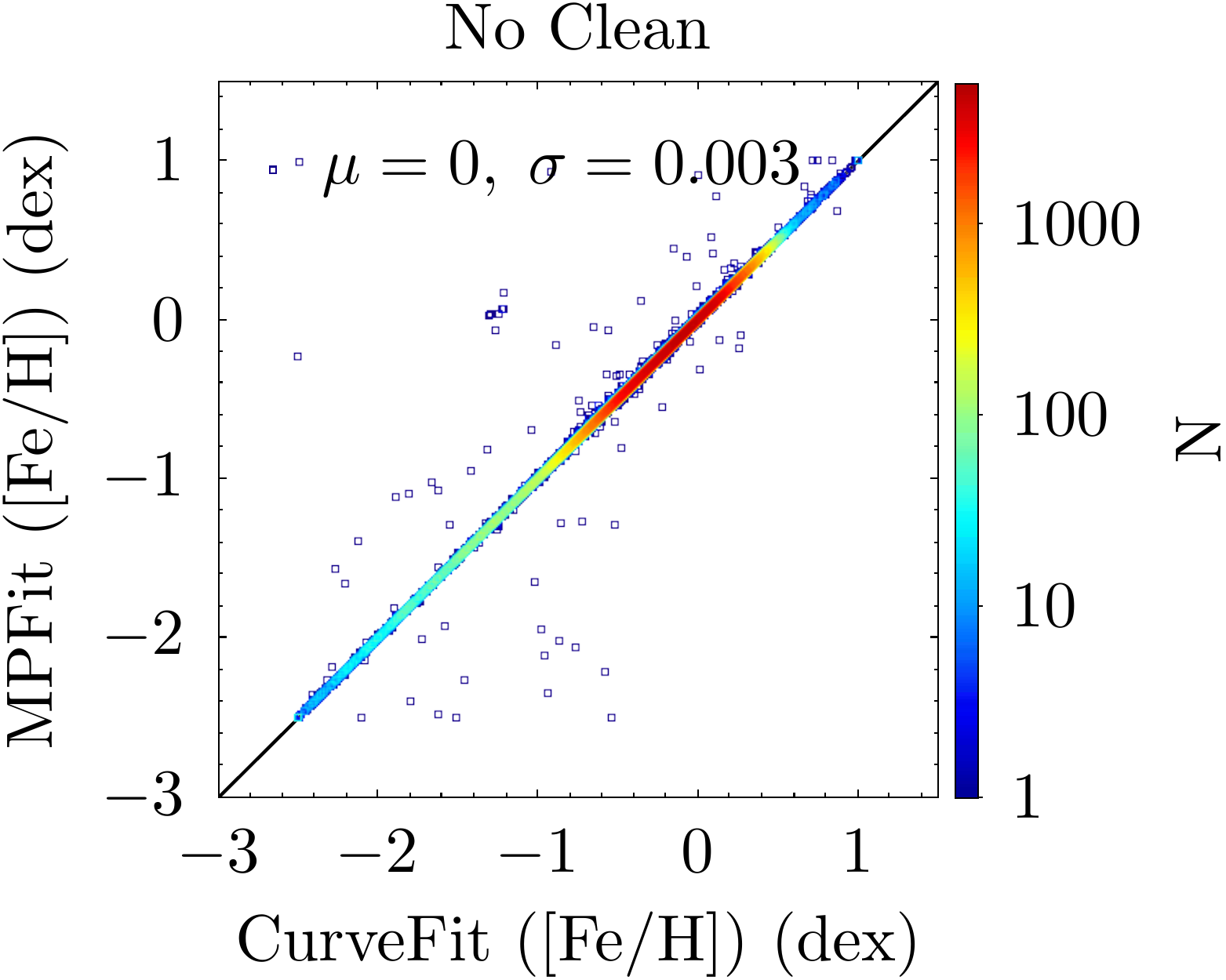}}
		
		\vspace{0.1mm}
		\subfigure{\includegraphics[width=0.25\textwidth, height=0.19\textwidth]{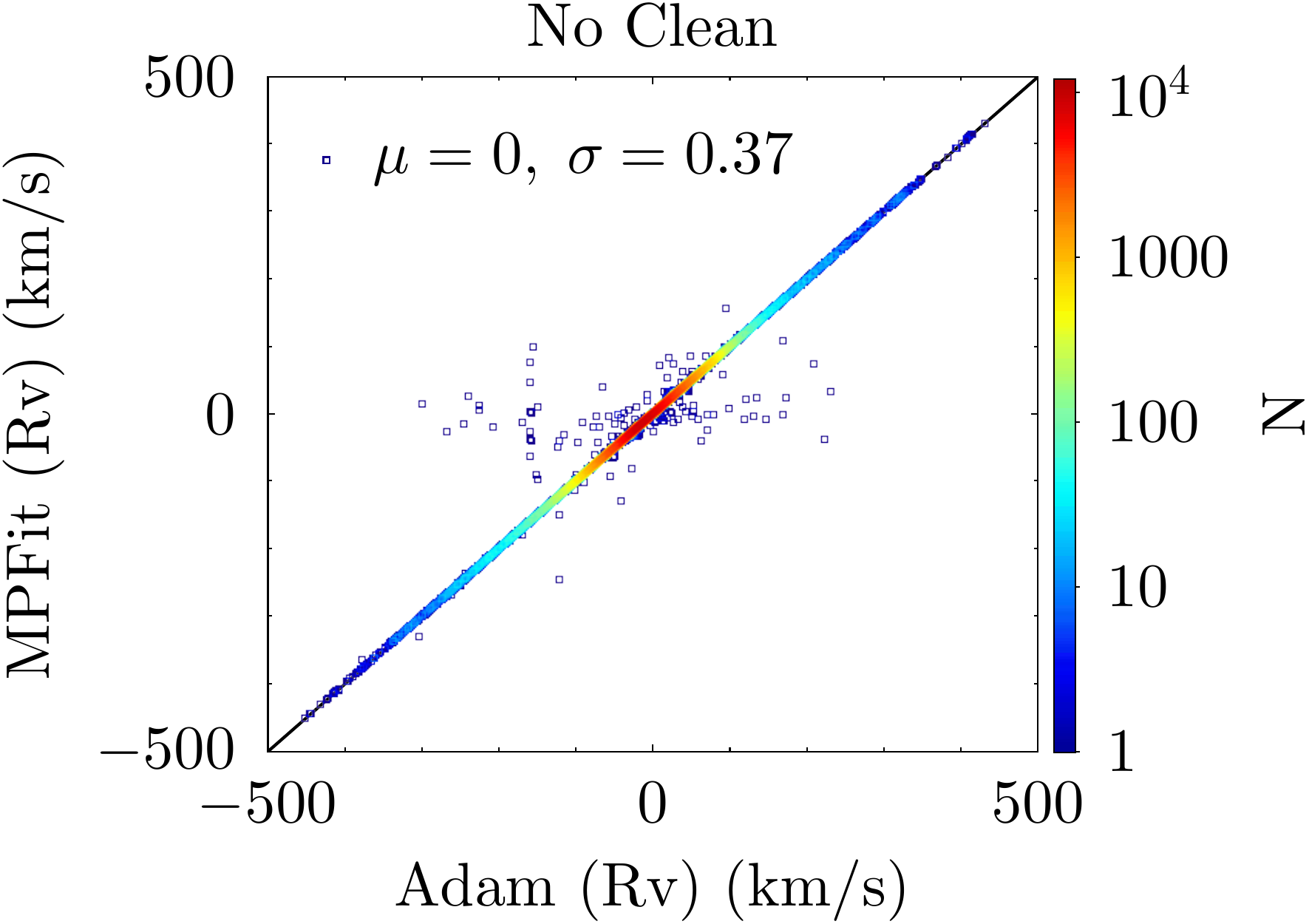}}
		\subfigure{\includegraphics[width=0.25\textwidth, height=0.19\textwidth]{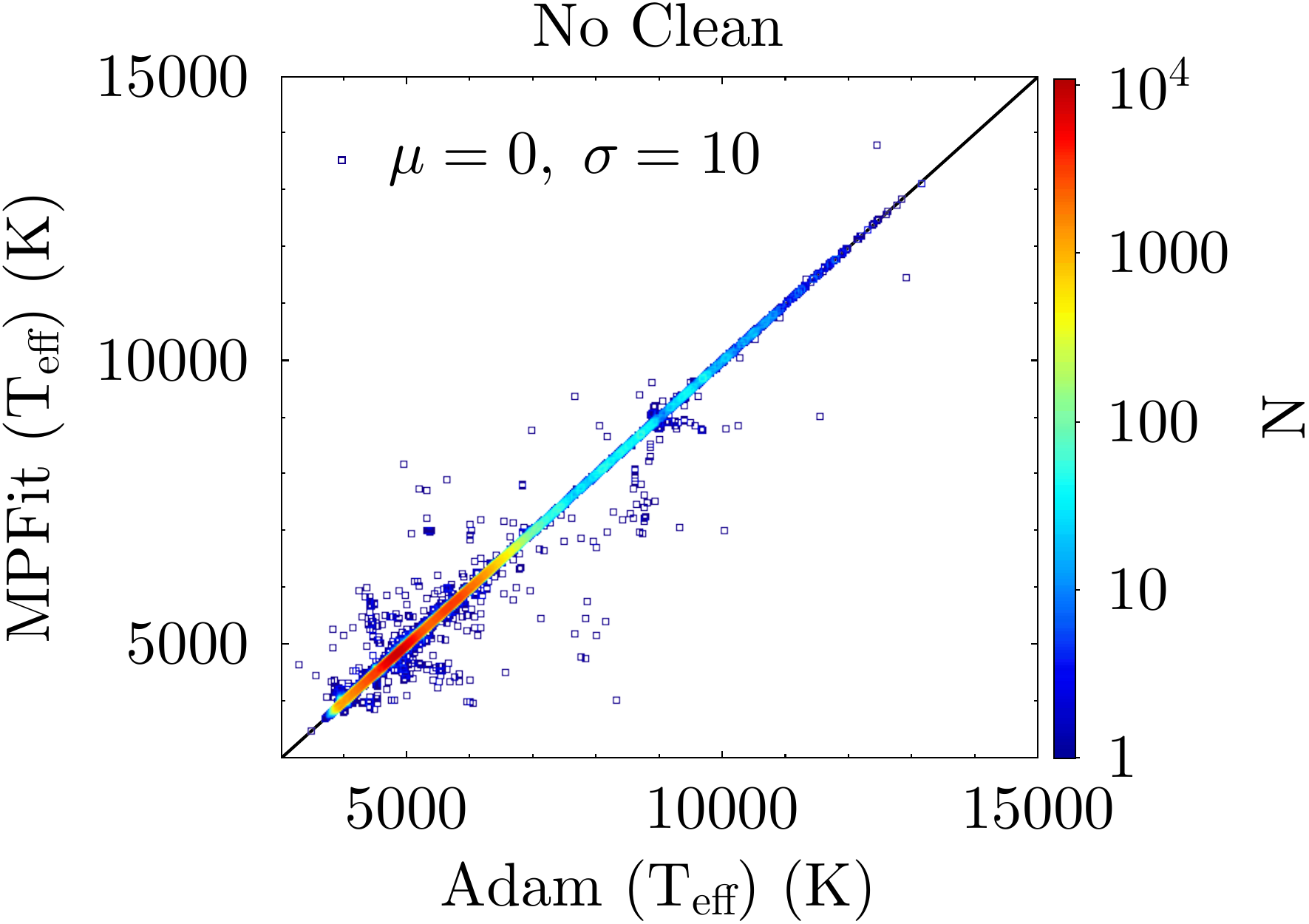}}
		\subfigure{\includegraphics[width=0.242\textwidth, height=0.19\textwidth]{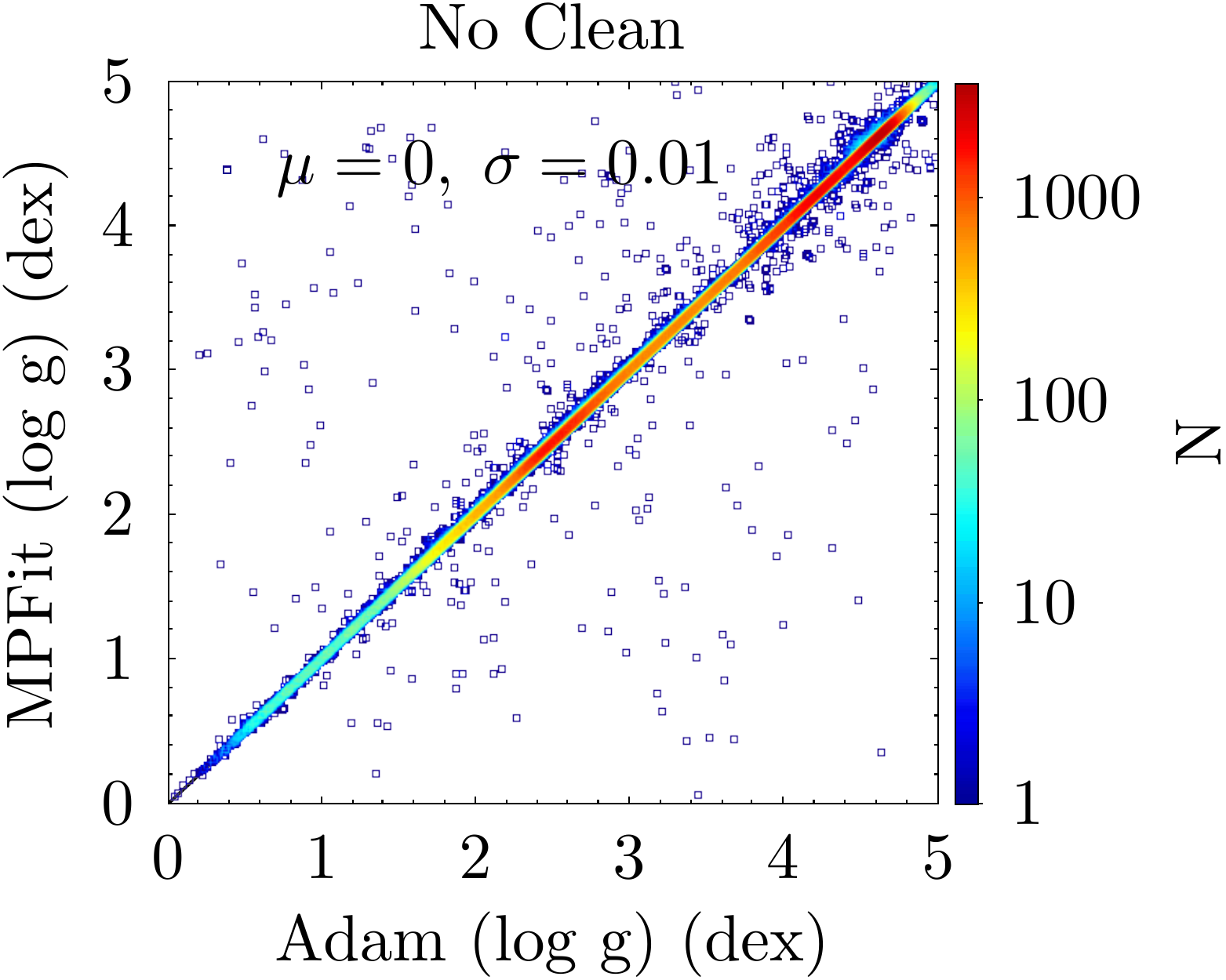}}
		\subfigure{\includegraphics[width=0.242\textwidth, height=0.19\textwidth]{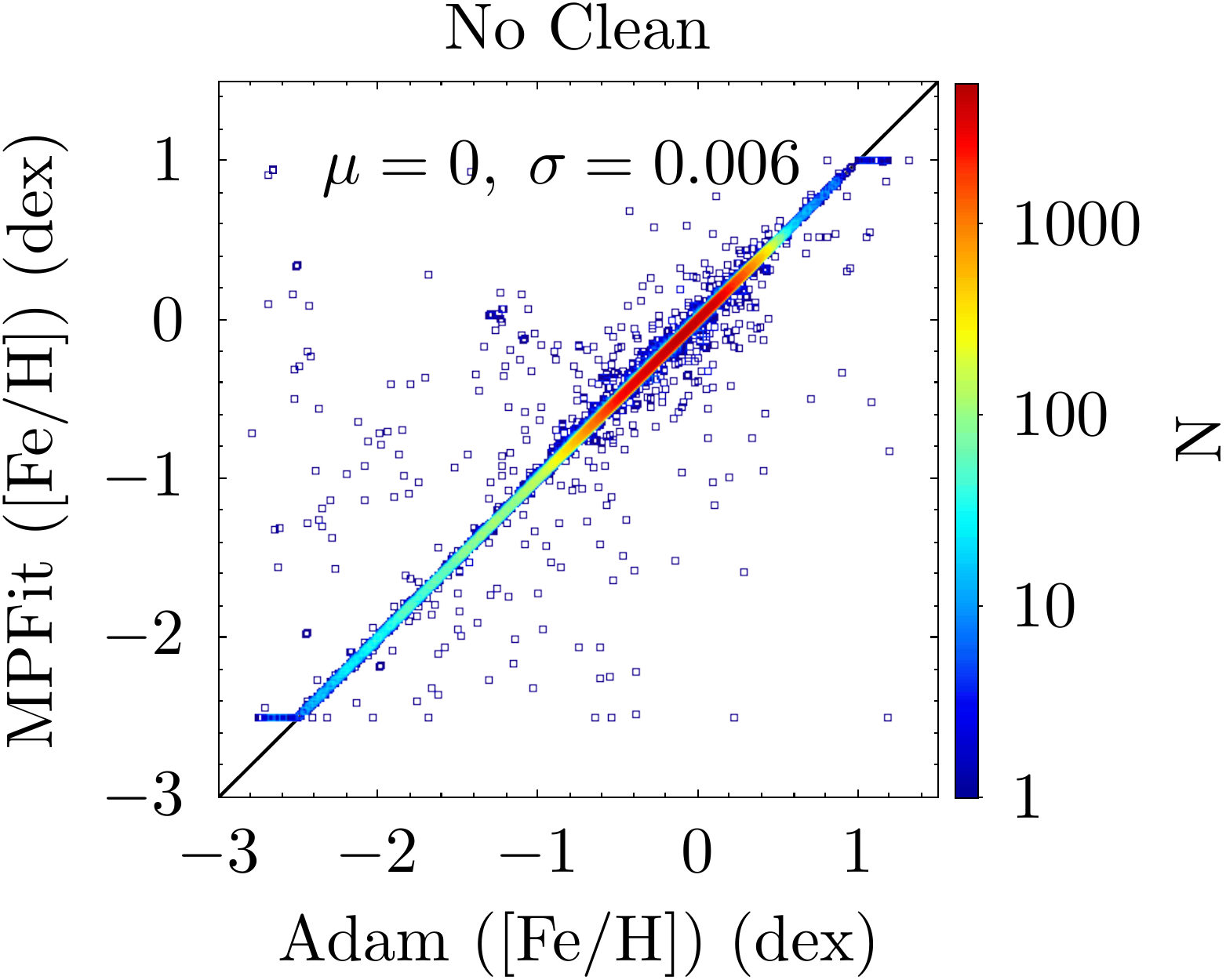}}
		
		\vspace{0.1mm}
		\subfigure{\includegraphics[width=0.25\textwidth, height=0.19\textwidth]{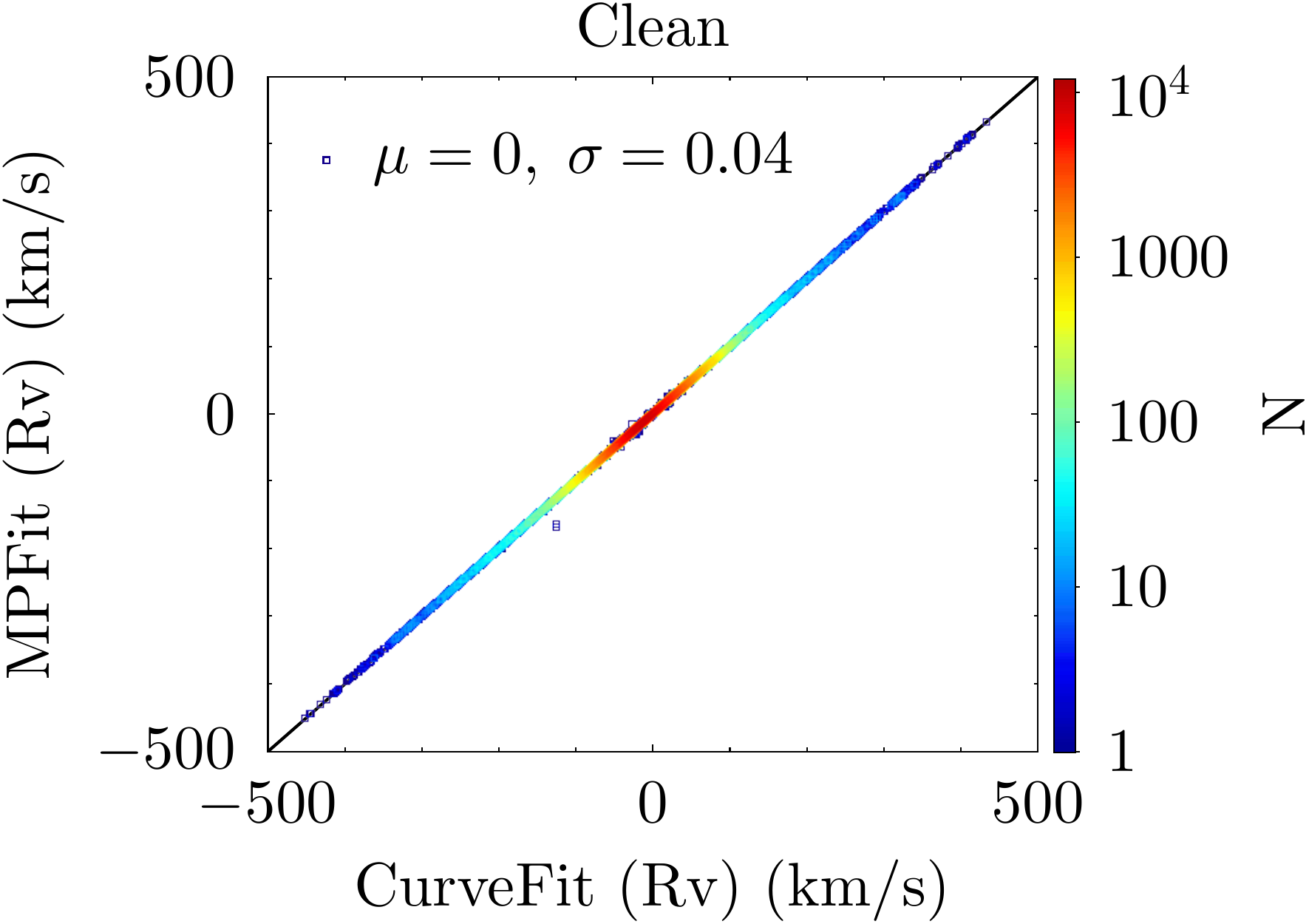}}
		\subfigure{\includegraphics[width=0.25\textwidth, height=0.19\textwidth]{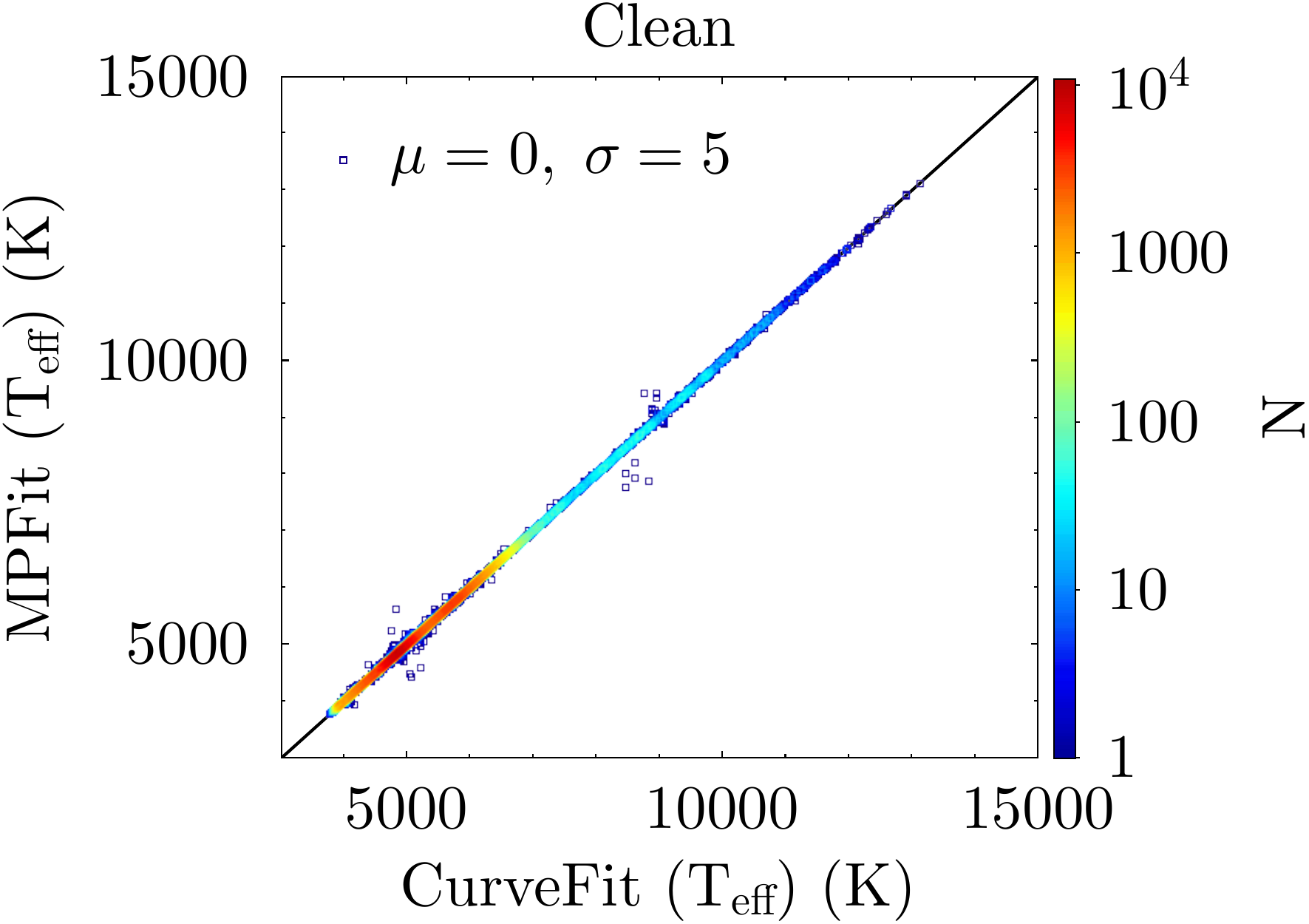}}
		\subfigure{\includegraphics[width=0.242\textwidth, height=0.19\textwidth]{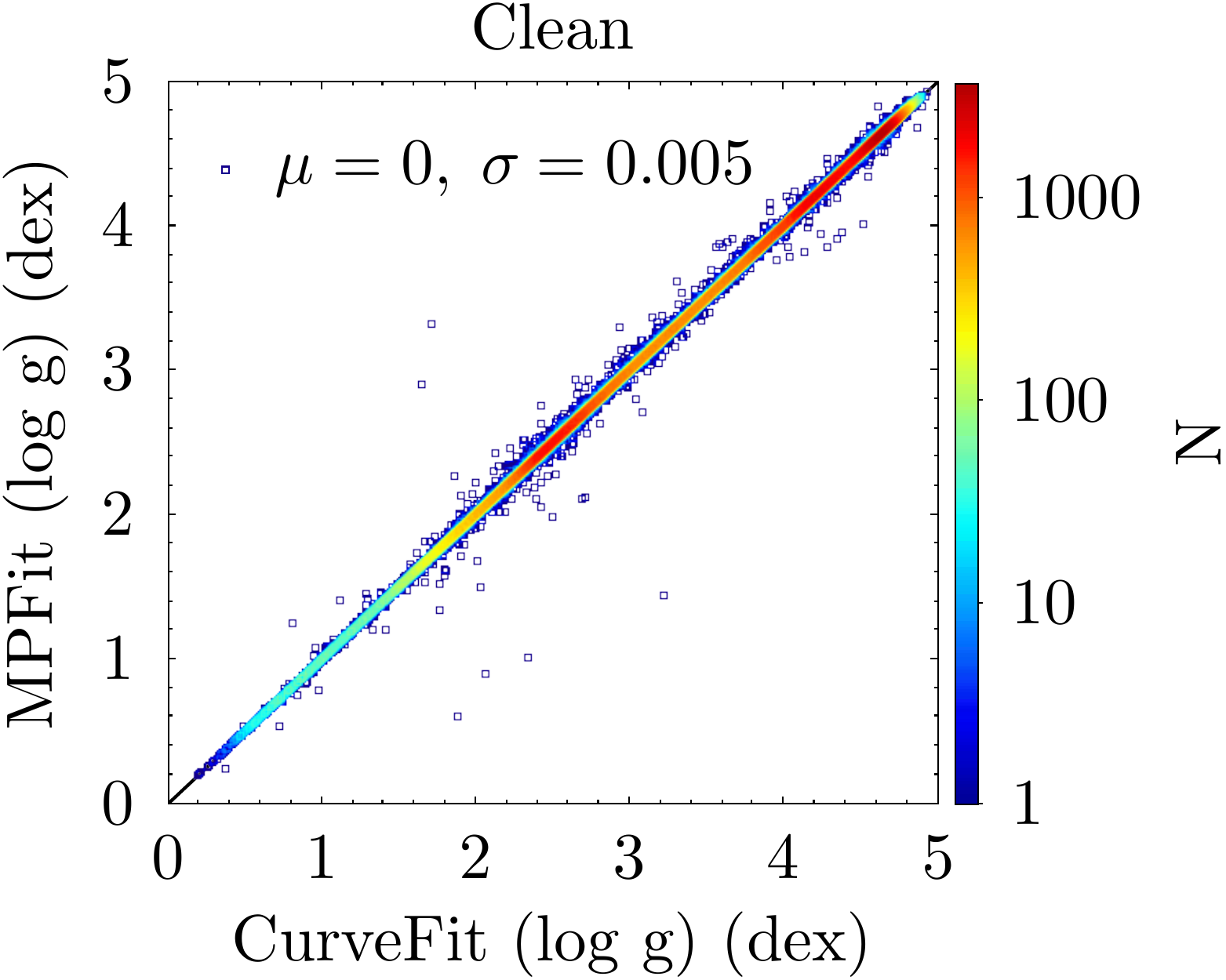}}
		\subfigure{\includegraphics[width=0.242\textwidth, height=0.19\textwidth]{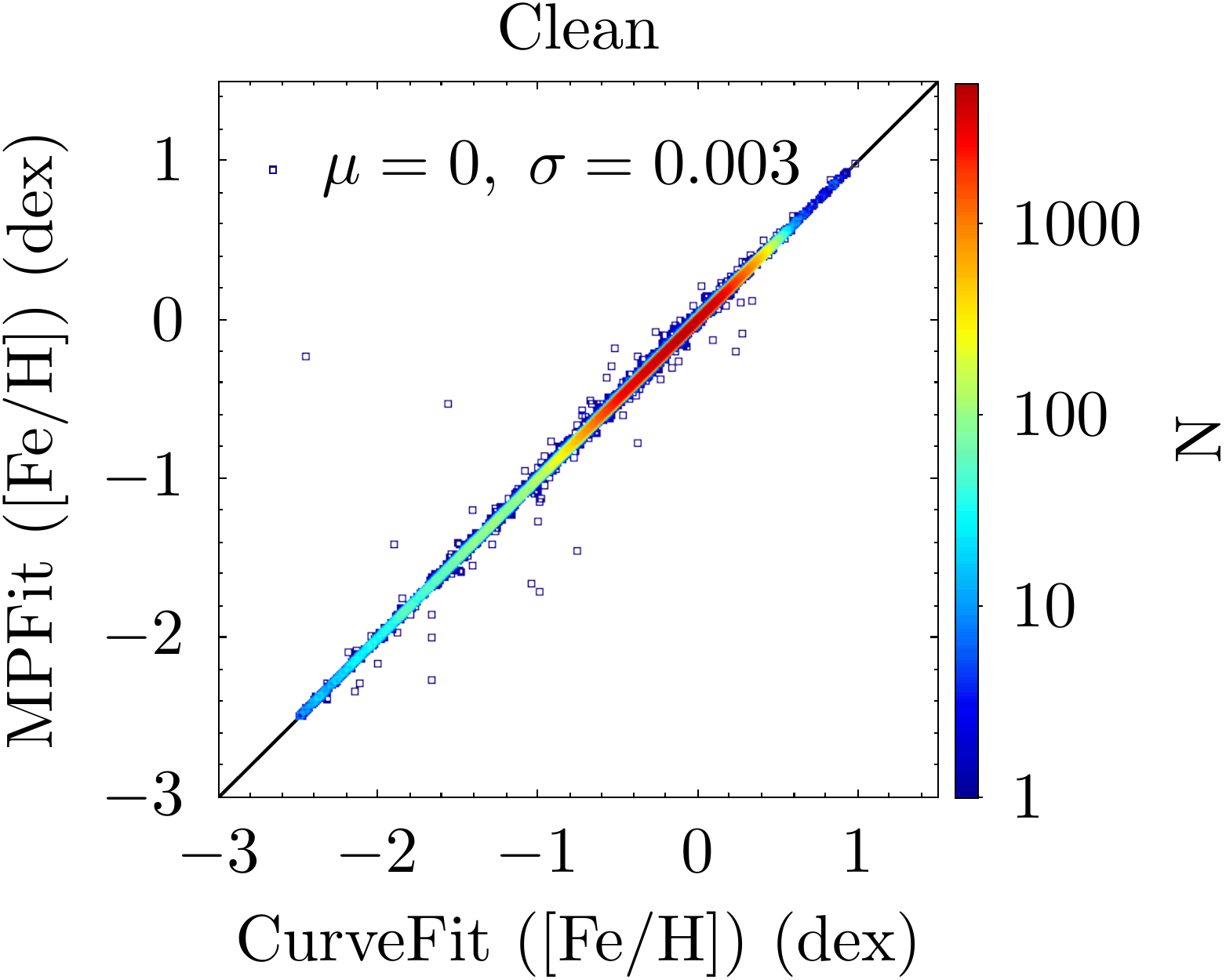}}
		
		\vspace{0.1mm}
		\subfigure{\includegraphics[width=0.25\textwidth, height=0.19\textwidth]{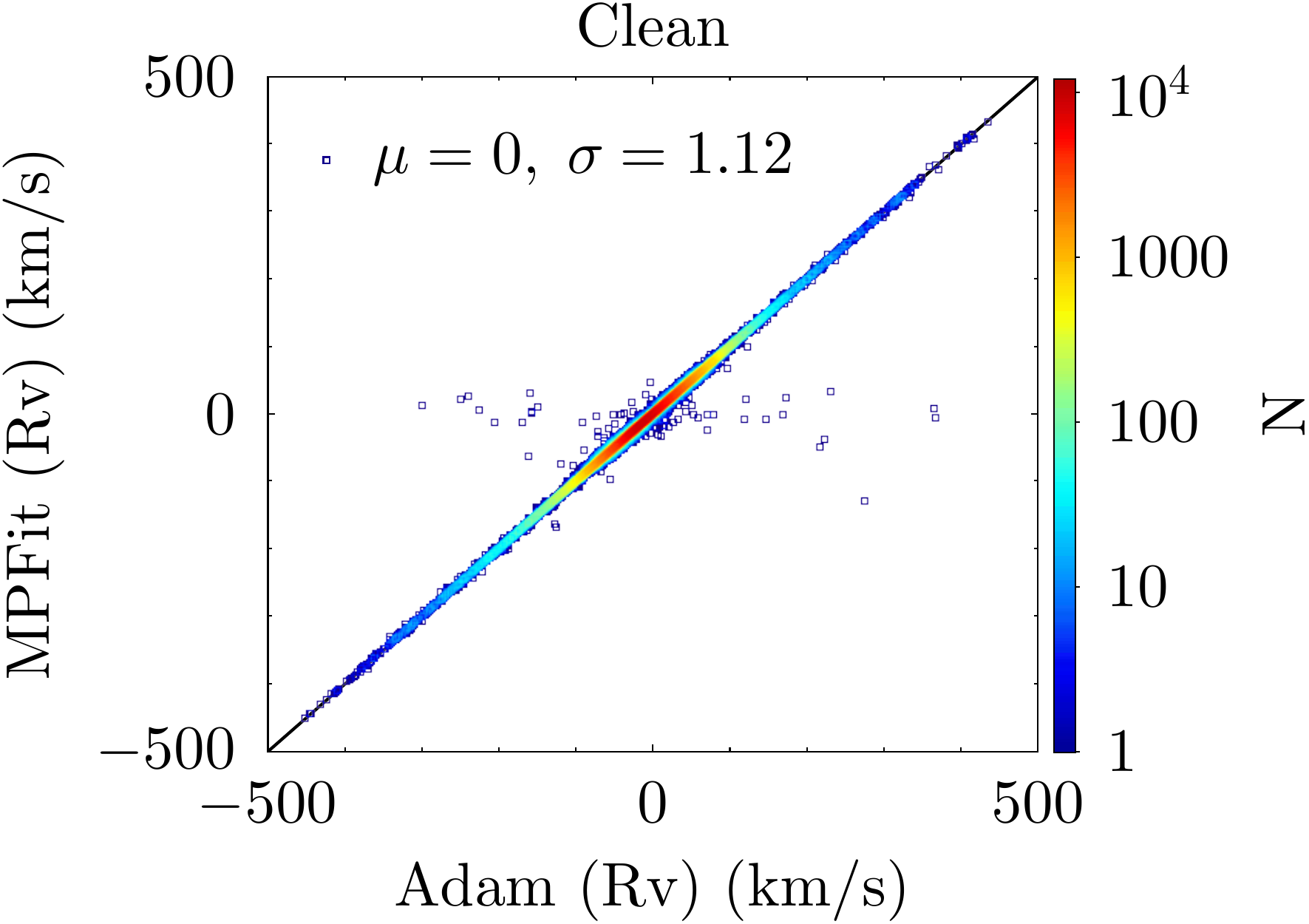}}
		\subfigure{\includegraphics[width=0.25\textwidth, height=0.19\textwidth]{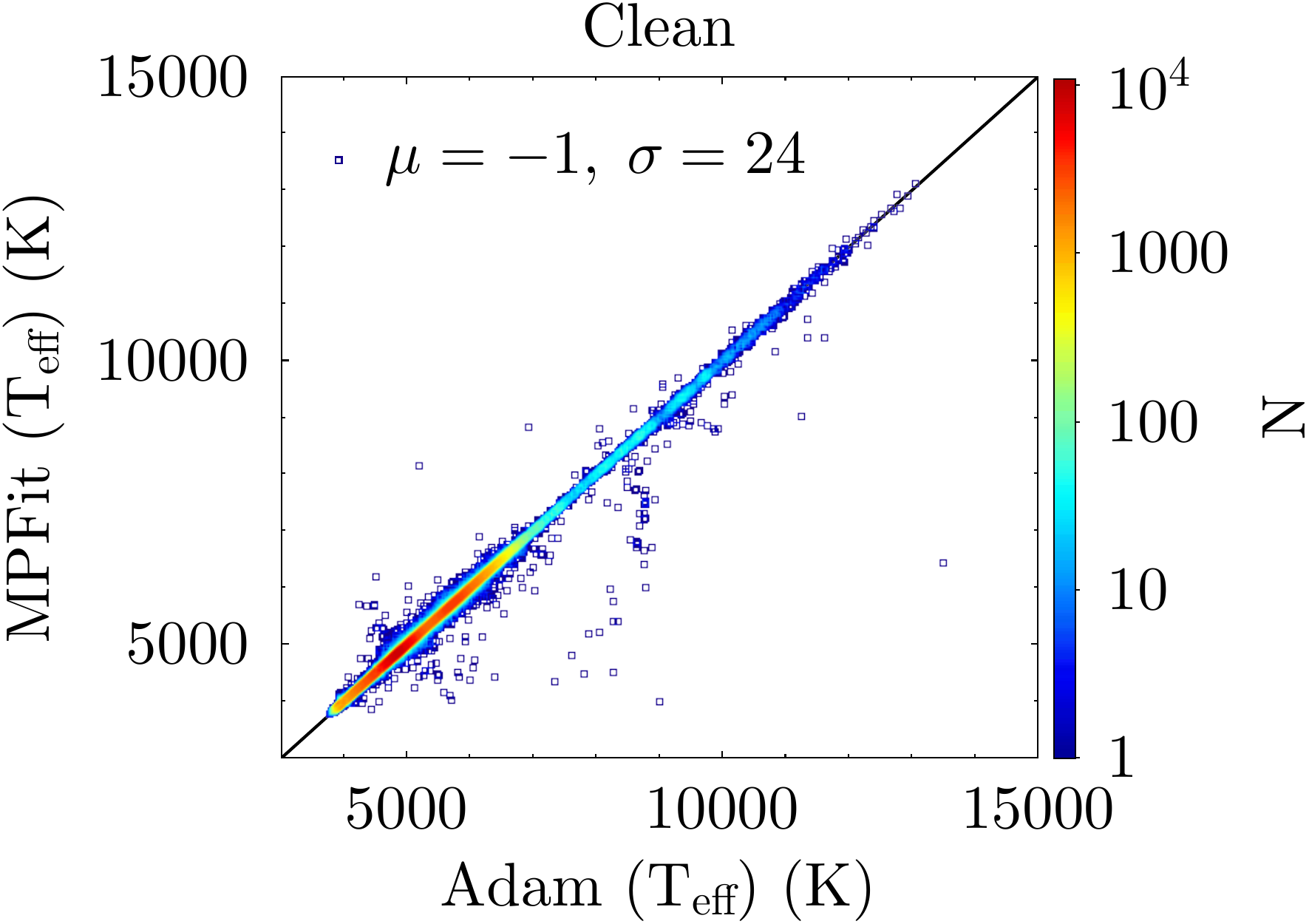}}
		\subfigure{\includegraphics[width=0.242\textwidth, height=0.19\textwidth]{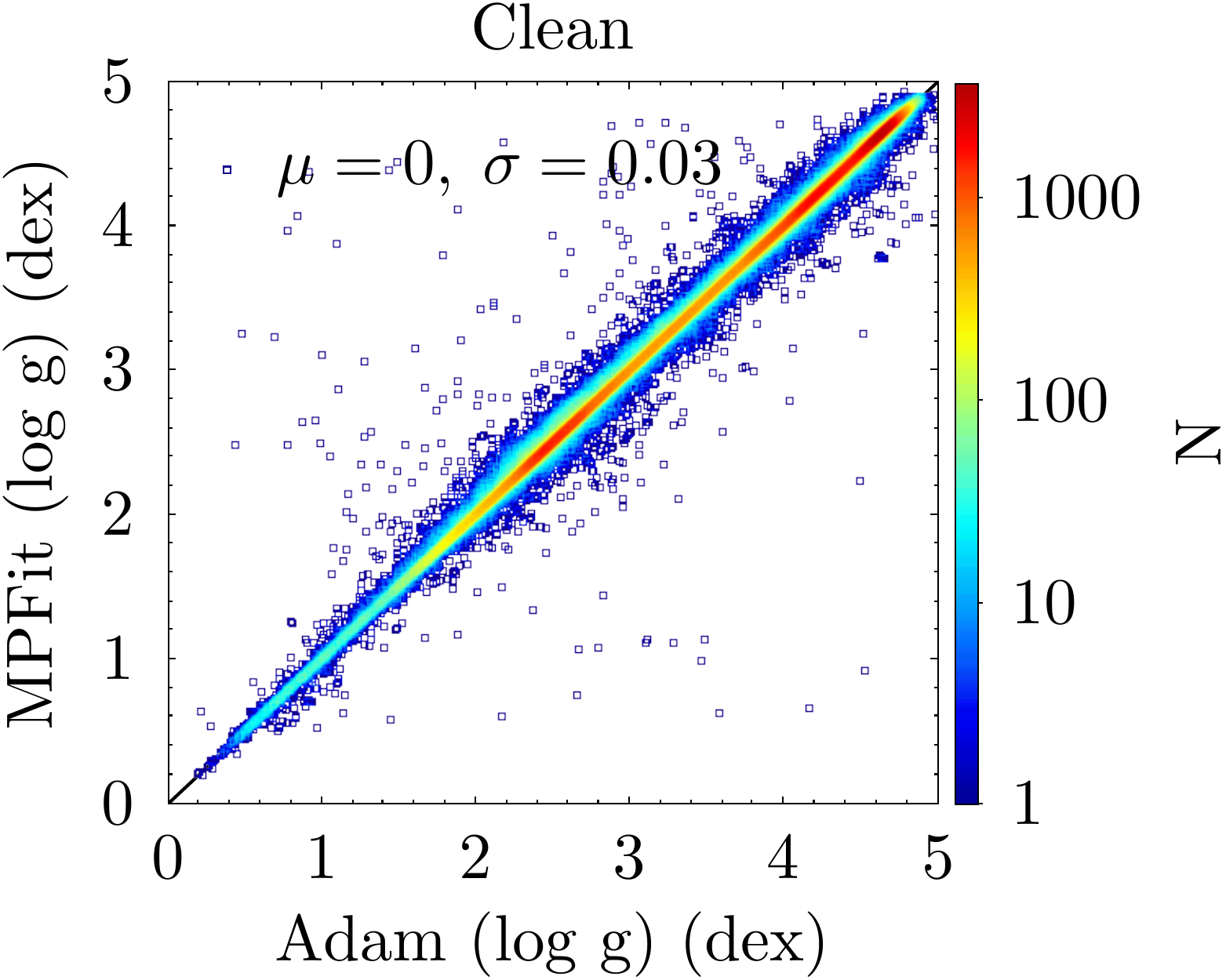}}
		\subfigure{\includegraphics[width=0.242\textwidth, height=0.19\textwidth]{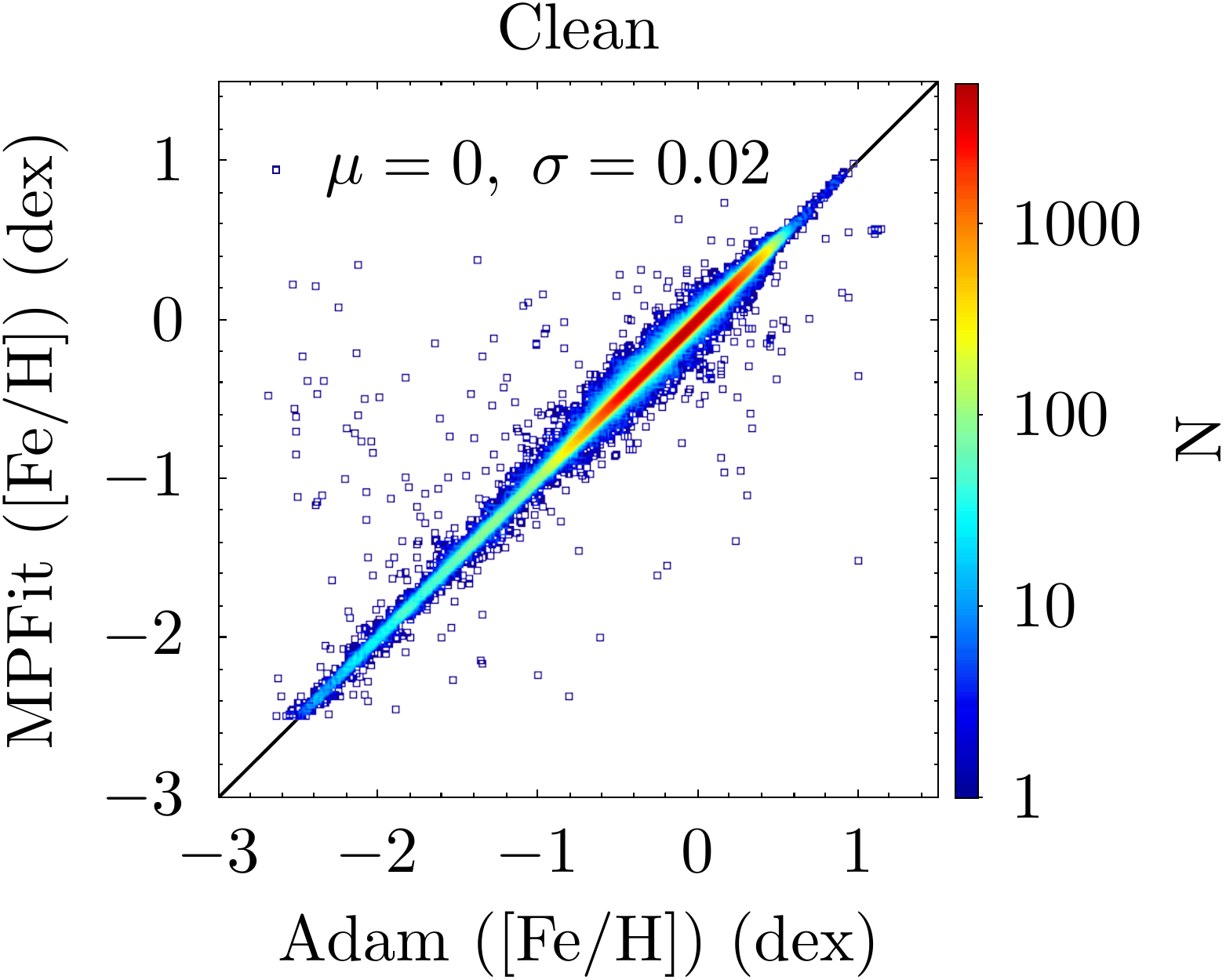}}
		
		\caption{Comparison of $\mathrm{RV}$ (column 1), $T_\mathrm{eff}$ (column 2), $\log g$ (column 3), and [Fe/H] (column 4), inferred using LASP-CurveFit, LASP-Adam-GPU, and LASP-MPFit under both the No Clean and Clean strategies. From top to bottom, the four rows correspond to LASP-CurveFit versus  LASP-MPFit (No Clean), LASP-Adam-GPU versus LASP-MPFit (No Clean), LASP-CurveFit versus LASP-MPFit (Clean), and LASP-Adam-GPU versus LASP-MPFit (Clean). The mean ($\mu$) and standard deviation ($\sigma$) of the differences are calculated using \texttt{astropy.stats.sigma\_clipped\_stats} with \texttt{maxiters=1}, excluding values beyond $5\sigma$. The \lq LASP\rq\ prefix has been omitted in all figure labels for clarity.}
		\label{AllParamCompare}
	\end{figure*}
	
	\begin{figure*}[!htbp]
		\centering
		\includegraphics[width=0.245\linewidth, height=0.185\textwidth]{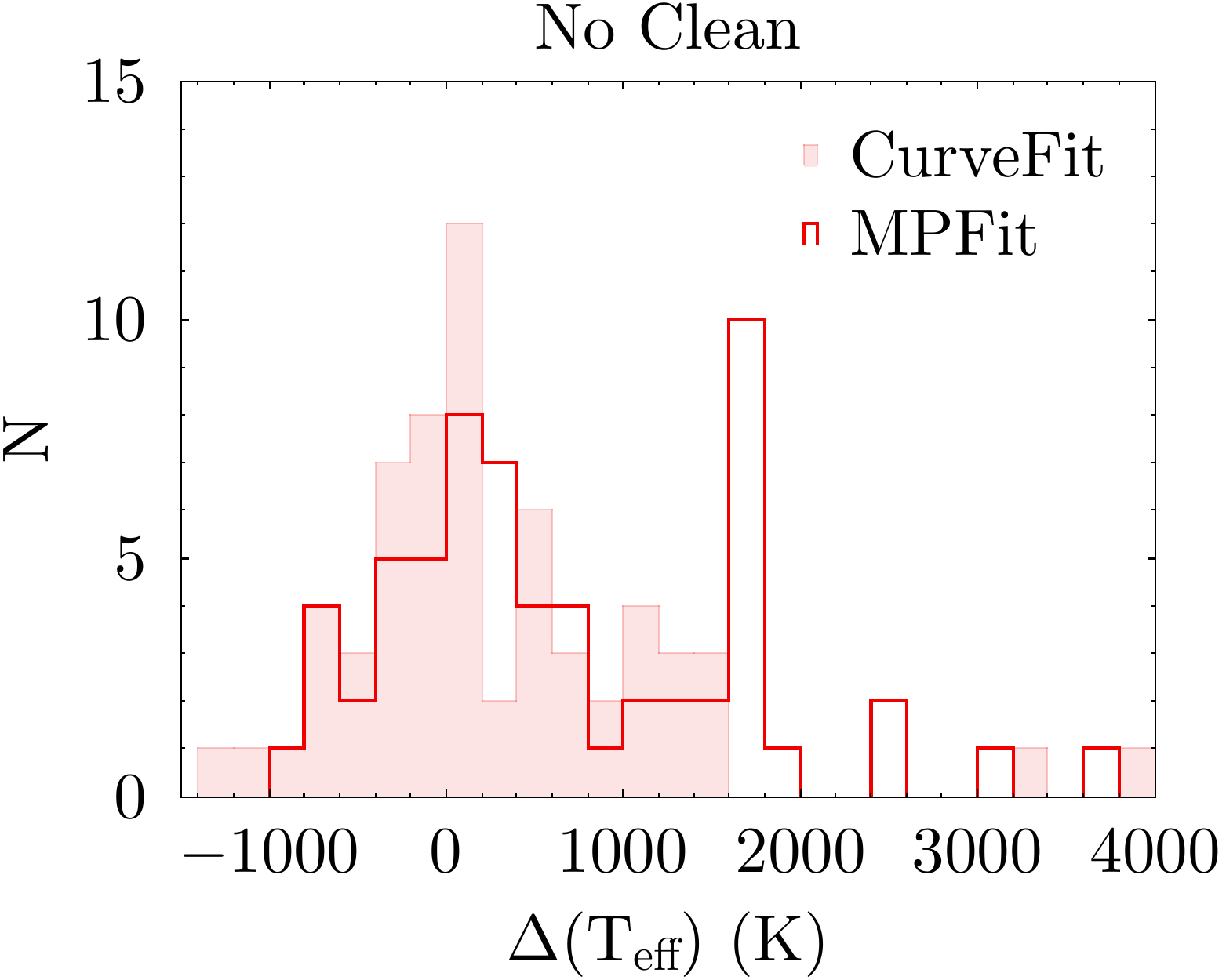}
		\includegraphics[width=0.245\linewidth, height=0.19\textwidth]{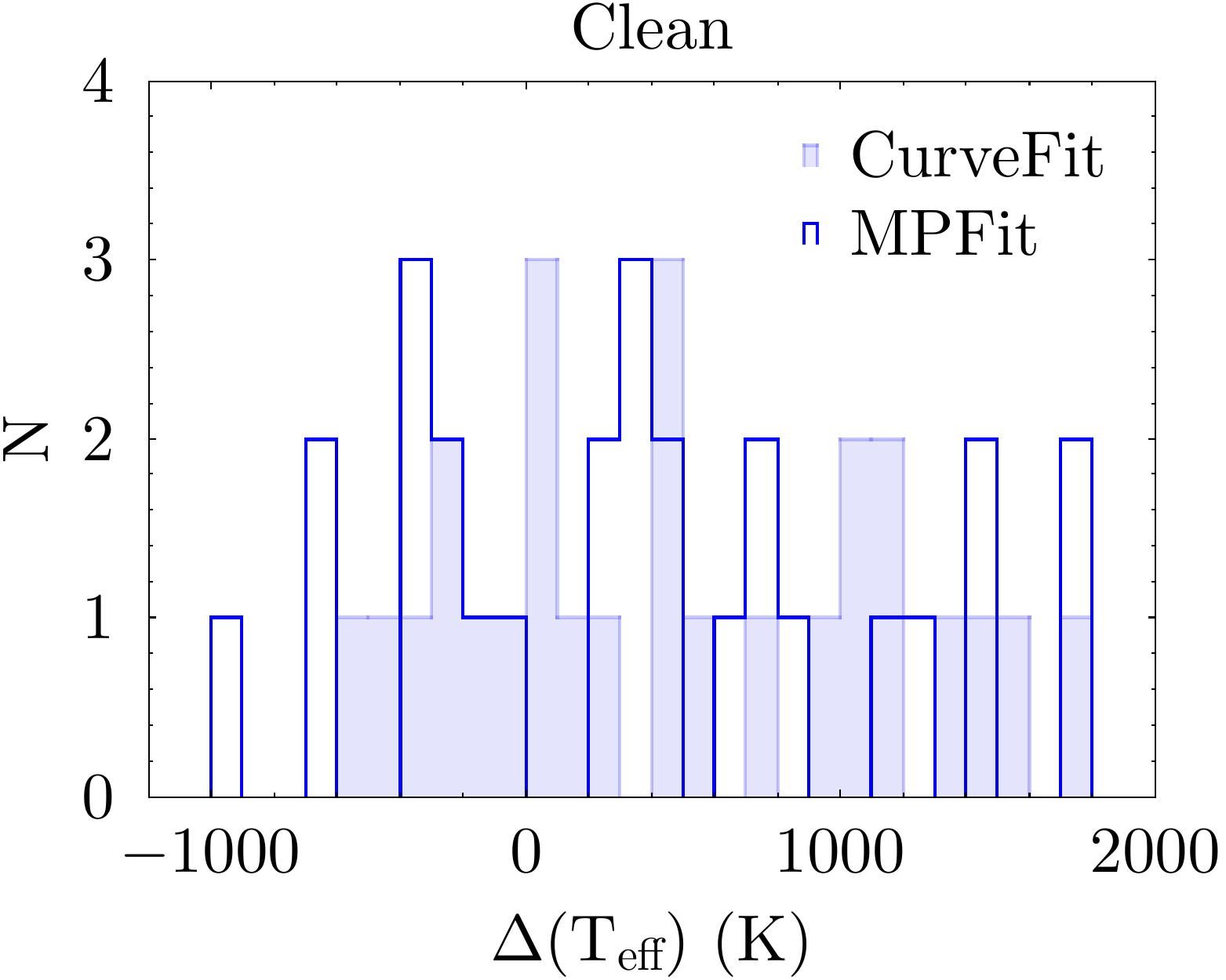}
		\includegraphics[width=0.245\linewidth, height=0.19\textwidth]{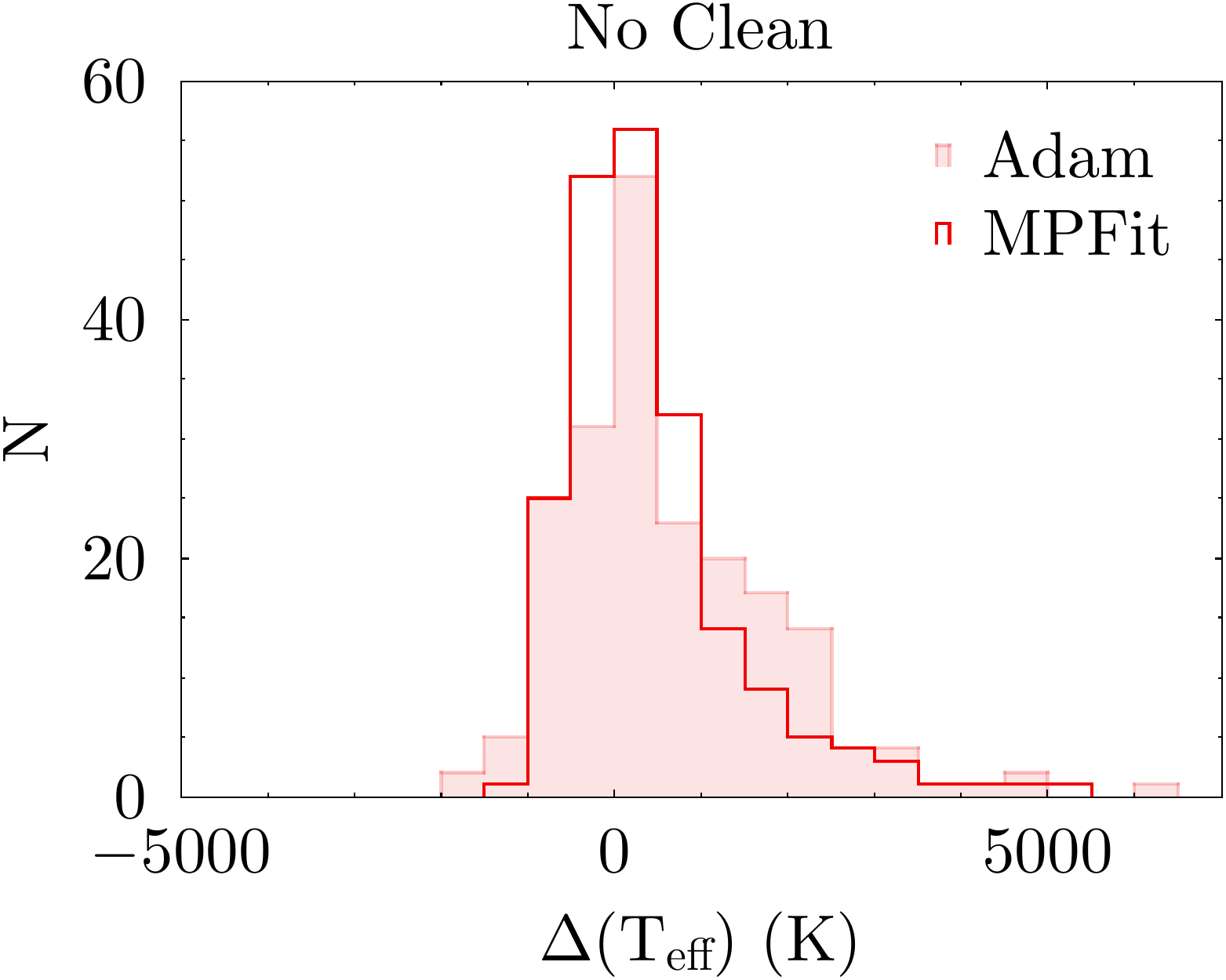} 
		\includegraphics[width=0.245\linewidth, height=0.185\textwidth]{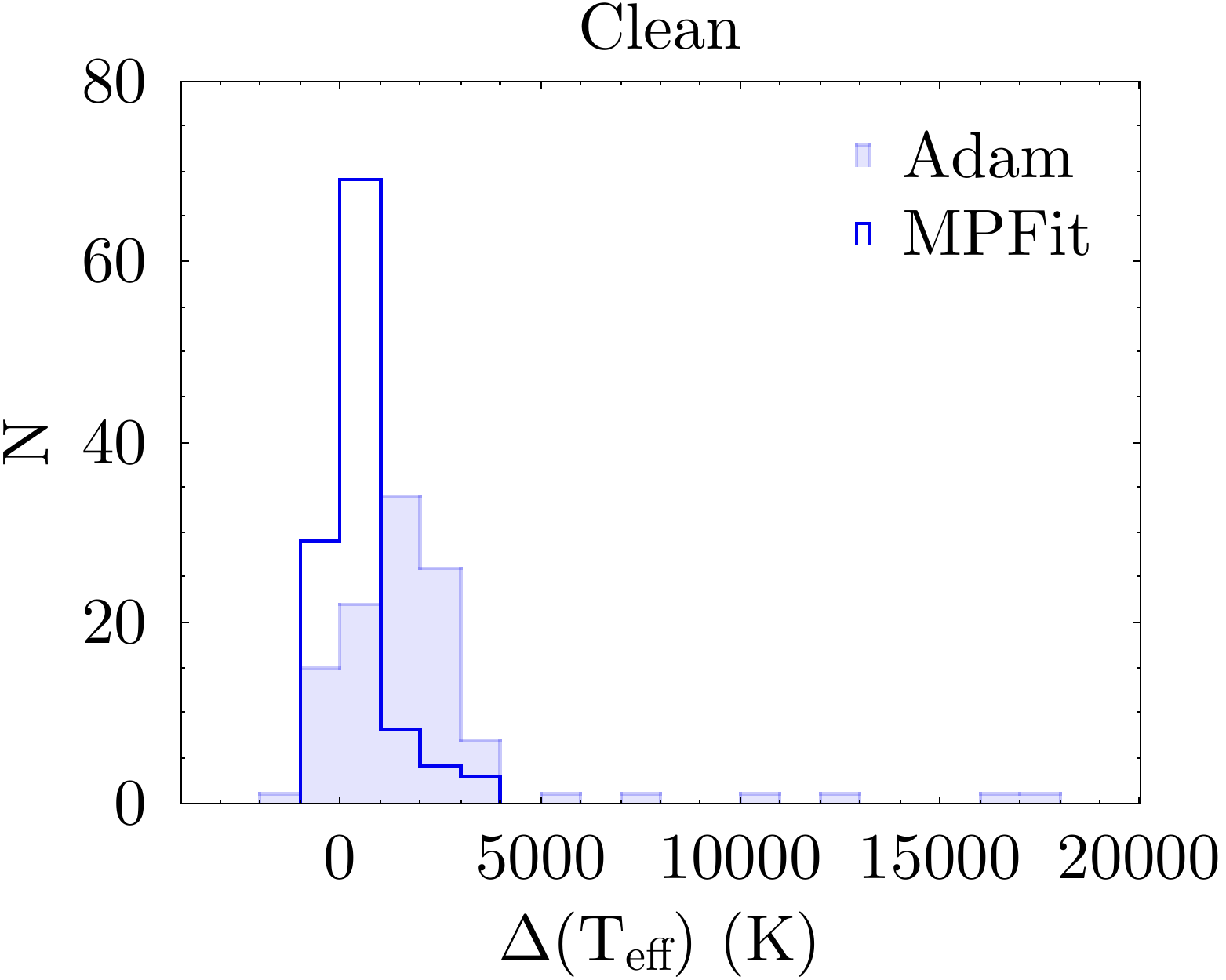}\\
		\includegraphics[width=0.245\linewidth, height=0.195\textwidth]{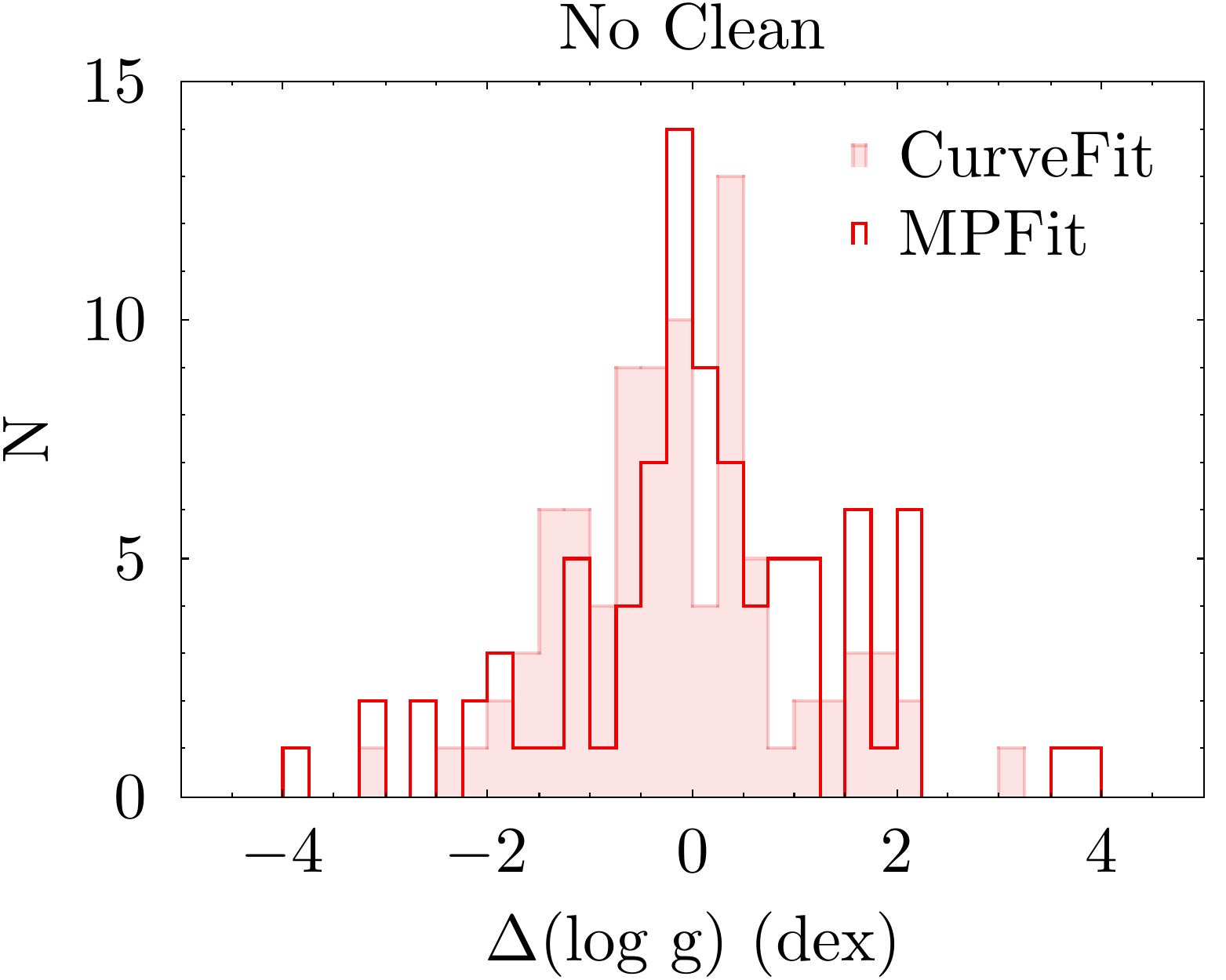}
		\includegraphics[width=0.245\linewidth, height=0.19\textwidth]{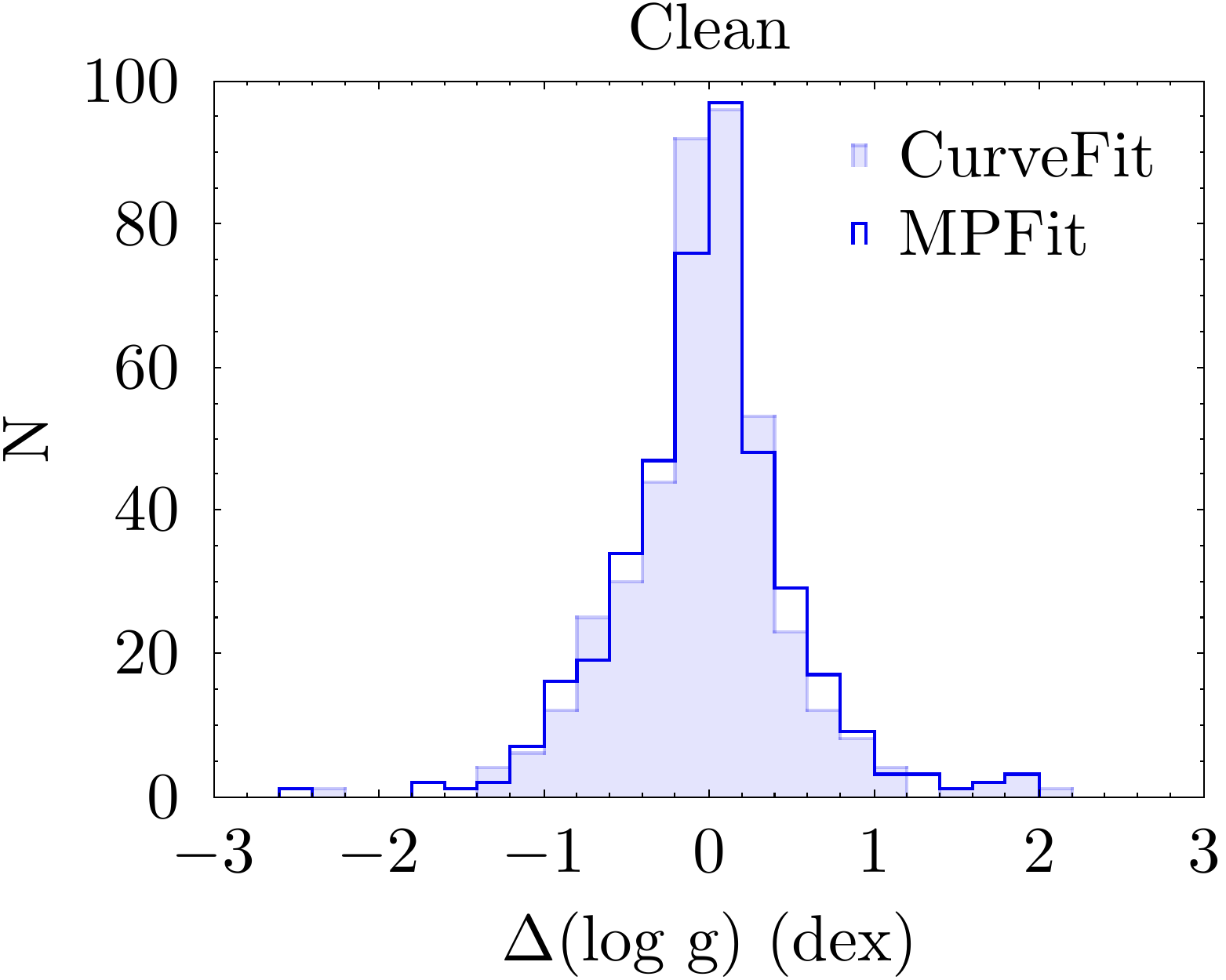}
		\includegraphics[width=0.245\linewidth, height=0.19\textwidth]{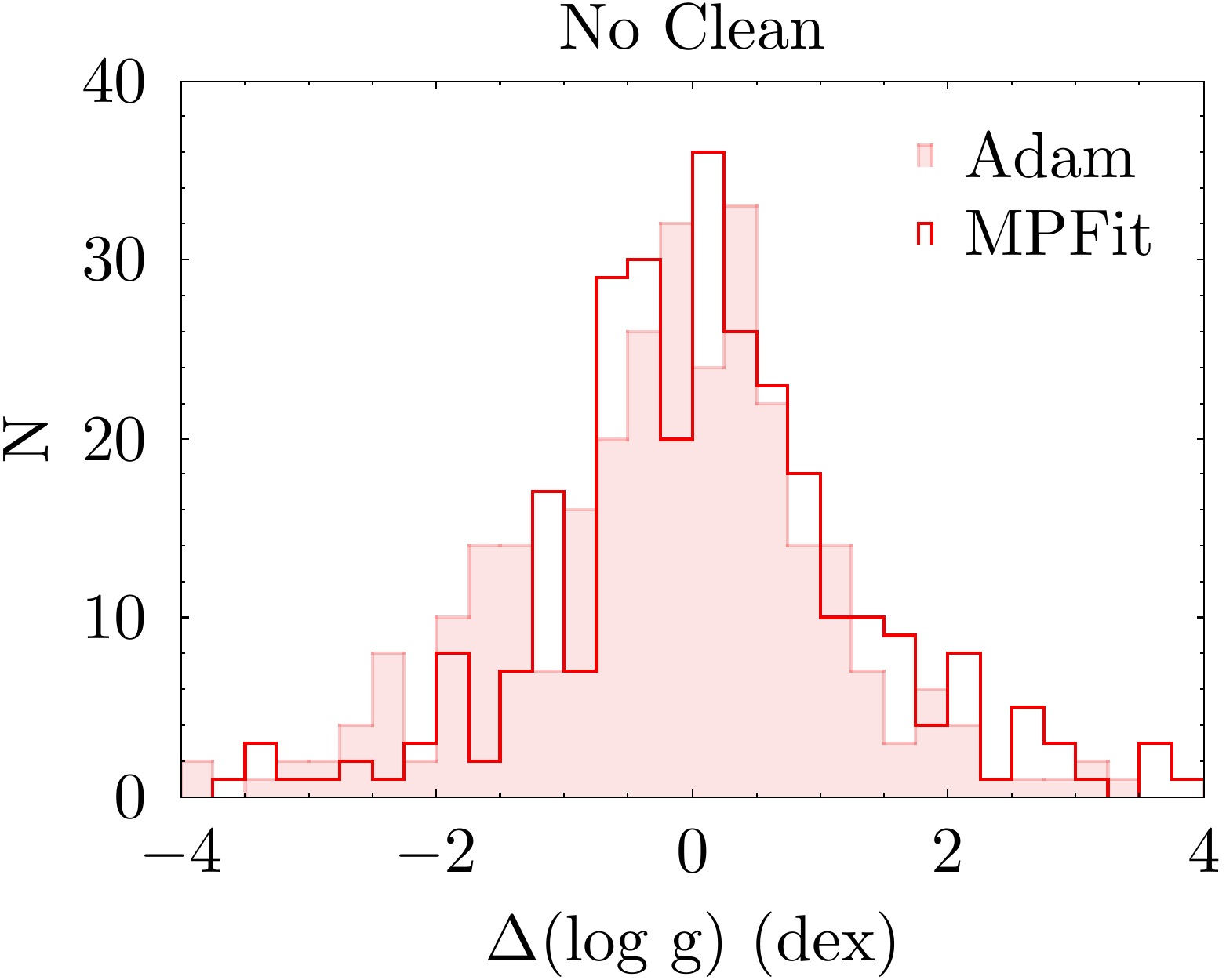}
		\includegraphics[width=0.245\linewidth, height=0.195\textwidth]{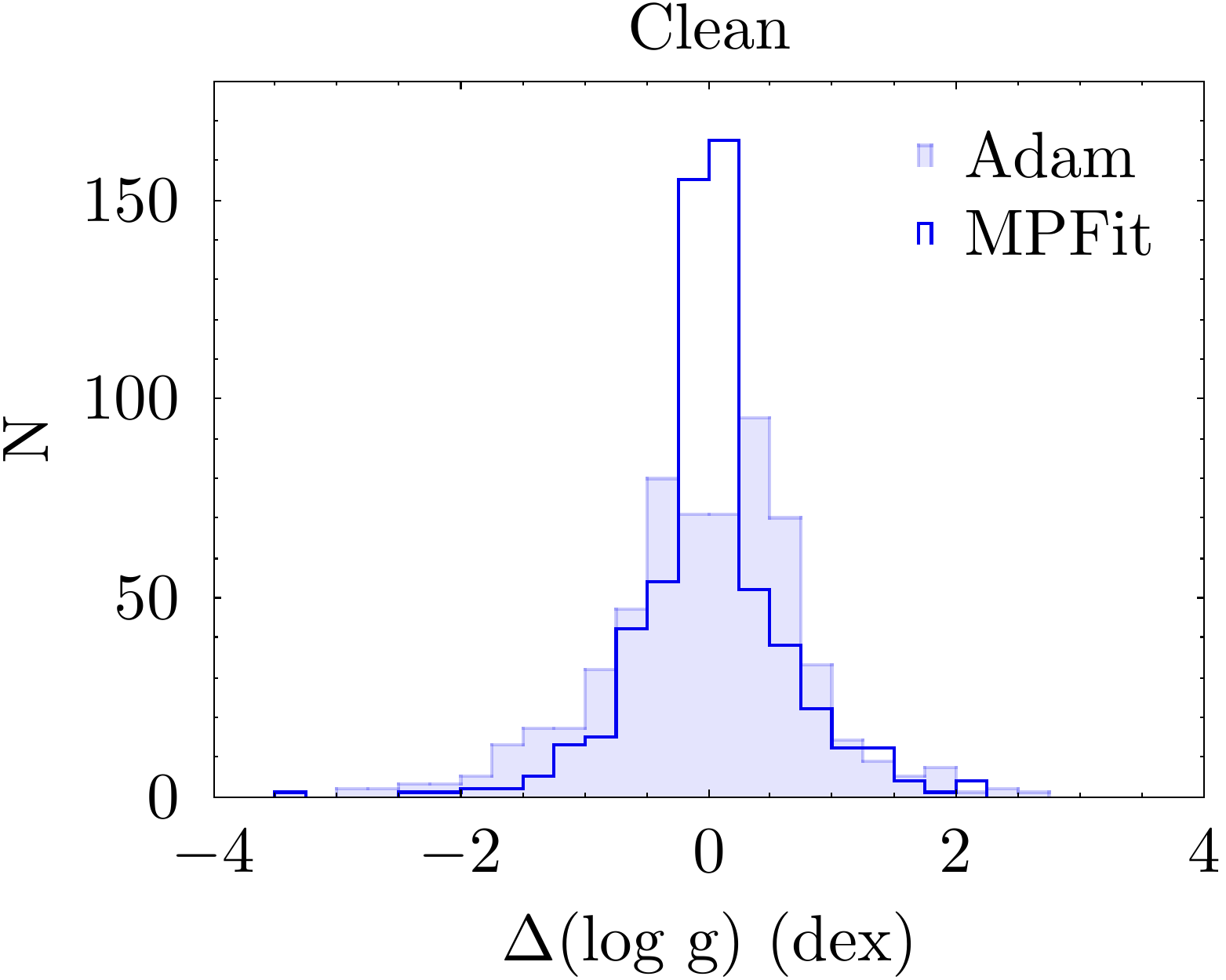}\\
		\includegraphics[width=0.245\linewidth, height=0.195\textwidth]{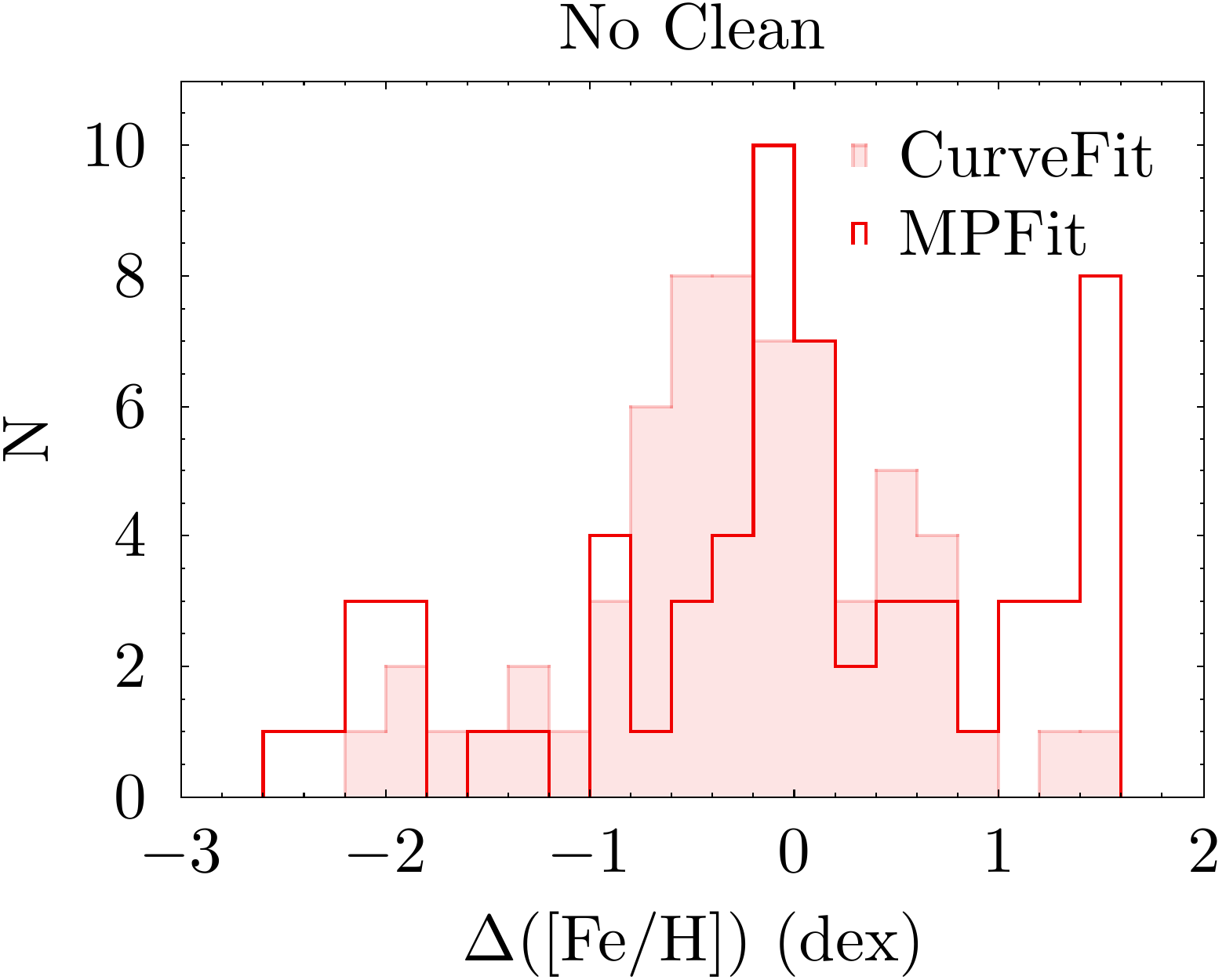}
		\includegraphics[width=0.245\linewidth, height=0.19\textwidth]{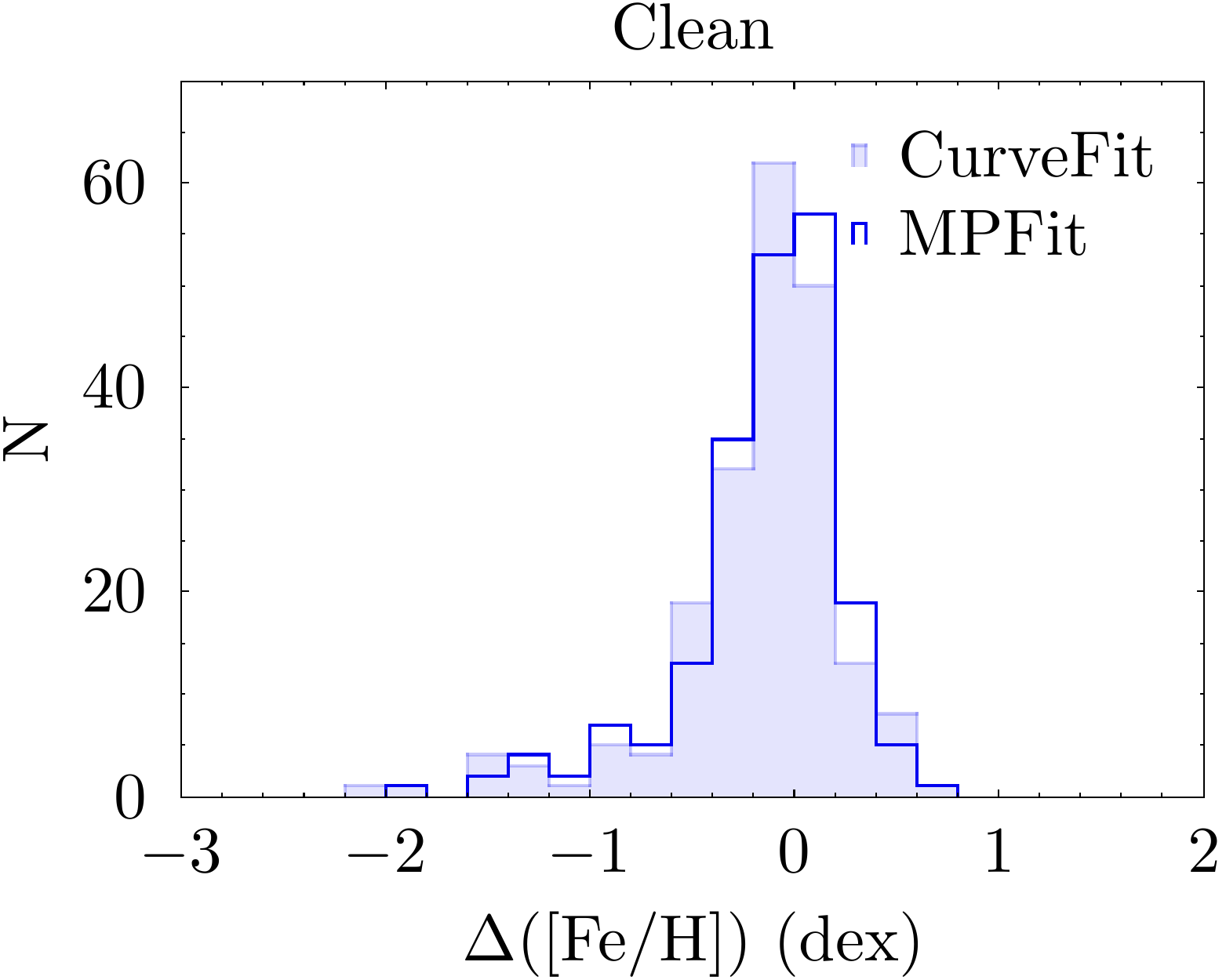}
		\includegraphics[width=0.245\linewidth, height=0.19\textwidth]{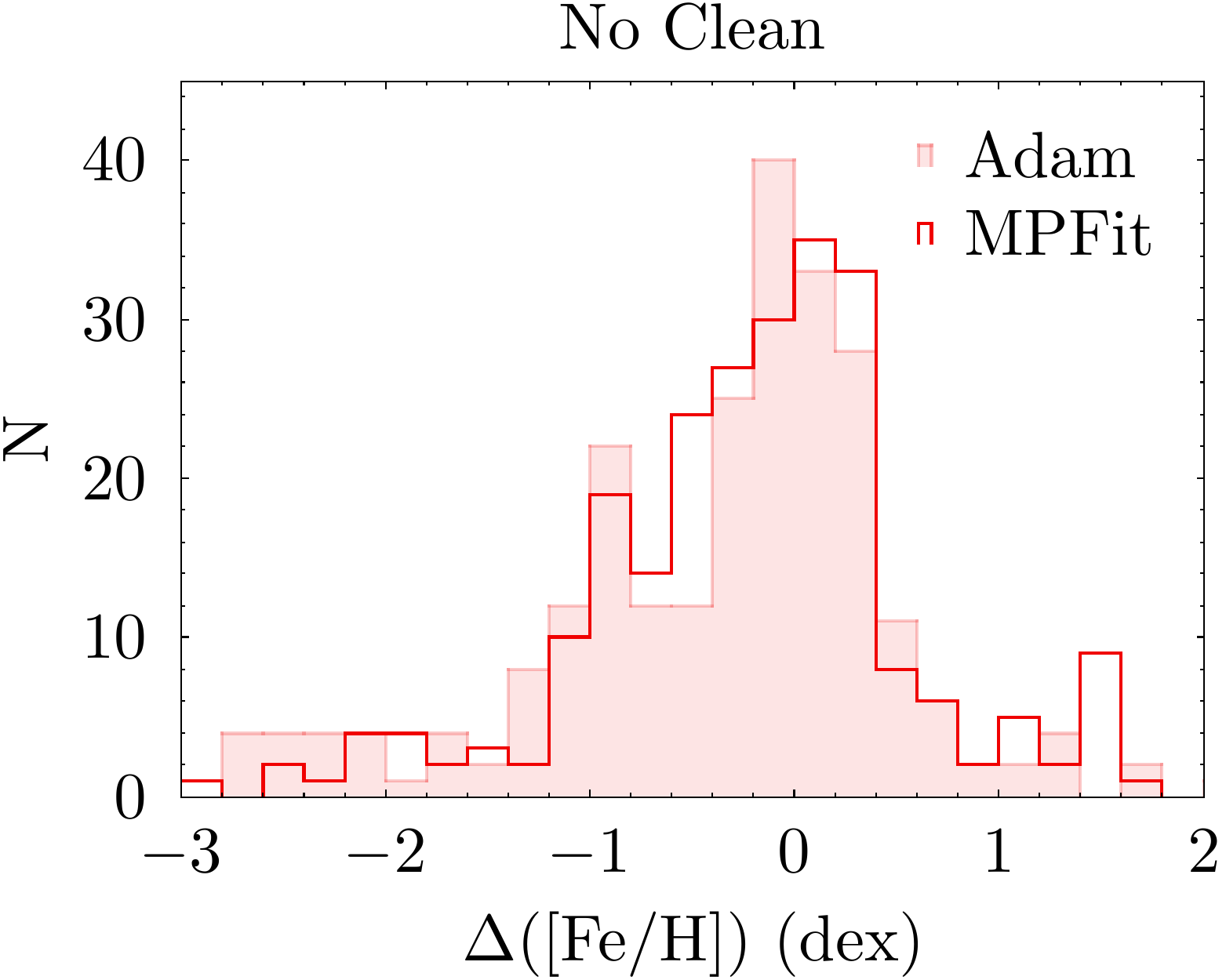}
		\includegraphics[width=0.245\linewidth, height=0.195\textwidth]{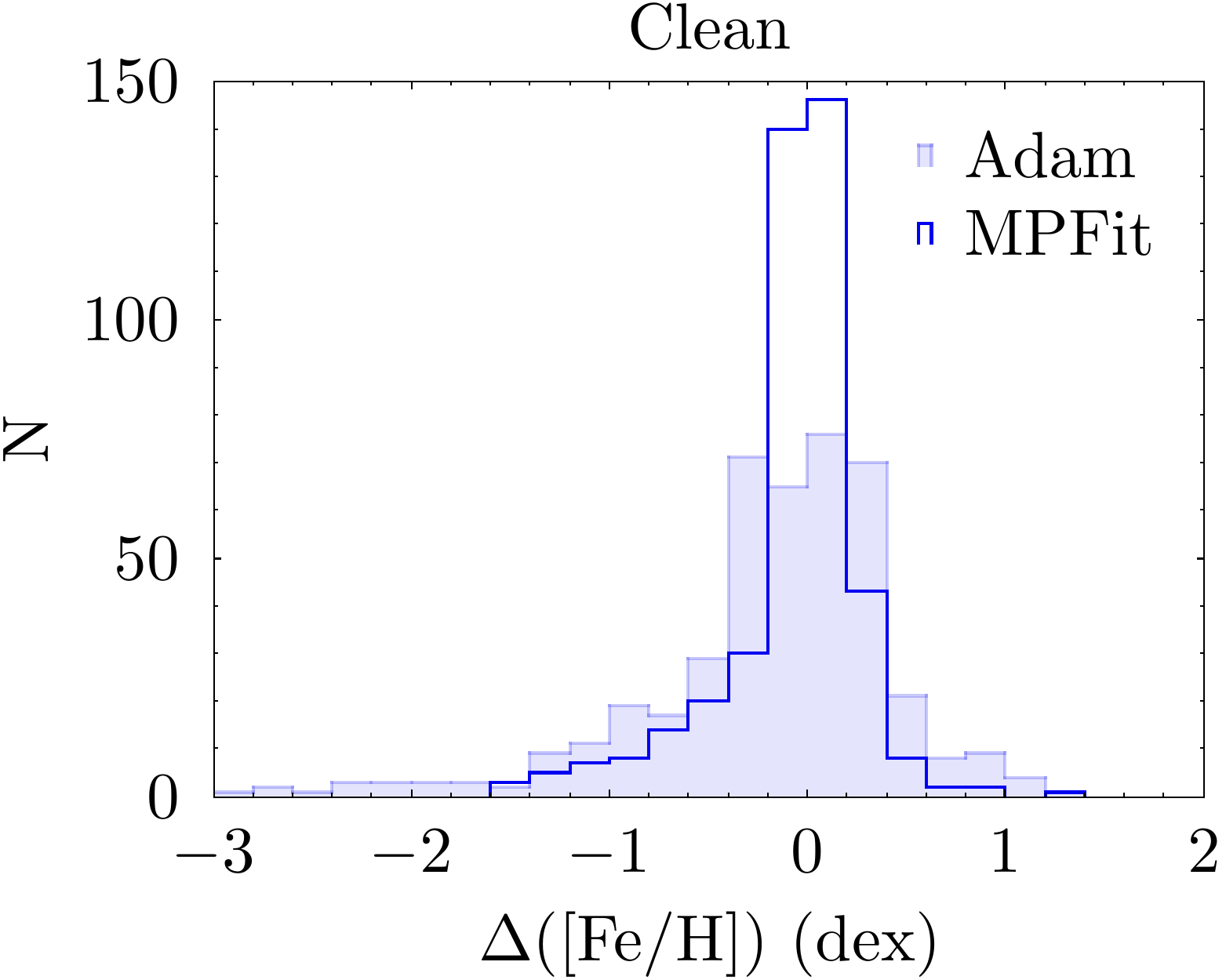}
		\includegraphics[width=0.245\linewidth, height=0.195\textwidth]{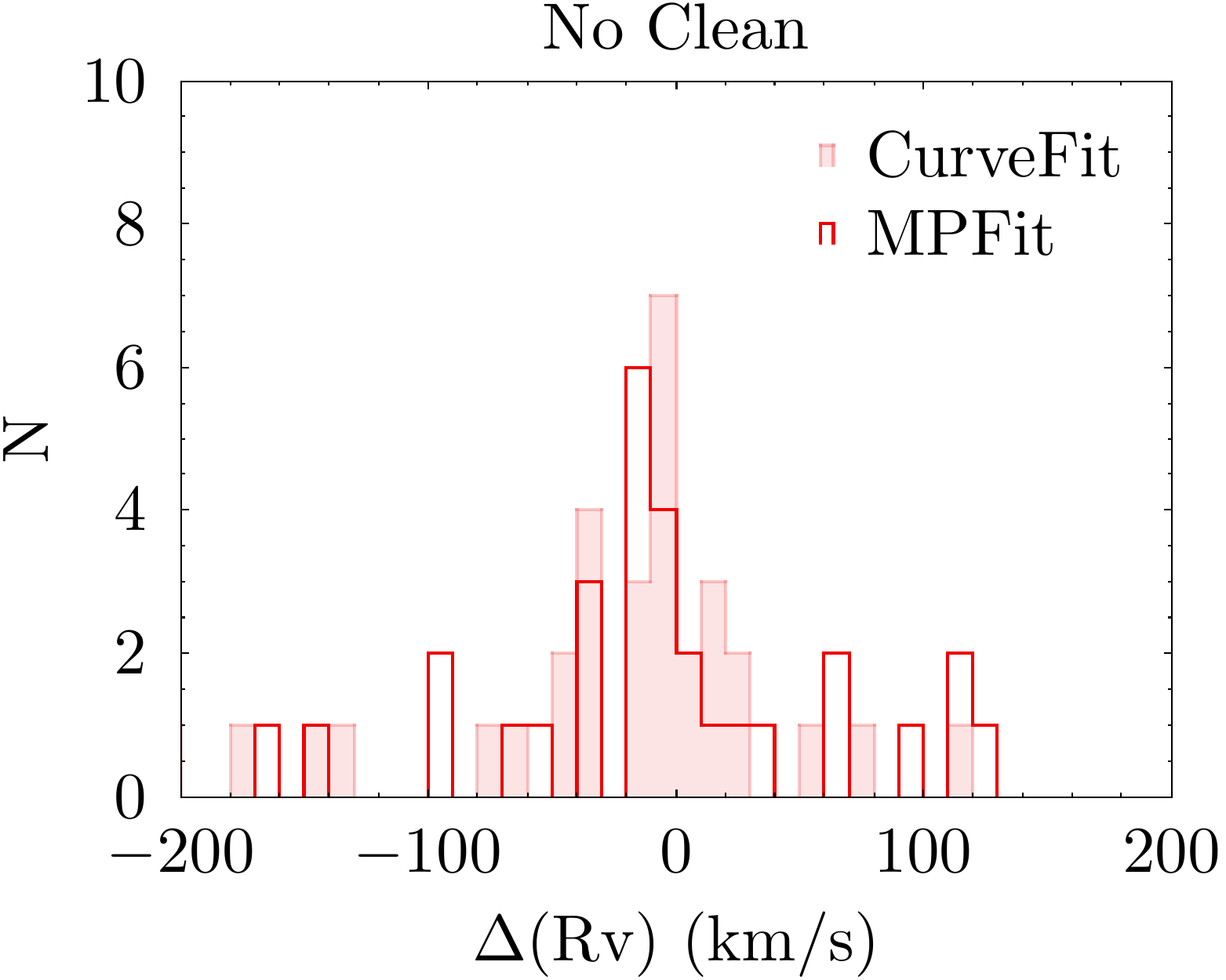}
		\includegraphics[width=0.245\linewidth, height=0.19\textwidth]{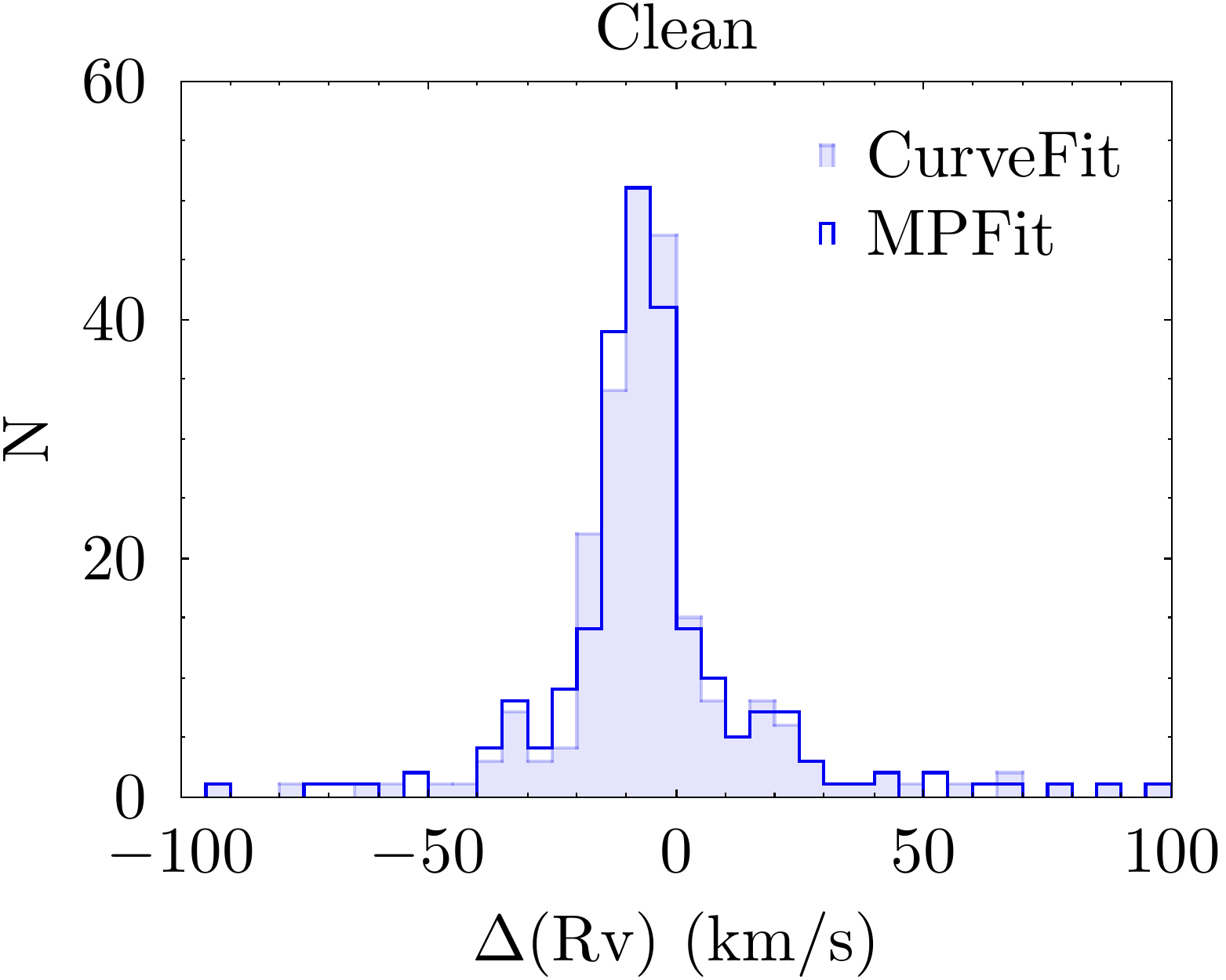}
		\includegraphics[width=0.245\linewidth, height=0.19\textwidth]{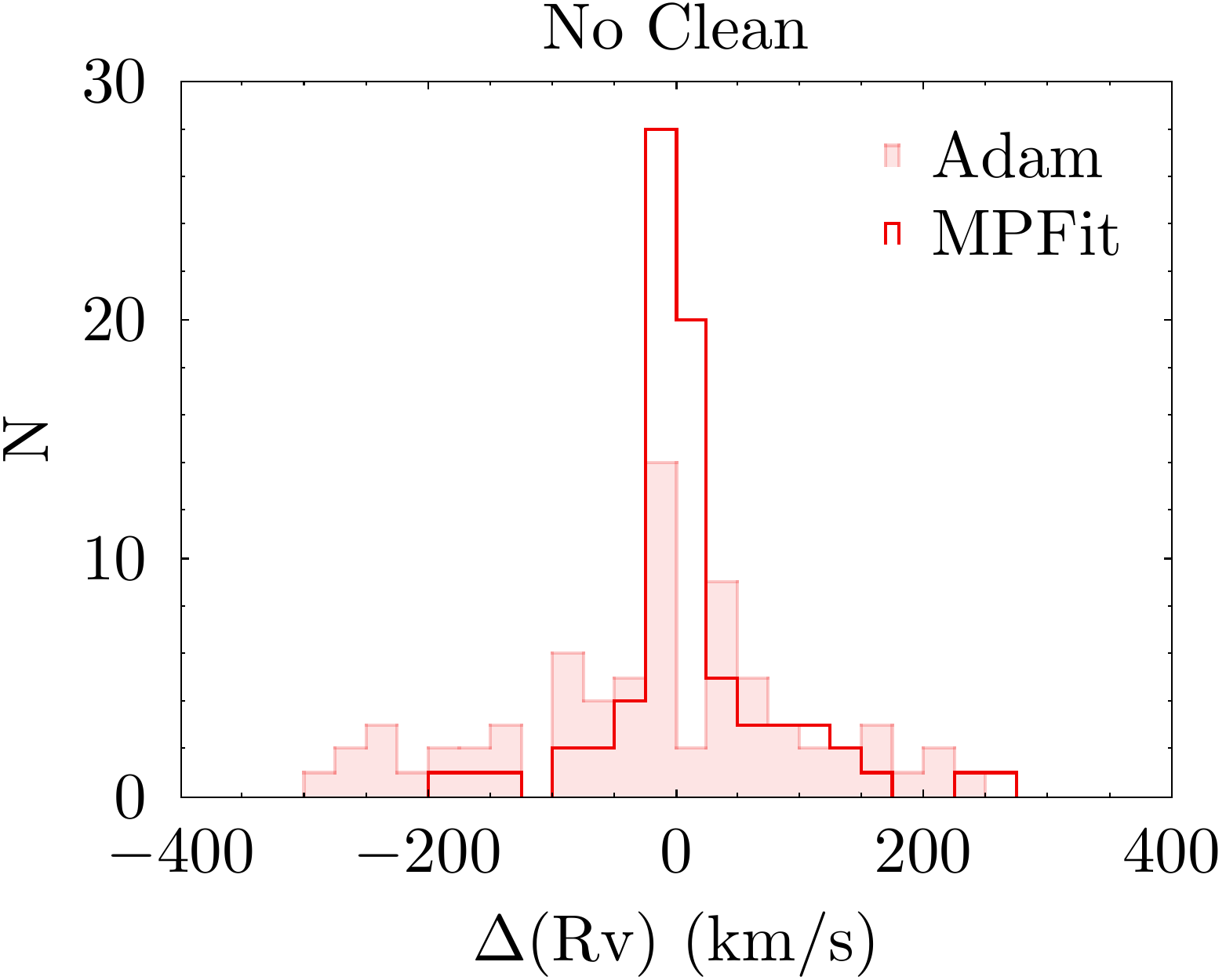}
		\includegraphics[width=0.245\linewidth, height=0.195\textwidth]{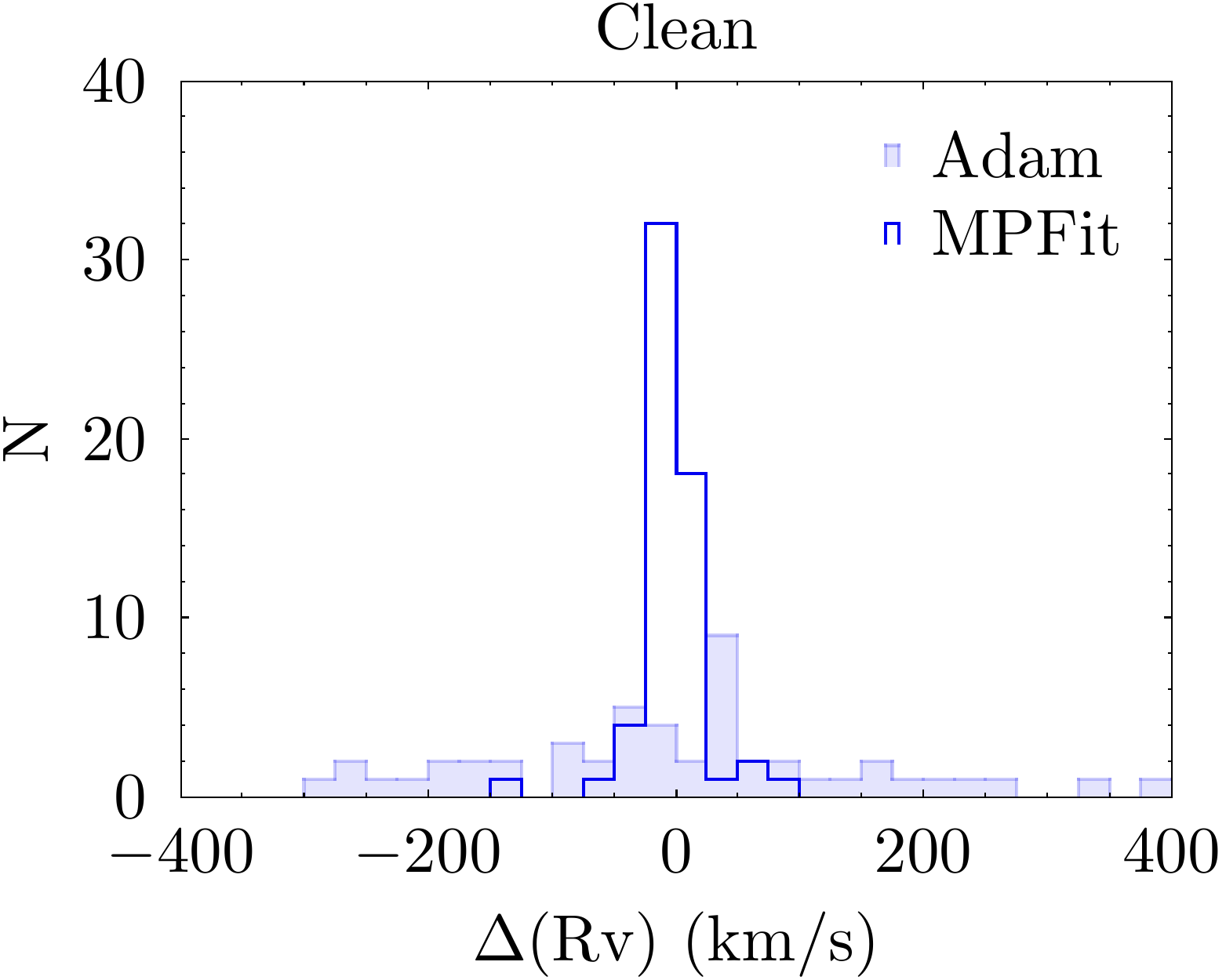}
		\caption{Residual distributions between LASP-derived parameters and APOGEE DR$16$ labels for outlier cases. Only parameters with internal discrepancies exceeding $5\sigma$ between the Python and IDL versions are included. Each row corresponds to a stellar parameter ($T_\mathrm{eff}$, $\log g$, $\mathrm{[Fe/H]}$, and ${\mathrm{RV}}$, from top to bottom), and each column represents a different LASP implementation and pixel masking method: LASP-CurveFit with No Clean, LASP-CurveFit with Clean, LASP-Adam-GPU with No Clean, and LASP-Adam-GPU with Clean (from left to right). In each panel, the residuals between APOGEE labels and the parameters derived by both LASP-MPFit and PyLASP are shown, allowing direct comparison of their consistency with APOGEE. For clarity, the \lq LASP\rq\ prefix is omitted from all legend labels.}
		\label{vsAPOGEE}
	\end{figure*}
	
	\begin{figure*}[htp!]
		\centering
		\includegraphics[width=0.98\linewidth]{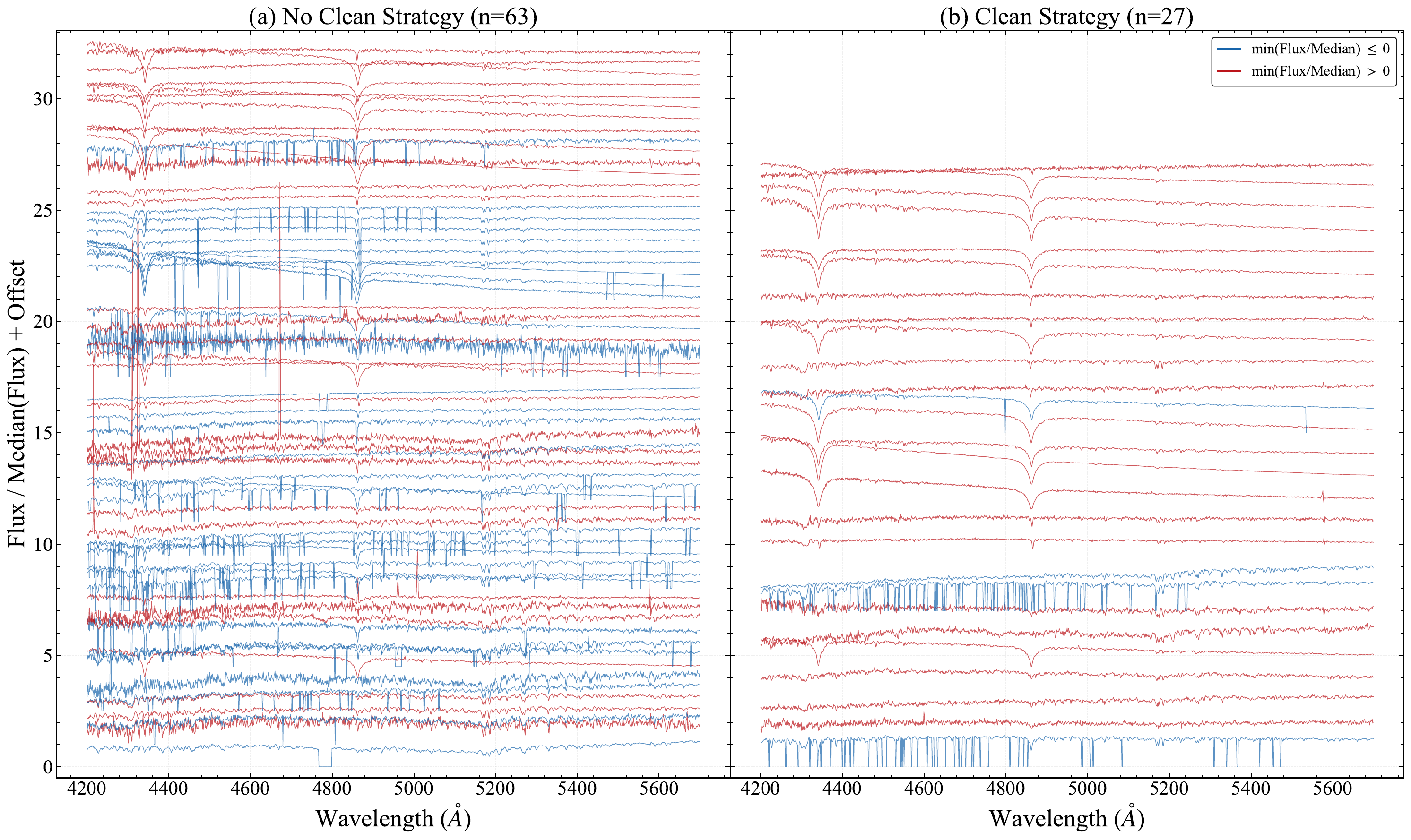}
		\caption{Spectral flux of stars with $T_{\mathrm{eff}}$ differences exceeding $5\sigma$ between LASP-CurveFit and LASP-MPFit. The left panel (a) shows the spectra under the No Clean strategy (n=63), while the right panel (b) shows the spectra under the Clean strategy (n=27). Blue lines represent spectra with negative flux values, while red lines represent spectra with no negative flux values.}
		\label{5std}
	\end{figure*}
	
	\section{Experiments} \label{Data Experiment}
	This section first introduces the efficiency-related parameters\footnote{There are two efficiency parameters: the number of parallel processes (\texttt{n\_jobs}) used in LASP-MPFit and LASP-CurveFit, and the number of spectra processed per group (\texttt{N}) in LASP-Adam-CPU and LASP-Adam-GPU, as defined in Equation~\ref{loss}.} in the Python and IDL versions of LASP, and evaluates the inference efficiency of each version using 10-million-scale spectroscopic datasets. After identifying the optimal efficiency configurations, we perform parameter inference on $177{,}848$ LAMOST stellar spectra and compare the results from LASP-CurveFit and LASP-Adam-GPU with those from LASP-MPFit, focusing on the consistency of $T_{\mathrm{eff}}$, $\log g$, [Fe/H], and $\mathrm{RV}$, as well as on the sources of discrepancies. We further assess the robustness of the Python implementation by testing different initial parameter settings. In addition, we compare the model-propagated errors from LASP-Adam-GPU with the empirical errors adopted in the official LASP-MPFit, which are estimated based on repeat observations and reduced $\chi^2$, to quantify the differences between the two error models. Finally, we apply LASP-CurveFit and LASP-Adam-GPU to a DESI DR$1$ dataset to further evaluate their cross-survey applicability and performance.
	
	\subsection{Setting efficiency parameters and evaluating inference efficiency} \label{Setting efficiency parameters}
	To determine the optimal efficiency settings for large-scale processing, we randomly select $10{,}000$ spectra from the $177{,}848$ LAMOST stellar spectra, and test the efficiency parameters of LASP-MPFit, LASP-CurveFit, LASP-Adam-CPU, and LASP-Adam-GPU across four computing platforms. The tested parameters include the number of parallel processes (\texttt{n\_jobs}) for CPU-based methods and the number of spectra simultaneously processed by the objective function (\texttt{N}) for Adam-based methods (see Table~\ref{laspdeviceconfig}). All tests are conducted under the No Clean strategy.
	
	As shown in Figure~\ref{IDL-Python-Time}, the inference time for LASP-MPFit and LASP-CurveFit decreases significantly with increasing \texttt{n\_jobs}, stabilizing once \texttt{n\_jobs} approaches half of the physical core count. On the same hardware, LASP-CurveFit achieves approximately $1.7$ times the efficiency of LASP-MPFit even at \texttt{n\_jobs=1}, and consistently outperforms it under all levels of parallelism, demonstrating the benefits of reconstruction-based optimization. For LASP-Adam, inference speed improves rapidly with increasing \texttt{N}. LASP-Adam-GPU reaches performance saturation around $\texttt{n\_jobs} {\approx} 2000$, and on an RTX 4060 achieves a speedup of $191$ times at \texttt{N=2500} compared to \texttt{N=1}, highlighting its strong parallel processing capability. LASP-Adam-CPU also benefits from increased \texttt{N}, performing optimally in the range $\texttt{N} {\in} [100,700]$ on a Ryzen 9 CPU and $\texttt{N} {\in} [20,500]$ on a Xeon platform, but remains significantly slower than the GPU version. Based on the shortest inference time across all configurations, we identify the optimal efficiency settings and provide recommended ranges for future use (see Table~\ref{laspdeviceconfigBest}).
	
	Using these optimal settings, we extrapolate the total time required to process $10$ million spectra. On a laptop with a Ryzen 9 7945HX CPU and an RTX 4060 GPU, LASP-Adam-GPU completes the task in approximately $26$ hr, outperforming LASP-CurveFit ($48$ hr) and LASP-MPFit ($84$ hr). LASP-Adam-CPU is the slowest, requiring $134$ hr and is therefore not suitable for large-scale inference tasks. On high-performance platforms, LASP-Adam-GPU achieves excellent scalability: it completes $10$ million spectra in about $8$ hr on an RTX 3090 and $7$ hr on an NVIDIA A100. Under the Clean strategy, inference time increases slightly due to the additional iterations required to dynamically mask outliers. For example, LASP-CurveFit (Ryzen 9) requires $63$ hr, while LASP-Adam-GPU takes $76$, $23$, and $19$ hr on RTX 4060, RTX 3090, and A100, respectively.
	
	In summary, LASP-Adam-GPU is highly suitable for large-scale stellar parameter inference. LASP-CurveFit offers an efficient CPU-based alternative, while LASP-Adam-CPU is not recommended due to its limited performance scalability.
	
	\subsection{Evaluating parameter consistency and robustness} \label{Evaluating parameter consistency and robustness}
	\subsubsection{Comparison with the baseline method} \label{Comparison with the baseline method}
	We evaluate the consistency between the parameters inferred by PyLASP and the baseline method LASP-MPFit, focusing on the performance of LASP-CurveFit and LASP-Adam-GPU. To ensure a valid comparison, we exclude the following cases in which parameter inference fails: (1) those with negative multiplicative shape correction factors (see Step~5 in Section~\ref{LASP-Python-GPU})\footnote{LASP-CurveFit treats negative multiplicative correction factors as failed fits by default. Note that both LASP-MPFit and LASP-CurveFit can dynamically adjust the order of the Legendre polynomial to reduce such failures.}, (2) those where the optimization does not converge within the maximum number of iterations, and (3) those for which the CFI-provided initial values fall outside the parameter space allowed by the ELODIE library. The number of valid samples after filtering is listed in Table~\ref{nsample}.
	
	As shown in Figure~\ref{AllParamCompare}, the parameter offsets between the Python and IDL versions of LASP are close to zero across all four parameters: $\mathrm{RV}$, $T_{\mathrm{eff}}$, $\log g$, and [Fe/H]. Under both the No Clean and Clean strategies, the standard deviations of the parameter differences between LASP-CurveFit and LASP-MPFit are 0.04/0.04\,km\,s$^{-1}$, 4/5\,K, 0.005/0.005\,dex, and 0.003/0.003\,dex, respectively. In comparison, LASP-Adam-GPU exhibits slightly larger discrepancies: 0.37/1.12\,km\,s$^{-1}$, 10/24\,K, 0.01/0.03\,dex, and 0.006/0.02\,dex. These differences primarily stem from two sources. First, the current version of LASP-Adam-GPU prioritizes computational efficiency and scalability, and does not yet implement the same failure detection mechanisms used in LASP-CurveFit and LASP-MPFit (e.g., masking spectra with anomalous multiplicative corrections). Second, under the Clean strategy, LASP-Adam-GPU employs a uniform early-stopping criterion for all spectra in a group, rather than dynamically determining convergence on a per-spectrum basis. Future versions will improve accuracy by refining failure detection and control strategies. 
	
	\begin{figure*}[!htbp]
		\centering
		\includegraphics[width=0.245\linewidth]{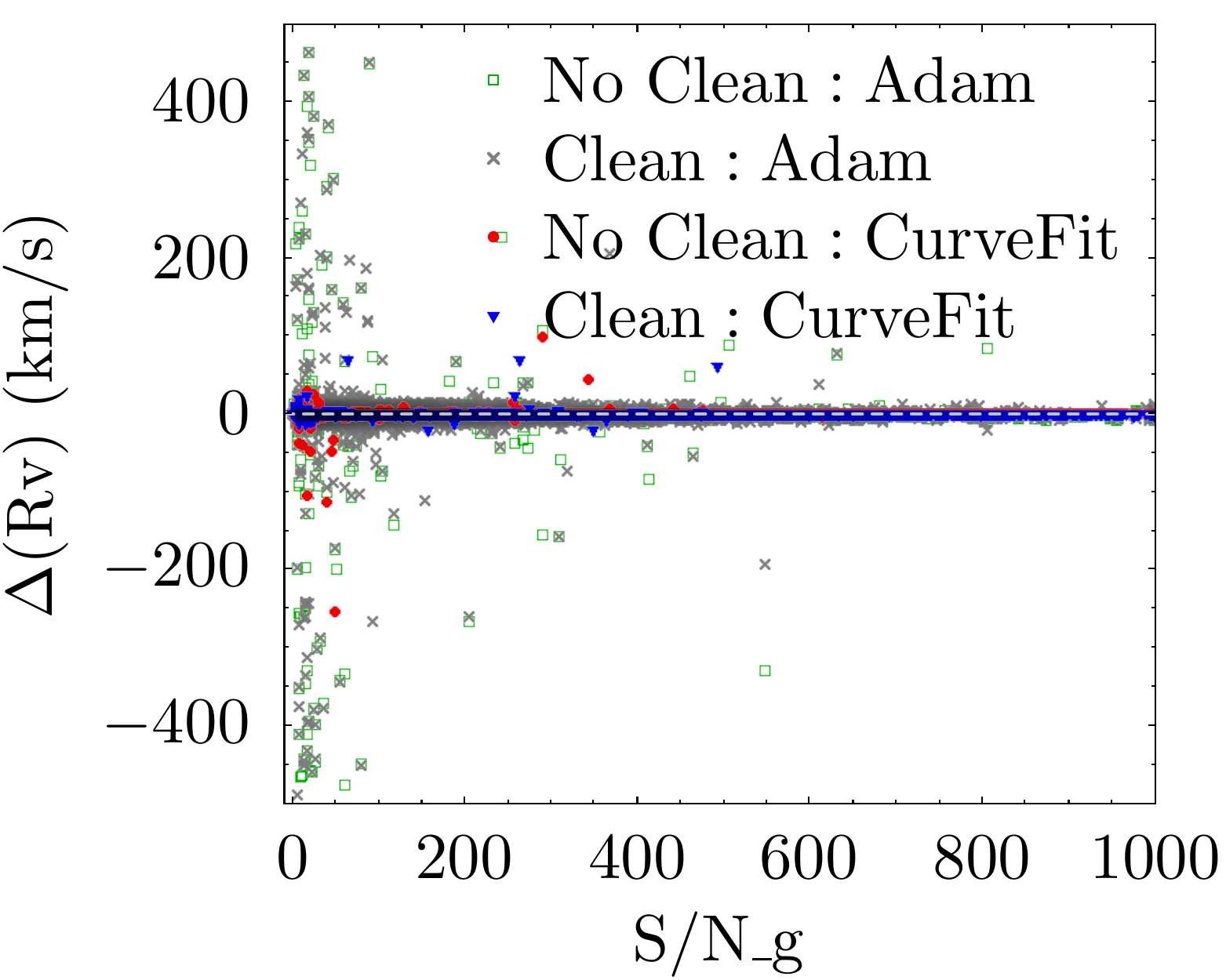}
		\includegraphics[width=0.245\linewidth]{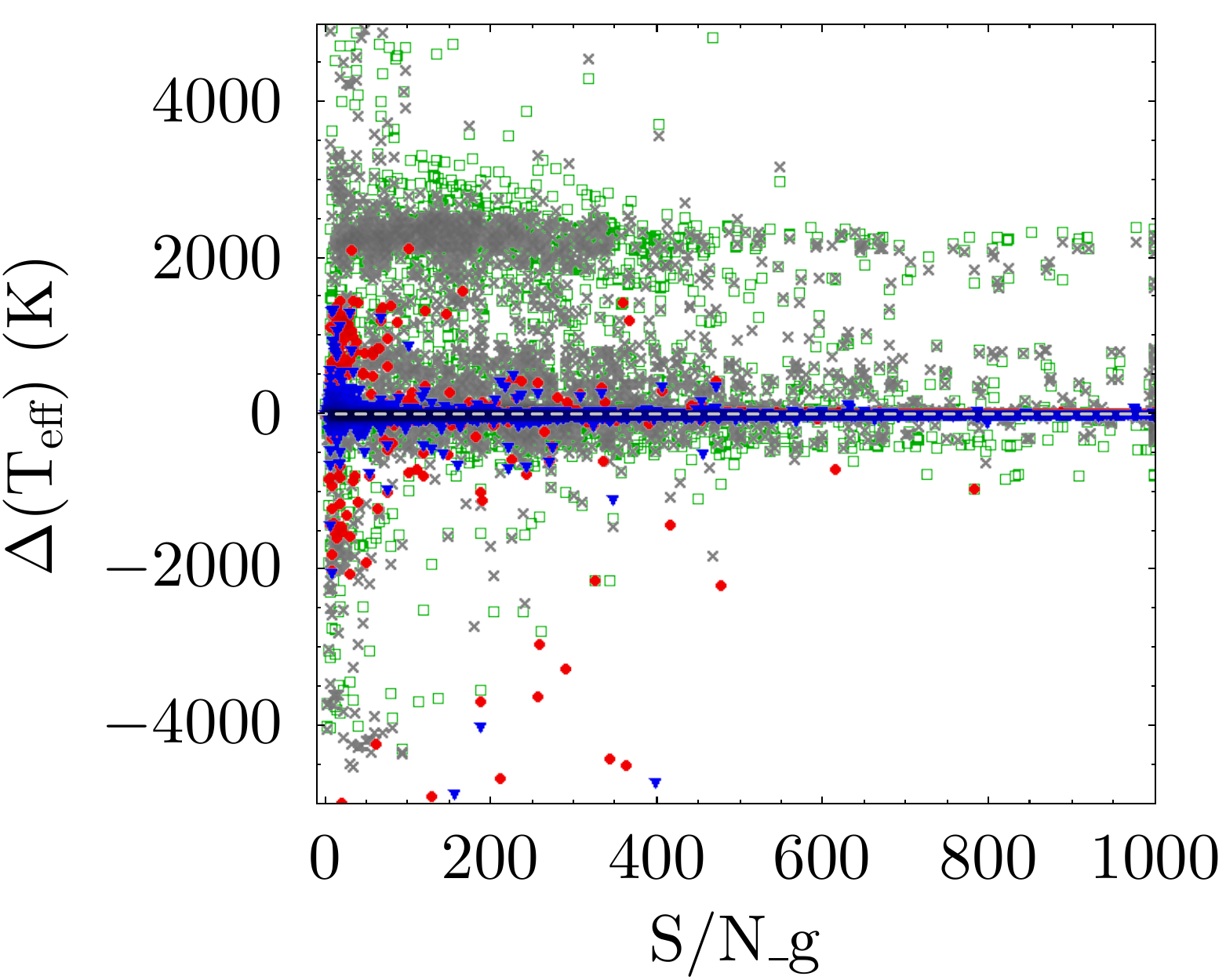}
		\includegraphics[width=0.245\linewidth]{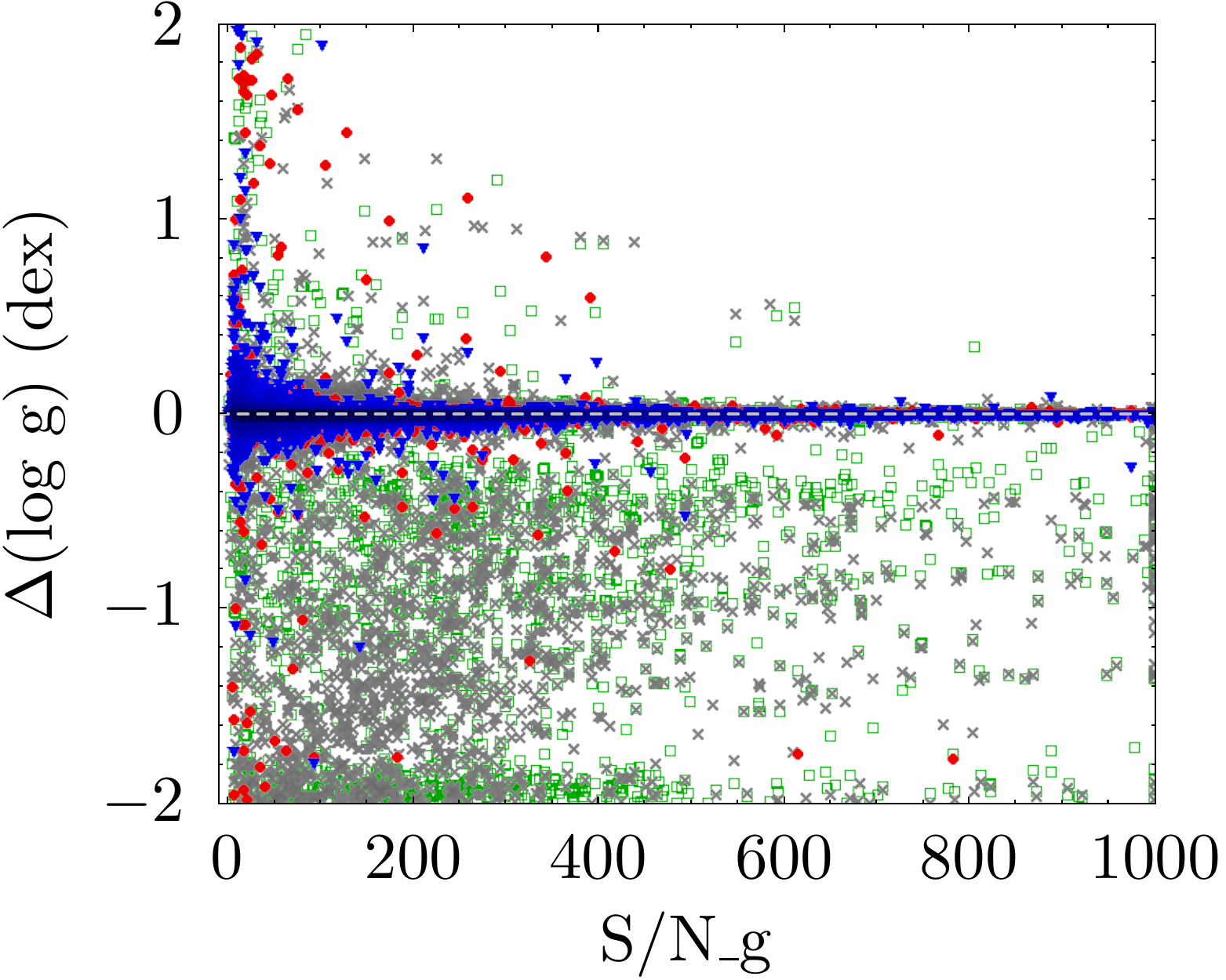}
		\includegraphics[width=0.245\linewidth]{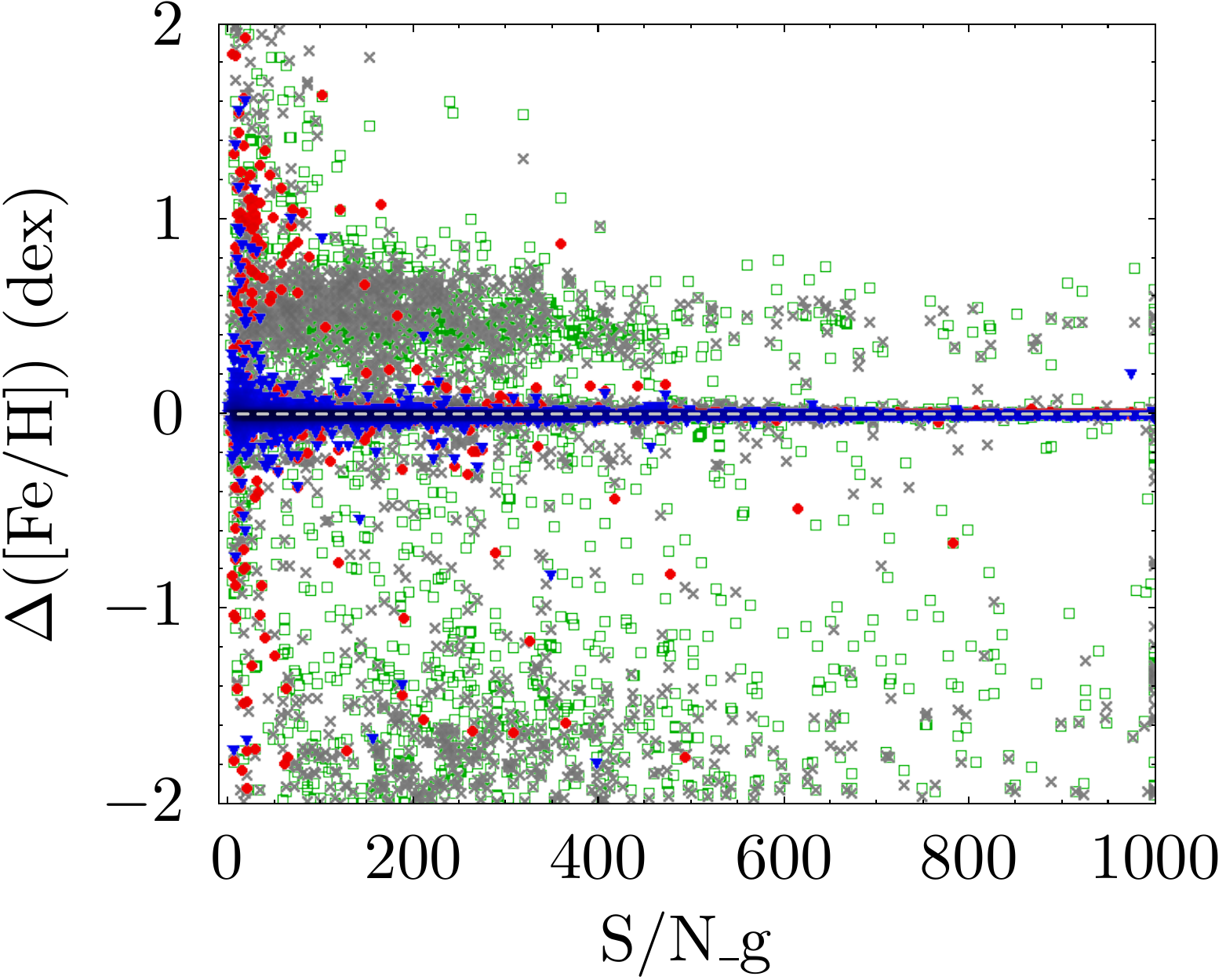}
		\caption{Stability test of initial values in PyLASP. $\Delta$ denotes the difference between stellar parameters inferred using fixed initial values $(T_{\mathrm{eff}}, \log g, \mathrm{[Fe/H]}) = (5000 \ \mathrm{K}, 3 \ \mathrm{dex}, -0.5 \ \mathrm{dex})$ and those inferred using CFI-derived initial values. For clarity, the legend is shown only in the first panel, and the \lq LASP\rq \ prefix is omitted from all legend labels.}
		\label{NoCFIvsCFI}
	\end{figure*}
	
	\begin{figure*}[!htbp]
		\centering
		\includegraphics[width=0.245\linewidth]{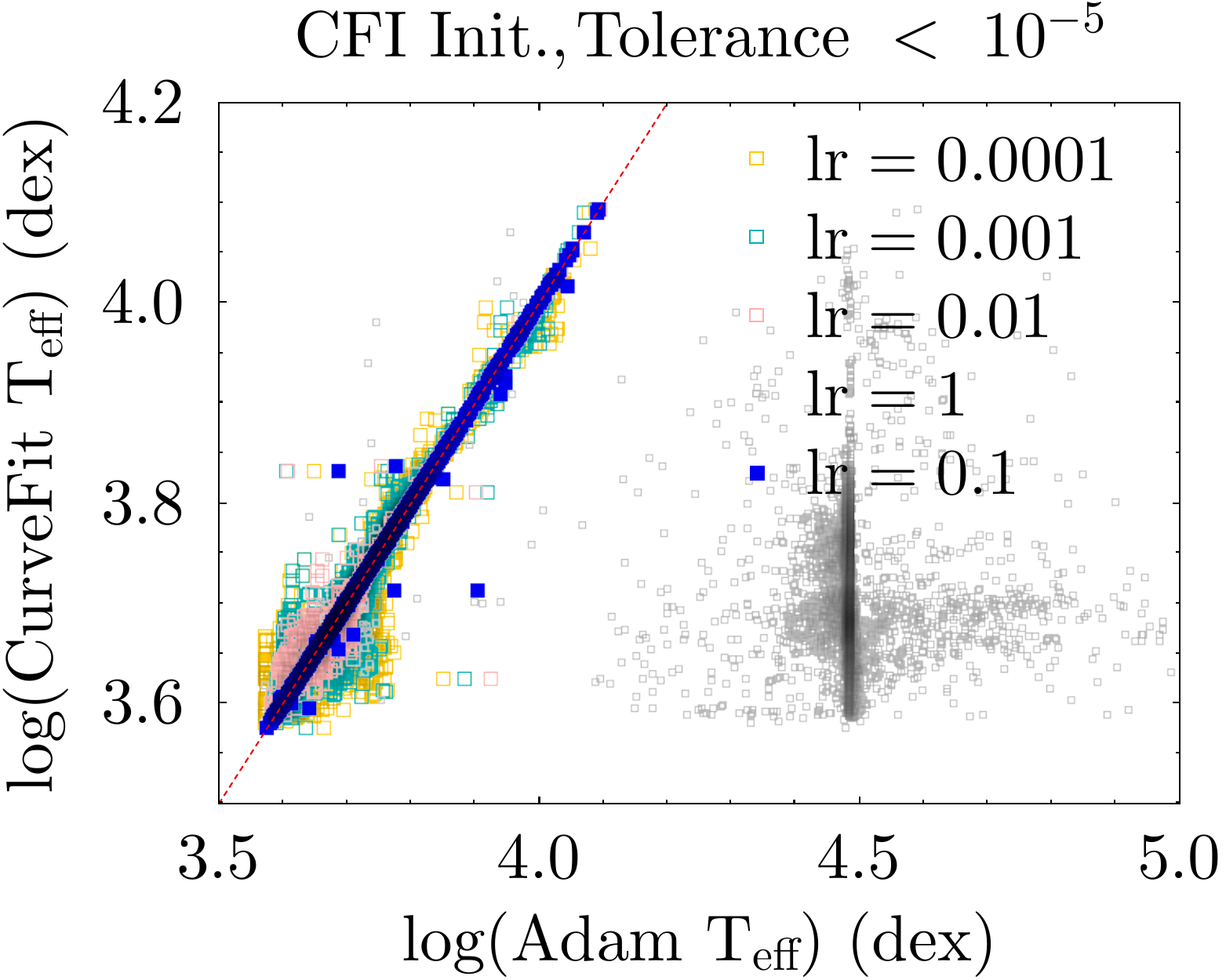}
		\includegraphics[width=0.245\linewidth]{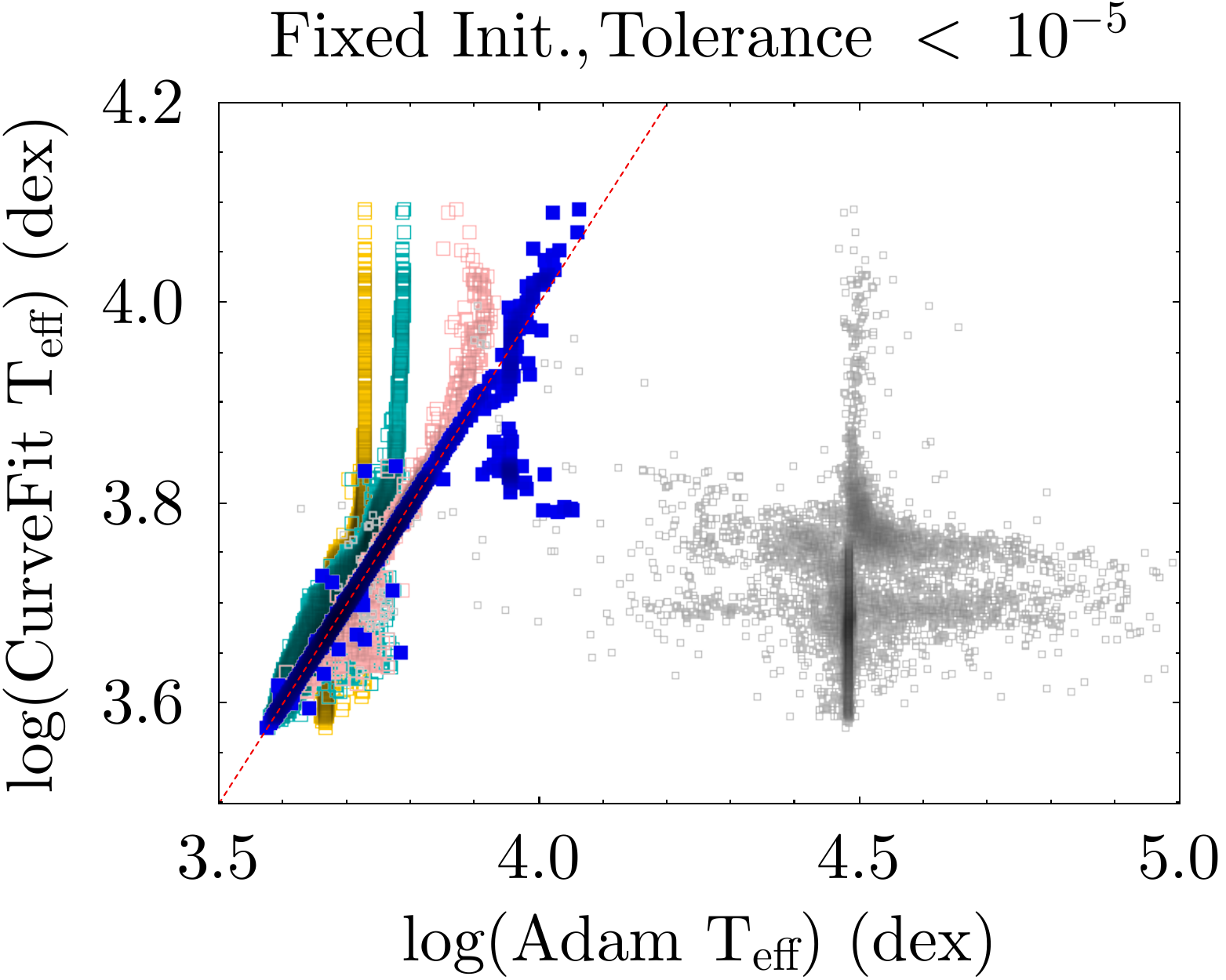}
		\includegraphics[width=0.245\linewidth]{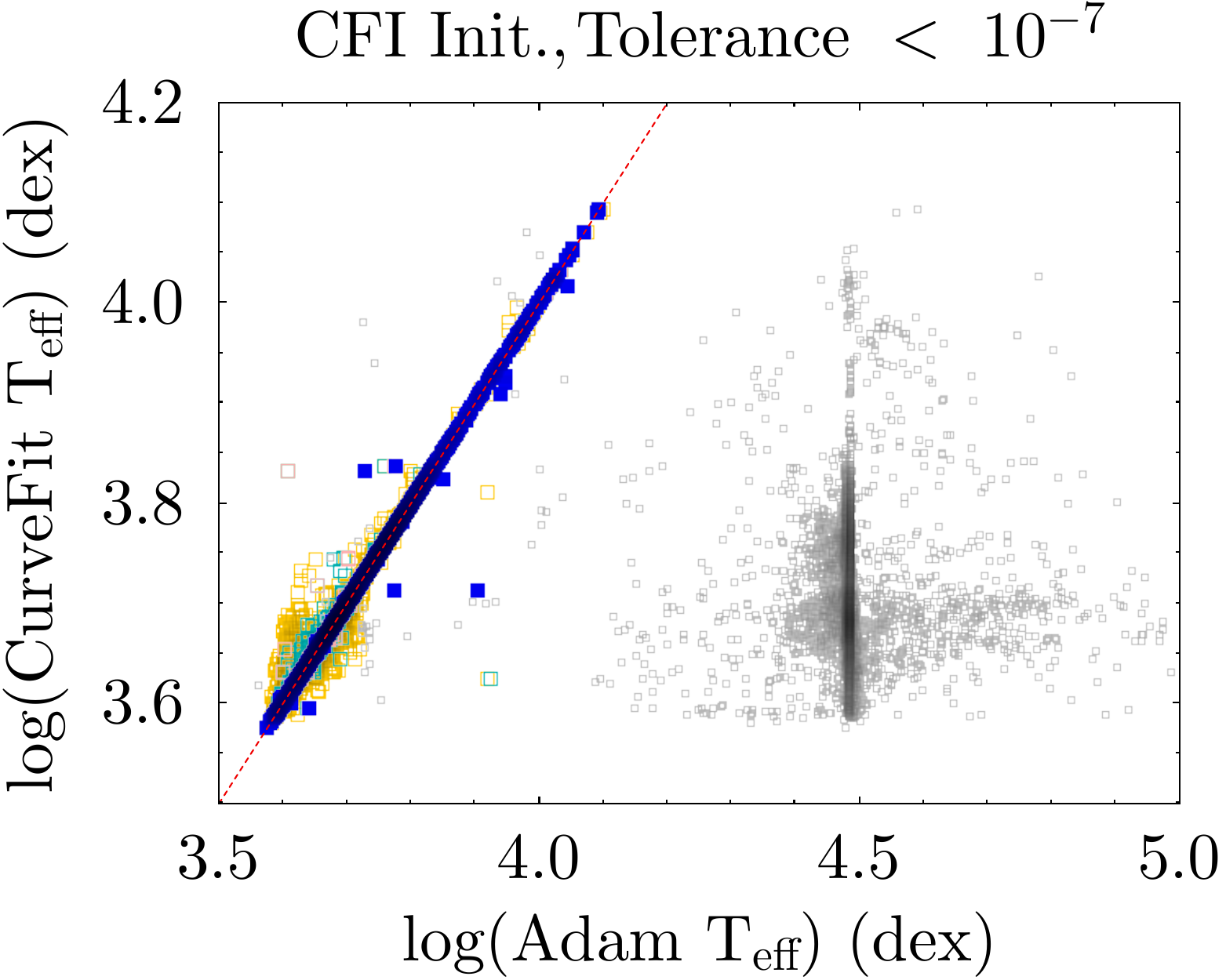}
		\includegraphics[width=0.245\linewidth]{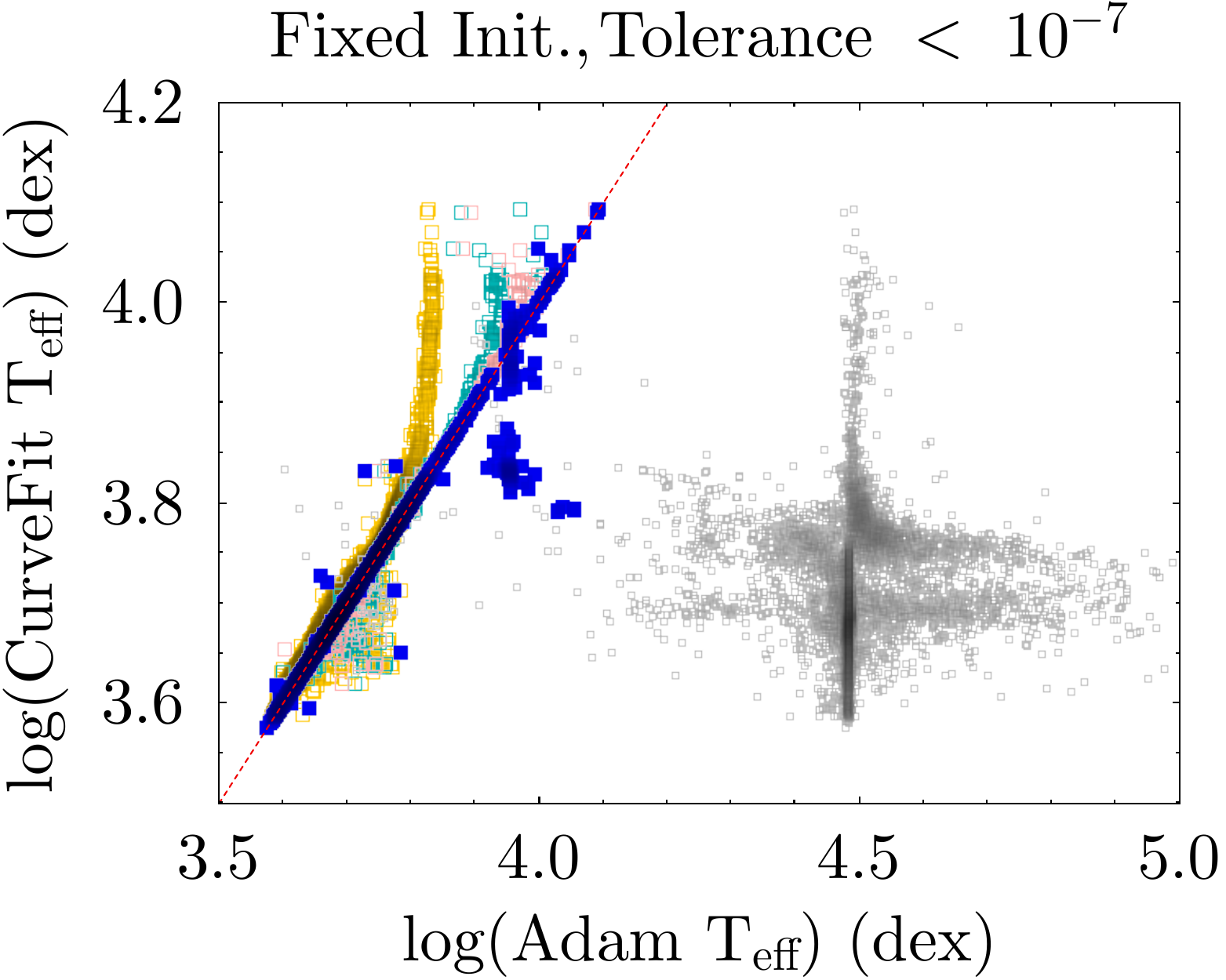}
		\caption{Comparison of $T_{\mathrm{eff}}$ between LASP-Adam-GPU and LASP-CurveFit under varying initialization schemes, based on $10{,}000$ randomly selected spectra. LASP-Adam-GPU uses either CFI-derived values (first and third panels) or fixed initial values $(5000\ \mathrm{K},\ 3\ \mathrm{dex},\ -0.5\ \mathrm{dex})$ (second and fourth panels), while LASP-CurveFit consistently uses CFI-derived values. The first two panels use a convergence threshold of $10^{-5}$, and the last two use $10^{-7}$. Colors of the scatter points indicate different learning rates (lr). The $x$-axis shows LASP-Adam-GPU results, and the $y$-axis shows LASP-CurveFit results. For clarity, the legend is shown only in the first panel, and the \lq LASP\rq \ prefix is omitted in all figure labels.}
		\label{lr}
	\end{figure*}

	To determine which method yields more reliable parameter estimates in cases of significant disagreement, we identify spectra where any of the four parameters differs by more than $5\sigma$ between the Python and IDL versions of LASP. This subset, accounting for no more than $0.35\%$ of the total sample (see Table~\ref{nsample}), is compared against APOGEE reference labels (Figure~\ref{vsAPOGEE}) to evaluate which version is closer to the external standard. Most of these outliers are associated with low-quality LAMOST spectra (Figure~\ref{5std}). For $T_{\mathrm{eff}}$ under the No Clean strategy, LASP-CurveFit exhibits a deviation distribution more consistent with APOGEE than LASP-MPFit: the proportion of samples with $\Delta(T_{\mathrm{eff}}) \in [-500, 500]$ K is approximately $18\%$ higher for LASP-CurveFit, while that with $\lvert\Delta(T_{\mathrm{eff}})\rvert > 1000$ K is about $32\%$ lower, indicating enhanced robustness to outliers. The underlying cause of this phenomenon remains unclear, but it may be related to the scale-adjusting effect of median-based normalization, differences in optimizer convergence behavior, or the numerical stability of operations such as matrix inversion during parameter inference. Under the Clean strategy, the deviation distributions of LASP-CurveFit and LASP-MPFit relative to APOGEE become comparable, suggesting that Clean effectively improves robustness. By contrast, LASP-Adam-GPU shows slightly worse agreement with APOGEE in the $\Delta(T_{\mathrm{eff}}) {=} 1000{-}3000$ K range. This is likely due to the absence of failure detection mechanisms and the use of a fixed early-stopping criterion for all spectra within a group, which may limit optimization depth for complex cases.
	
	In summary, PyLASP achieves parameter consistency with LASP-MPFit for approximately $99.65\%$ of the sample. LASP-CurveFit exhibits higher parameter accuracy for low-quality spectra, whereas LASP-Adam-GPU, by design, trades off some accuracy for computational efficiency. Given that the Clean strategy has minimal impact on the parameter consistency between LASP-CurveFit and LASP-MPFit, we speculate that introducing a dynamic per-spectrum Clean early-stopping strategy within LASP-Adam-GPU could substantially reduce the dispersion in parameter differences relative to LASP-MPFit, potentially restoring the consistency level observed under the No Clean condition.
	
	\subsubsection{Sensitivity to initial values} \label{Sensitivity}
	To evaluate the sensitivity of PyLASP implementations to initialization, we compare two approaches to $177{,}848$ spectra: (1) a fixed initialization of $(T_{\mathrm{eff}}, \log g, \mathrm{[Fe/H]}) = (5000\,\mathrm{K}, 3\,\mathrm{dex}, -0.5\,\mathrm{dex})$, simulating a worst-case scenario with no prior information, and (2) the use of CFI-derived initial values. As shown in Figure~\ref{NoCFIvsCFI}, LASP-CurveFit is insensitive to the choice of initial values: Under the No Clean strategy, only $0.09\%$, $0.11\%$, $0.10\%$, and $0.11\%$ of samples show differences exceeding $1$ km s$^{-1}$ in $\mathrm{RV}$, $100$ K in $T_{\mathrm{eff}}$, $0.1$ dex in $\log g$, and $0.1$ dex in [Fe/H], respectively. The corresponding fractions under the Clean strategy are $0.13\%$, $0.10\%$, $0.17\%$, and $0.08\%$. Most of these discrepancies are associated with low-quality spectra with signal-to-noise ratios (S/N) below $20$. In contrast, LASP-Adam-GPU is more sensitive to initialization for the three atmospheric parameters at $T_{\mathrm{eff}} > 8000$ K: Under the No Clean strategy, $76\%$, $80\%$, and $78\%$ of samples in this temperature range exhibit differences exceeding $100$ K in $T_{\mathrm{eff}}$, $0.1$ dex in $\log g$, and $0.1$ dex in [Fe/H], respectively; under the Clean strategy, the corresponding values decrease to $65\%$, $68\%$, and $66\%$. These deviations appear to be independent of S/N. At $T_{\mathrm{eff}} \leq 8000$ K, LASP-Adam-GPU shows robustness comparable to that of LASP-CurveFit, with low sensitivity to initialization under both masking strategies.
	
\begin{figure}[t]
		\centering
		\includegraphics[width=0.98\linewidth]{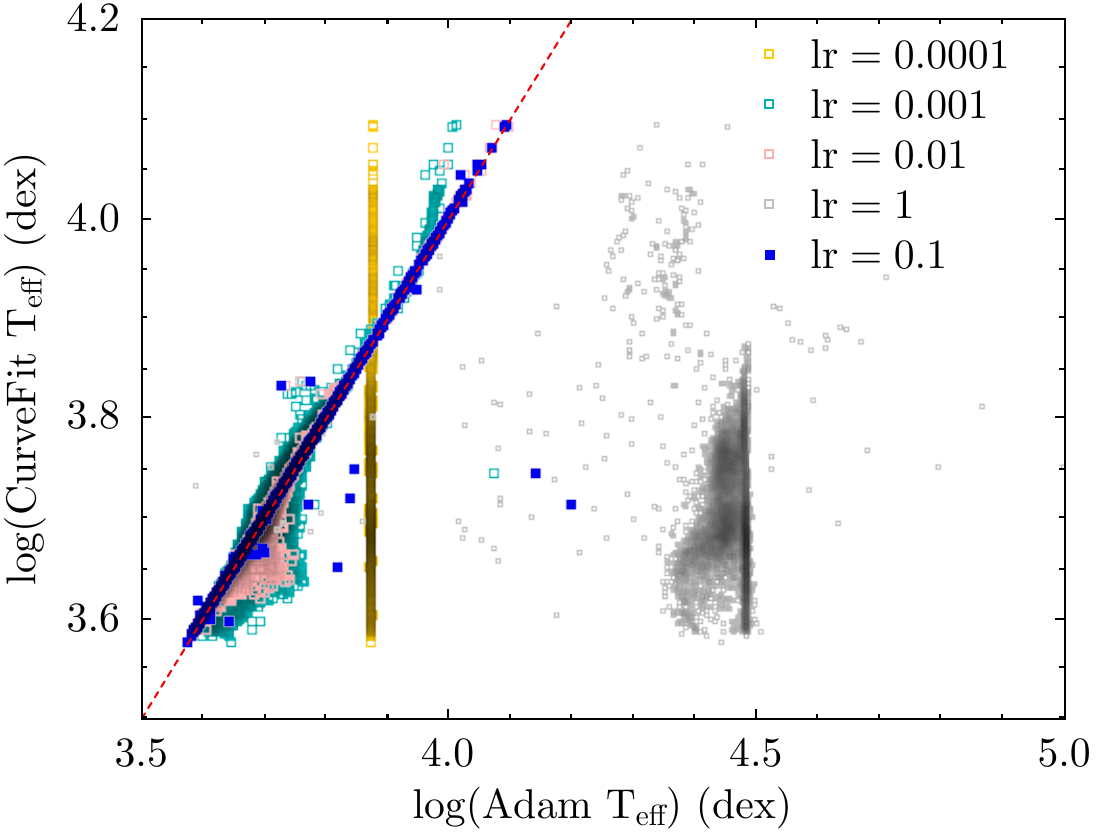}
		\caption{Comparison of $T_{\mathrm{eff}}$ between LASP-Adam-GPU and LASP-CurveFit under different initialization schemes, based on $10{,}000$ randomly selected spectra. LASP-Adam-GPU uses fixed initial values of $(7500\ \mathrm{K},\ 3\ \mathrm{dex},\ -0.5\ \mathrm{dex})$ and applies a convergence threshold of $10^{-5}$, while LASP-CurveFit uses CFI-derived initial values. Scatter point colors indicate different learning rates. The $x$-axis shows LASP-Adam-GPU results, and the $y$-axis shows LASP-CurveFit results. The \lq LASP\rq \ prefix is omitted in all figure labels for clarity.}
		\label{7500NoCFIvsCurveFitCFI}
	\end{figure}

	To investigate the causes of this behavior, we analyze the problem from two perspectives:
	\begin{enumerate}
		\item \textbf{Impact of optimizer learning rate and convergence threshold.}
		To evaluate the impact of optimizer configurations on the inferred parameters under different initialization strategies, we randomly select $10{,}000$ spectra and evaluate the consistency of LASP-Adam-GPU results against those from LASP-CurveFit (using CFI initialization), under a range of learning rates ($1$, $0.1$, $0.01$, $10^{-3}$, and $10^{-4}$) and two convergence criteria (requiring the loss to vary by less than $10^{-7}$ or $10^{-5}$ over $50$ consecutive iterations). As shown in \reffig{lr}, second and final panels (corresponding to fixed initialization),  the consistency of $T_{\mathrm{eff}}$ between LASP-Adam-GPU and LASP-CurveFit improves with stricter convergence thresholds (e.g., $10^{-7}$) at a fixed learning rate, and with smaller learning rates when the convergence threshold is held constant (for learning rates below $0.1$). However, these improvements come with a substantial computational cost: for instance, with a learning rate of $10^{-4}$ and a threshold of $10^{-7}$, the runtime is approximately $14$ times longer than that for a learning rate of $0.1$; even at a $10^{-5}$ threshold, the factor remains as high as $4$. Inappropriate learning rate settings can lead to specific clustering patterns in the inferred $\log T_{\mathrm{eff}}$ with LASP-Adam-GPU: when the learning rate is too small, potentially incomplete convergence within the 
		$5000$-iteration limit set in \refsubsection{LASP-Python-GPU} causes values to stagnate near the initialization point (e.g., $\log T_{\mathrm{eff}} {\approx} 3.7$); conversely, excessively large learning rates may cause the optimizer to overshoot the optimal region, with updates stepping beyond reasonable ranges and resulting in clustering near $\log T_{\mathrm{eff}} {\approx} 4.5$. At a learning rate of $0.1$, a small accumulation near $\log T_{\mathrm{eff}}{\approx}3.95$ is also observed. Similar trends are observed under CFI initialization (first and third panels): stricter convergence improves consistency, while clustering due to inappropriate learning rate settings (except $0.1$) persists. Therefore, we infer that the clustering at $\log T_{\mathrm{eff}}{\approx}3.95$ likely reflects inappropriate initialization rather than the learning rate or convergence threshold, whereas clustering at other locations is attributable to learning rate settings. Balancing computational efficiency and parameter consistency, the configuration with a learning rate of $0.1$ and a convergence threshold of $10^{-5}$ yields the best overall performance: in the first two panels, where LASP-Adam-GPU is initialized with CFI and fixed values, respectively, the fraction of spectra for which the $T_{\mathrm{eff}}$ difference between LASP-Adam-GPU and LASP-CurveFit exceeds $100$ K is $0.1\%$ and $1.9\%$---the lowest among all learning rate configurations at this convergence threshold. We therefore recommend this configuration as the default for LASP-Adam-GPU.
		\item \textbf{Connection between initialization bias and emulator structure.}
		To assess whether the stability of LASP-Adam-GPU is affected by the piecewise structure of the spectral emulator (Step $2$ in \refsection{LASP-Python-GPU}), we test an alternative fixed initialization of $(T_{\mathrm{eff}}, \log g, \mathrm{[Fe/H]}) = (7500\ \mathrm{K}, 3\ \mathrm{dex}, -0.5\ \mathrm{dex})$. As shown in \reffig{7500NoCFIvsCurveFitCFI}, similar to the behavior observed in \reffig{lr}, inappropriate learning rates cause $\log T_{\mathrm{eff}}$ values inferred by LASP-Adam-GPU to cluster near the initialization or around $3.7$ and $4.5$, indicating that such clustering arises primarily from the learning rate rather than from the initialization strategy. Importantly, under the alternative initialization with a learning rate of $0.1$, the previously observed clustering near the emulator’s piecewise boundary at $\log T_{\mathrm{eff}} {\approx} 3.95$ (in the second and final panels of \reffig{lr}) is notably suppressed in \reffig{7500NoCFIvsCurveFitCFI}. This result indicates that LASP-Adam-GPU exhibits directional sensitivity to the initial $T_{\mathrm{eff}}$: optimization from lower to higher temperatures tends to be less stable, whereas the reverse direction is relatively robust. This may reflect the local gradient discontinuities along the $T_{\mathrm{eff}}$ axis, introduced by the emulator’s piecewise design, which can interfere with the optimizer’s update trajectory.
	\end{enumerate}
	
	In summary, LASP-CurveFit and LASP-Adam-GPU exhibit distinct behaviors with respect to initialization sensitivity. LASP-CurveFit yields stable results across the entire parameter space and is therefore suitable for scientific applications that demand high robustness. For LASP-Adam-GPU, CFI-based initialization is preferred when available; if such initialization is not accessible, we recommend adopting $(7500 \ \mathrm{K}, 3  \ \mathrm{dex}, -0.5 \ \mathrm{dex})$ as the default initial guess to improve inference accuracy in high-temperature regions.
	
	\begin{figure*}[htp!]
		\centering
		\includegraphics[width=0.245\linewidth]{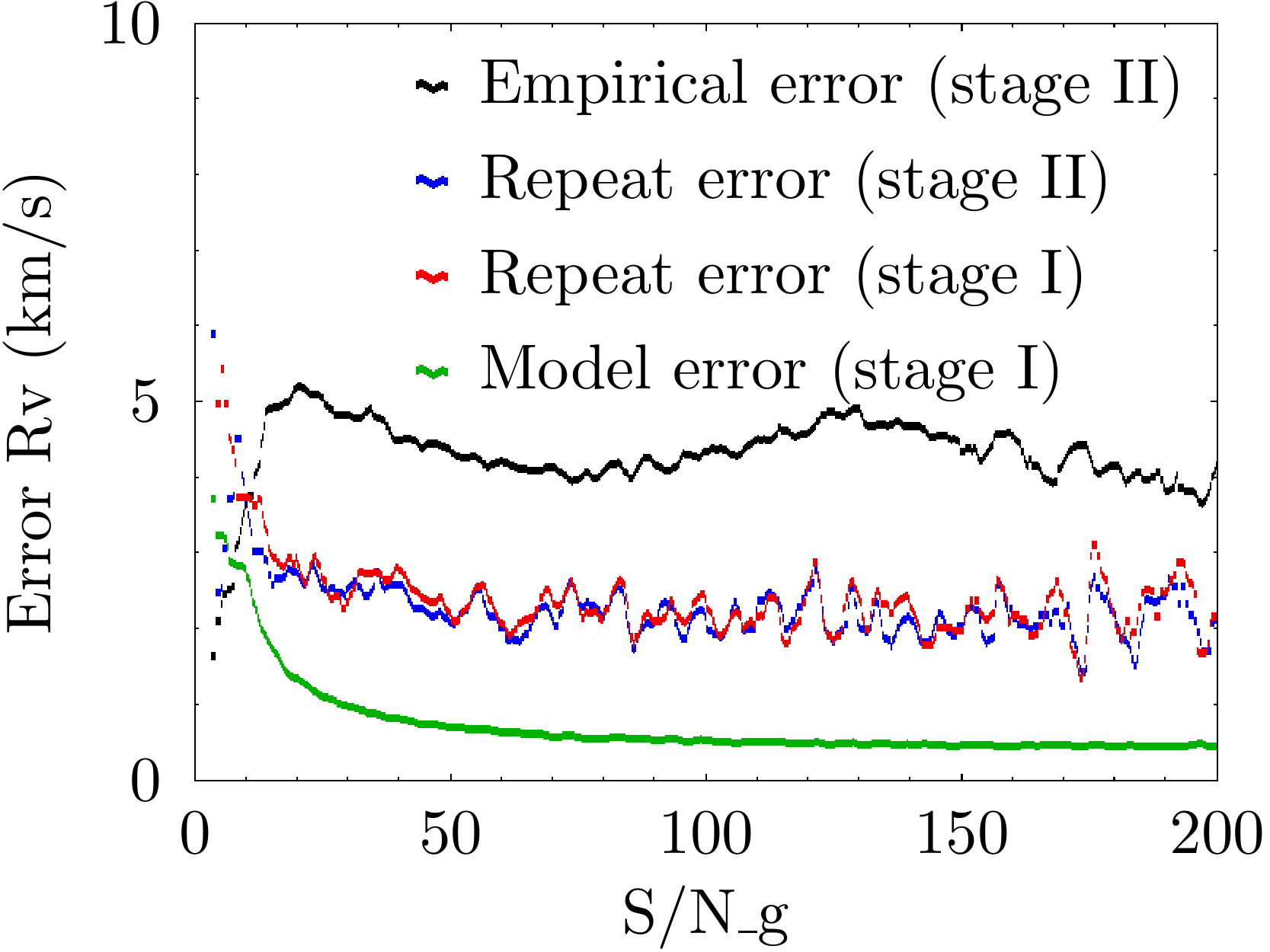}
		\includegraphics[width=0.245\linewidth]{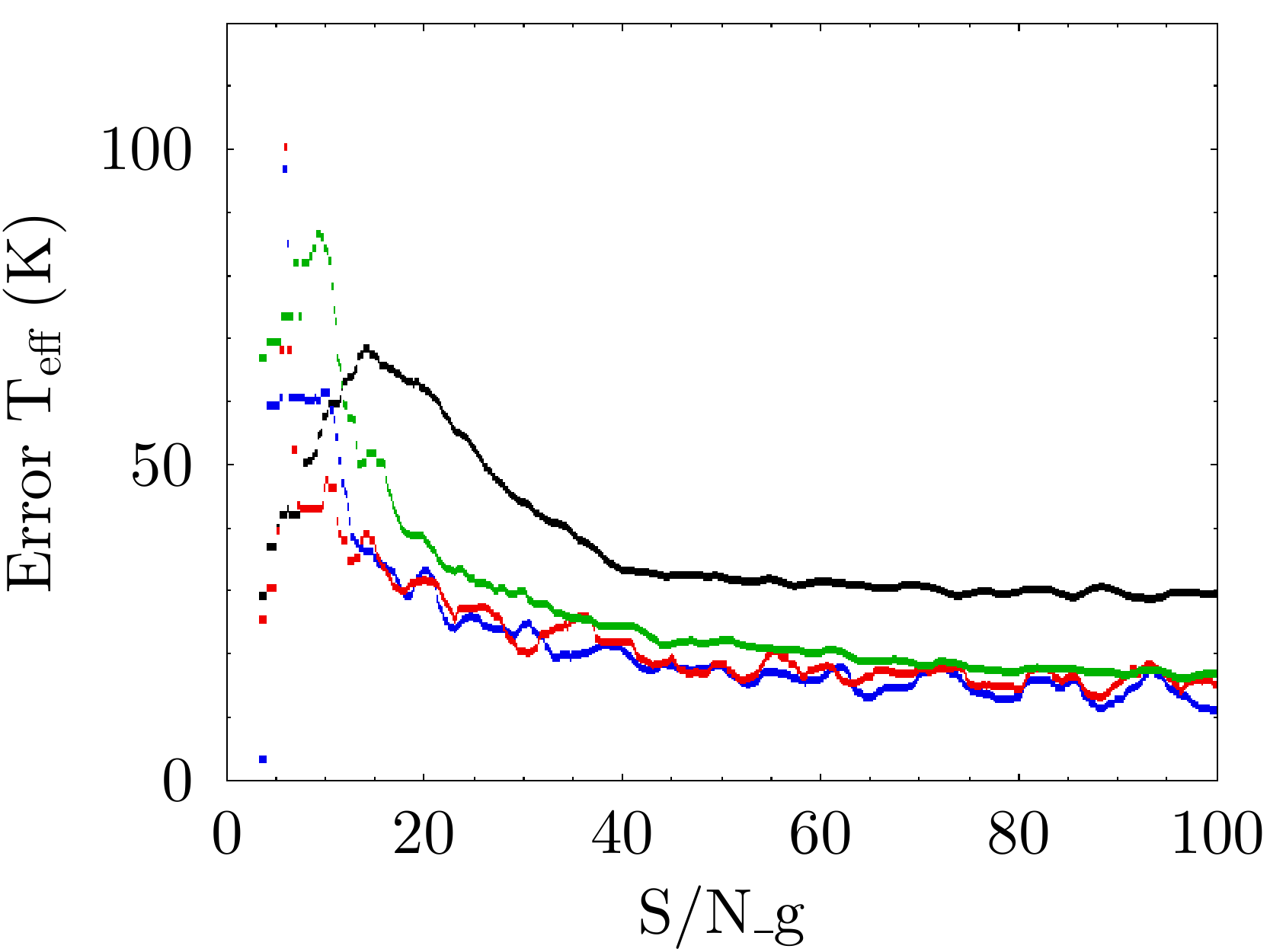}
		\includegraphics[width=0.245\linewidth]{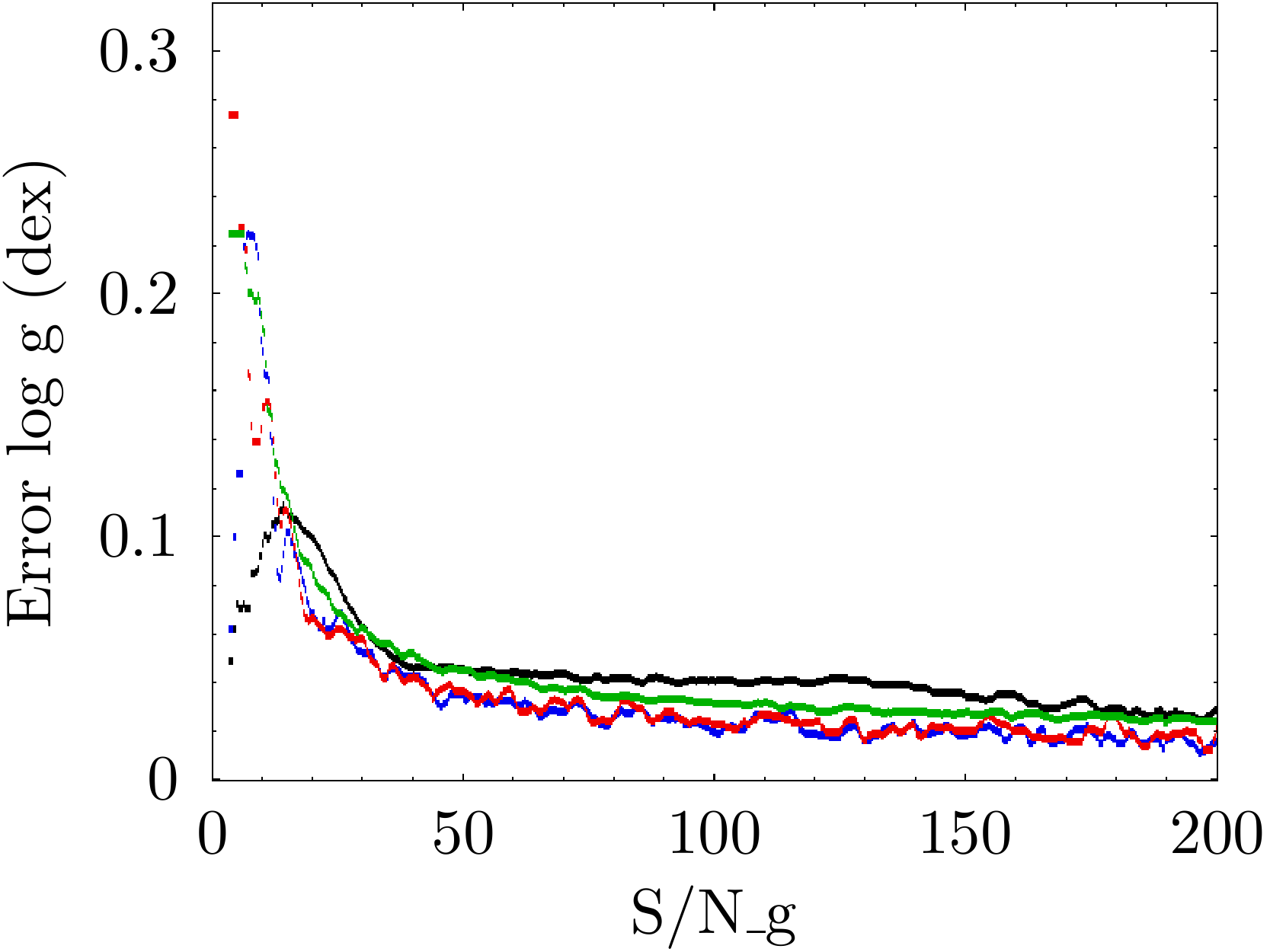}
		\includegraphics[width=0.245\linewidth]{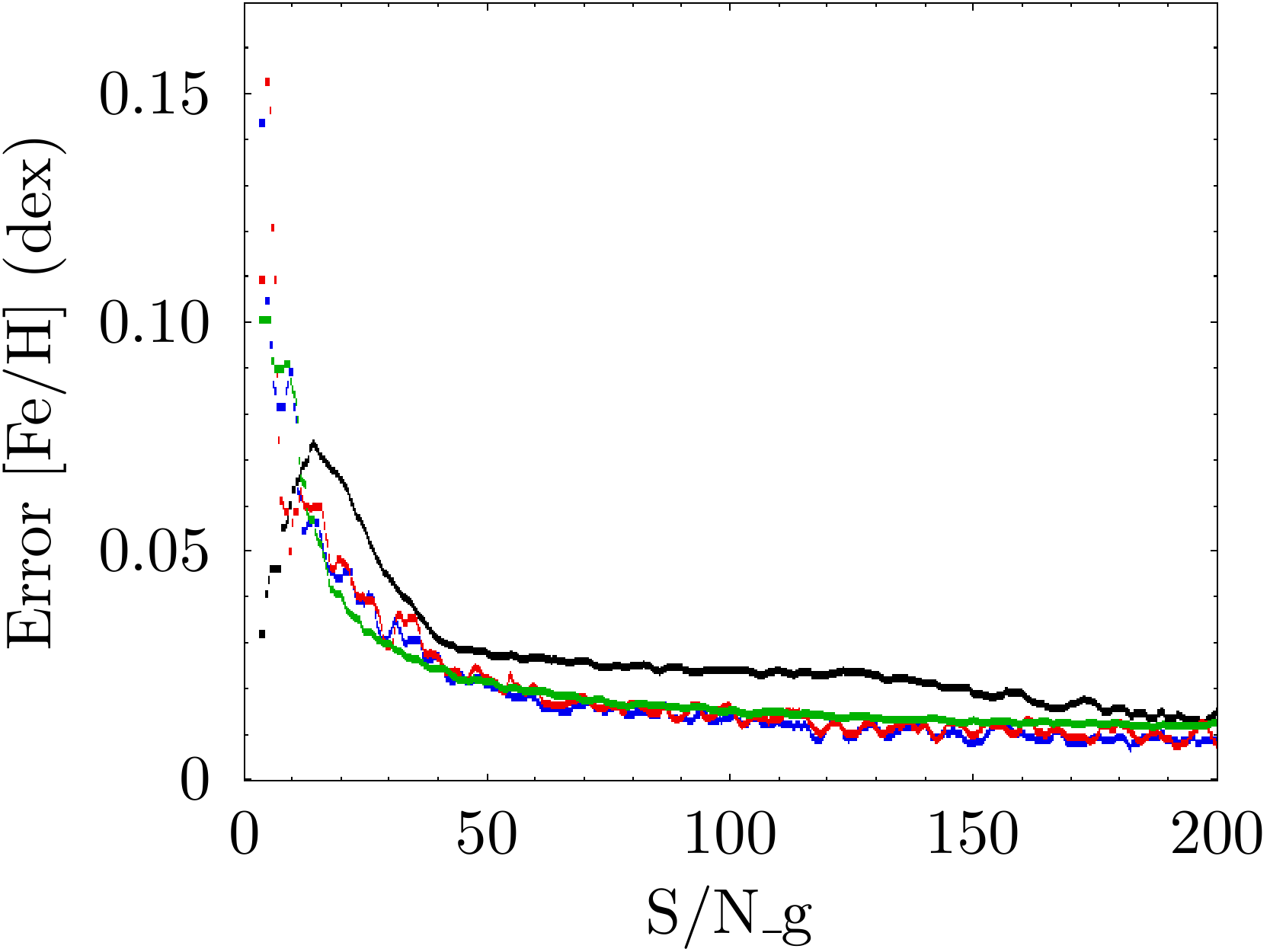}
		\caption{Comparison of the $0.5$ quantile (median) errors for LASP-Adam-GPU (stage I), repeat observation (stage I), LAMOST official empirical formula (stage II), and LAMOST official repeat observation (stage II) as a function of \texttt{S/N\_g}. The errors are smoothed using an Epanechnikov kernel (smoothing factor is set to $2$). For clarity, the legend is shown only in the first panel.}
		\label{LASPErr}
	\end{figure*}
	
	\begin{figure*}[htp]
		\centering
		\includegraphics[width=0.24\linewidth]{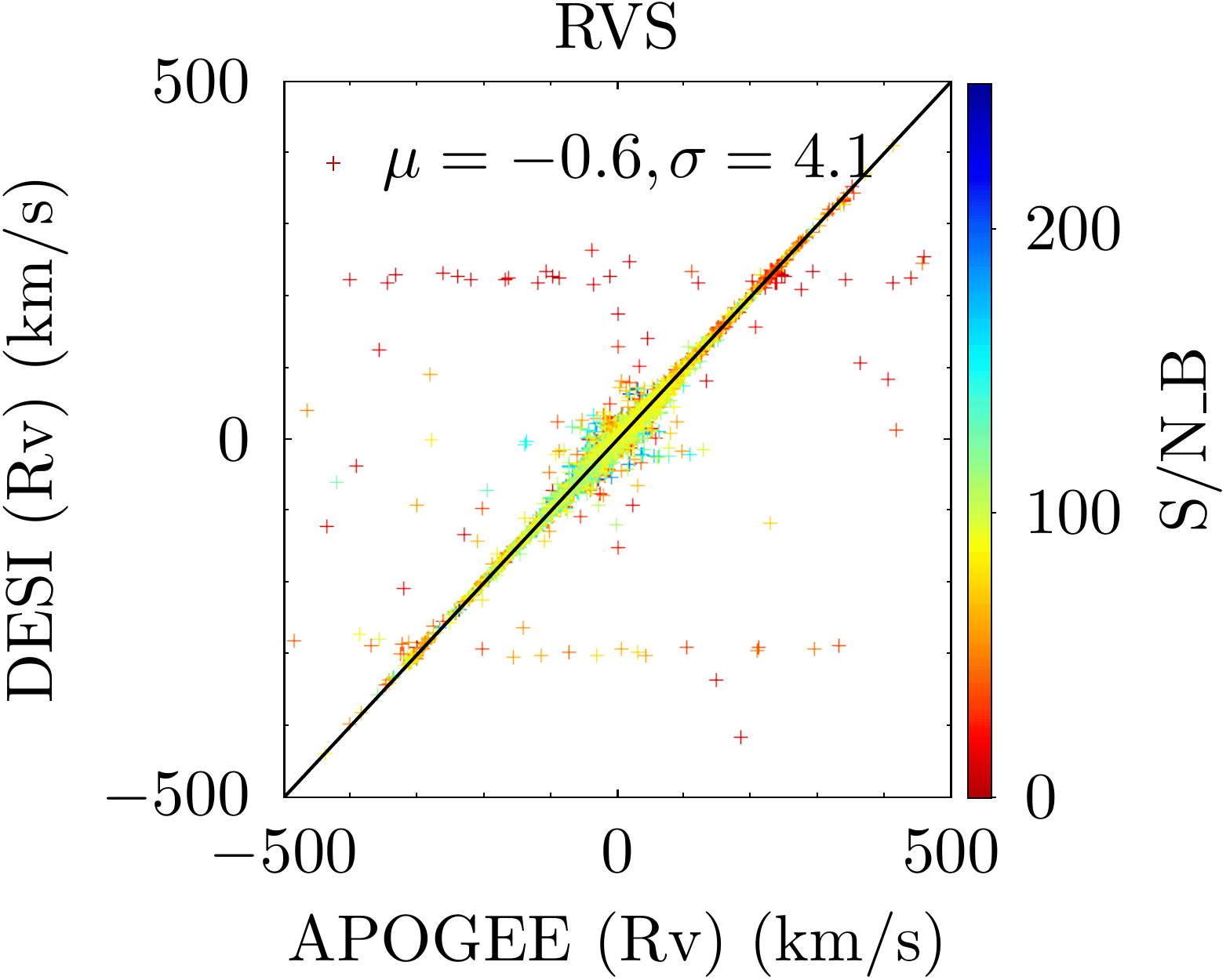}
		\includegraphics[width=0.24\linewidth]{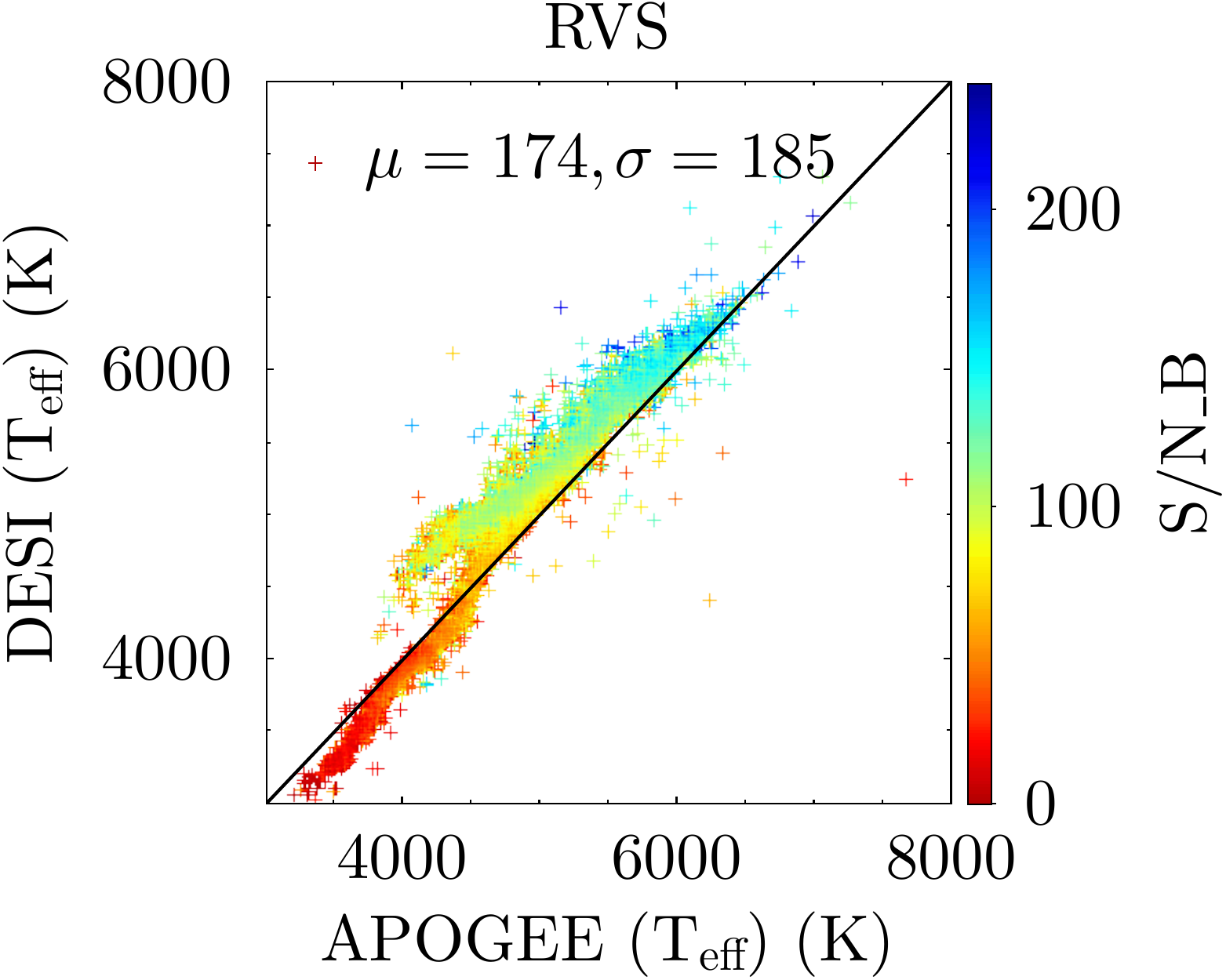}
		\includegraphics[width=0.24\linewidth]{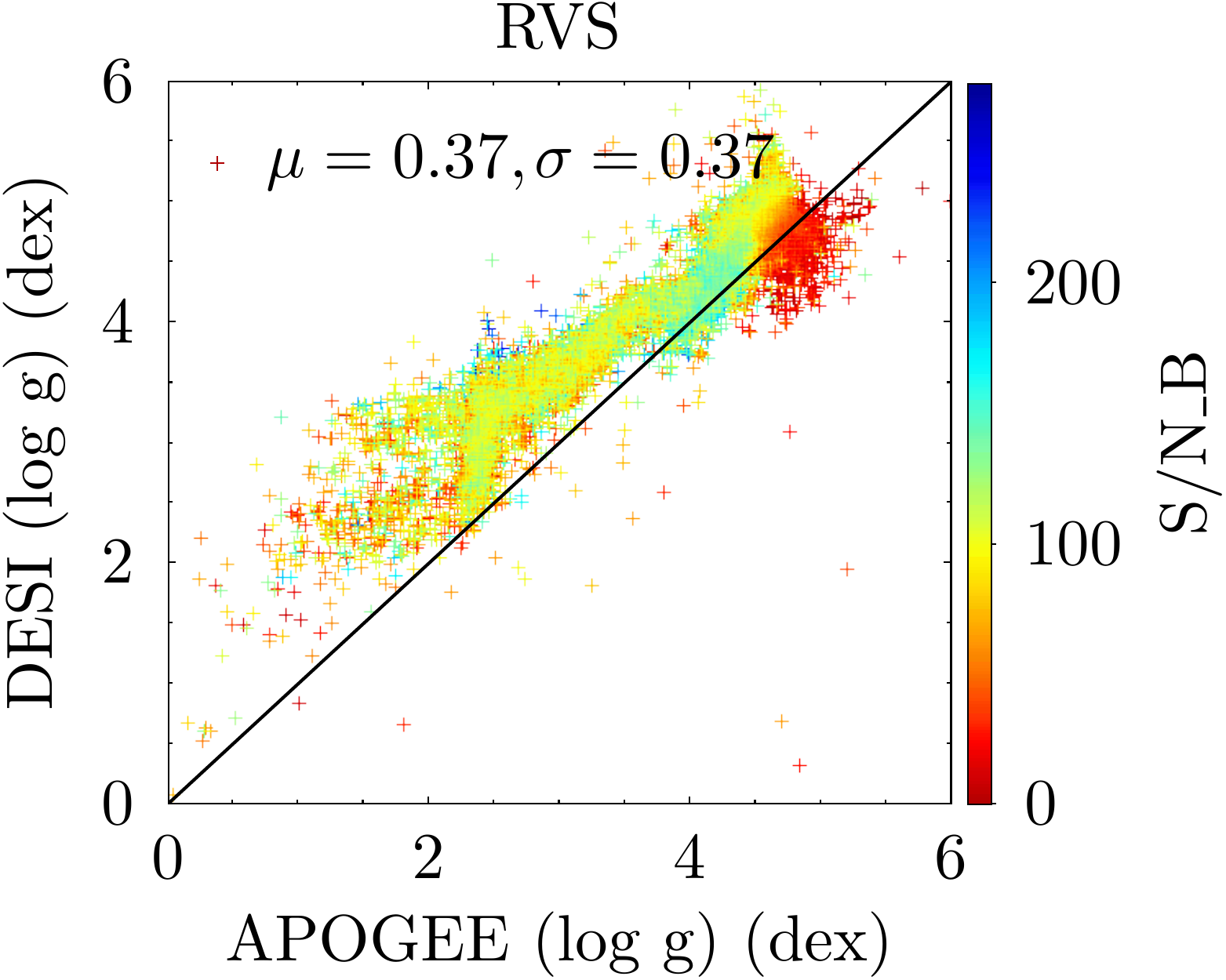}
		\includegraphics[width=0.24\linewidth]{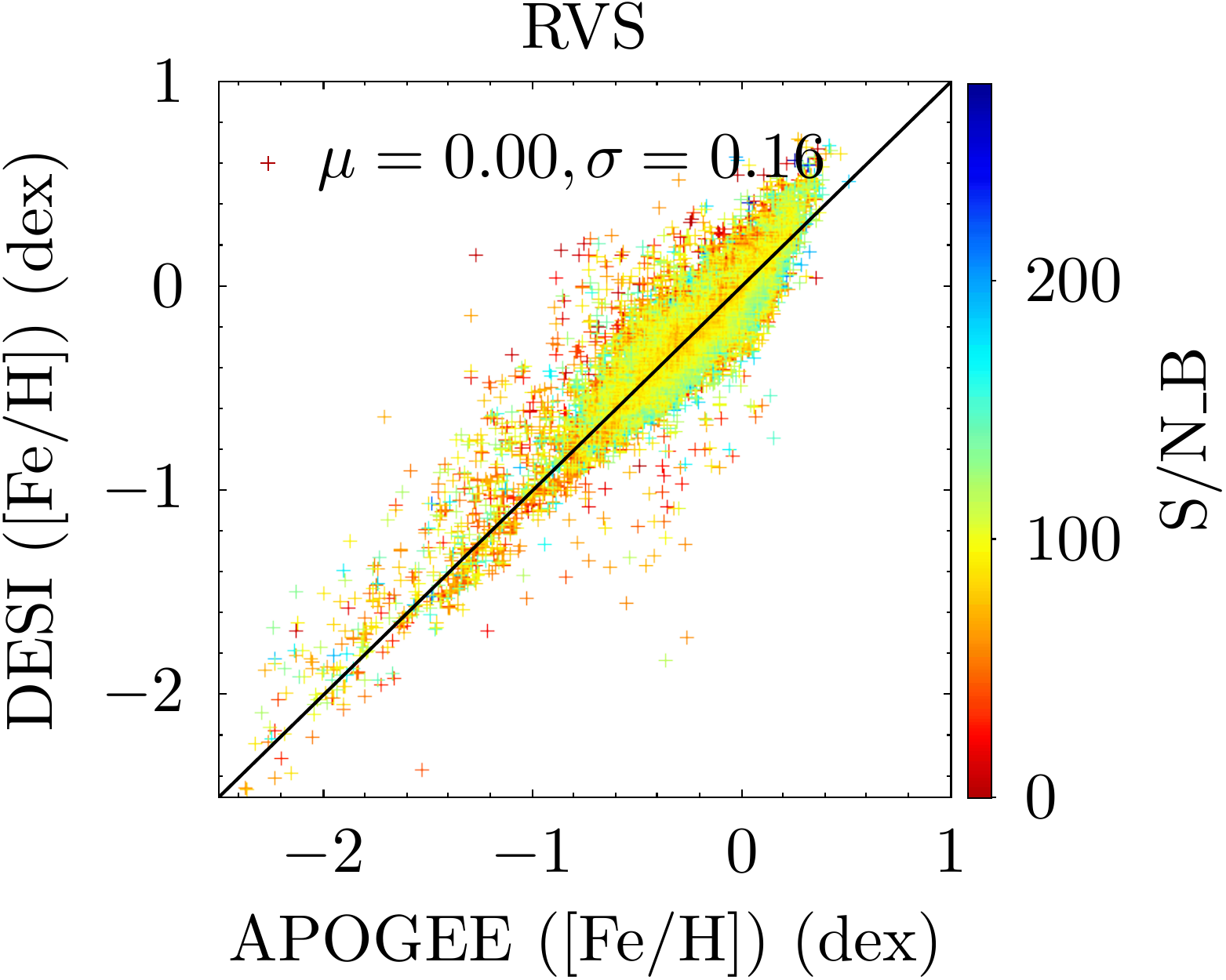} \\
		\includegraphics[width=0.24\linewidth]{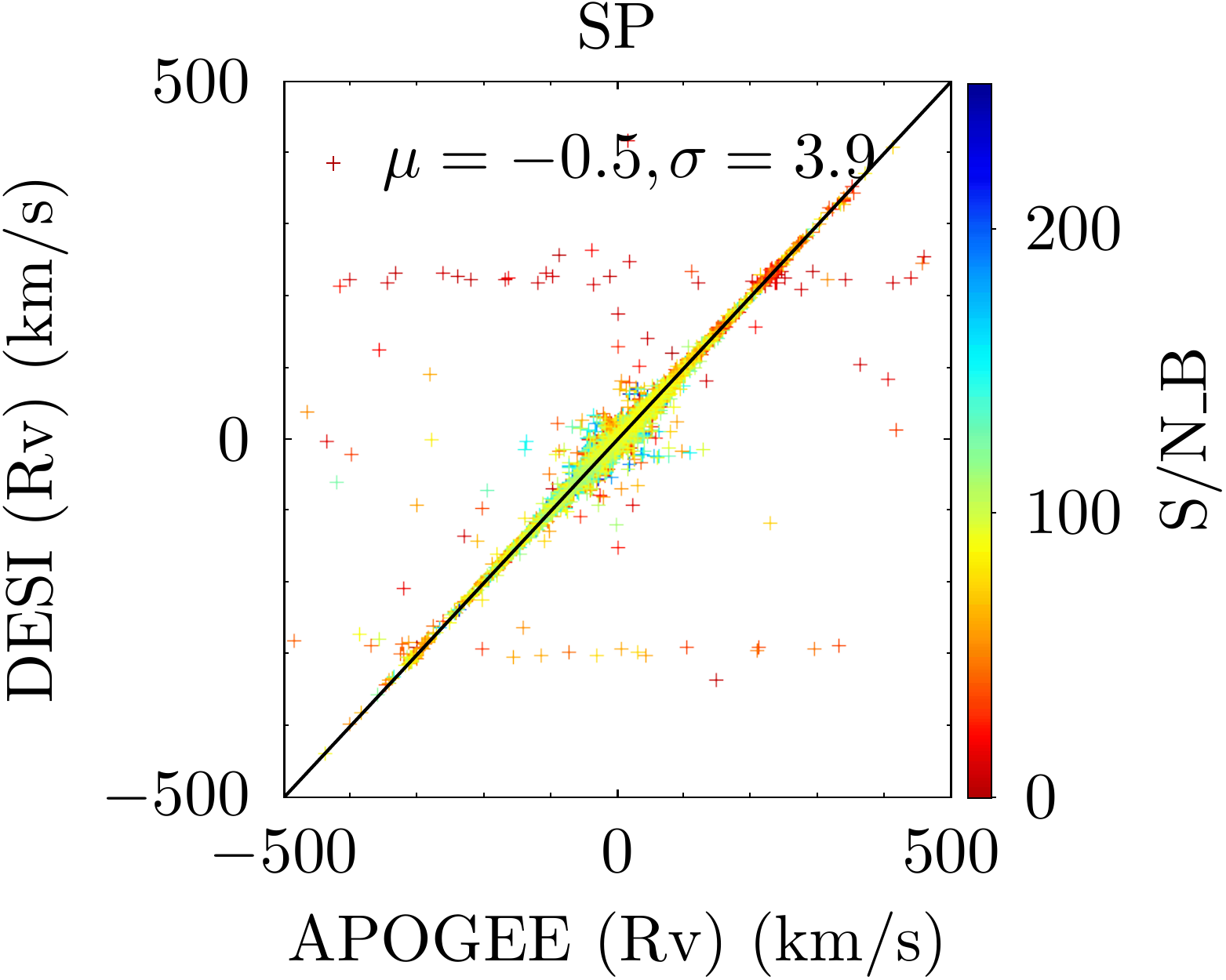}
		\includegraphics[width=0.24\linewidth]{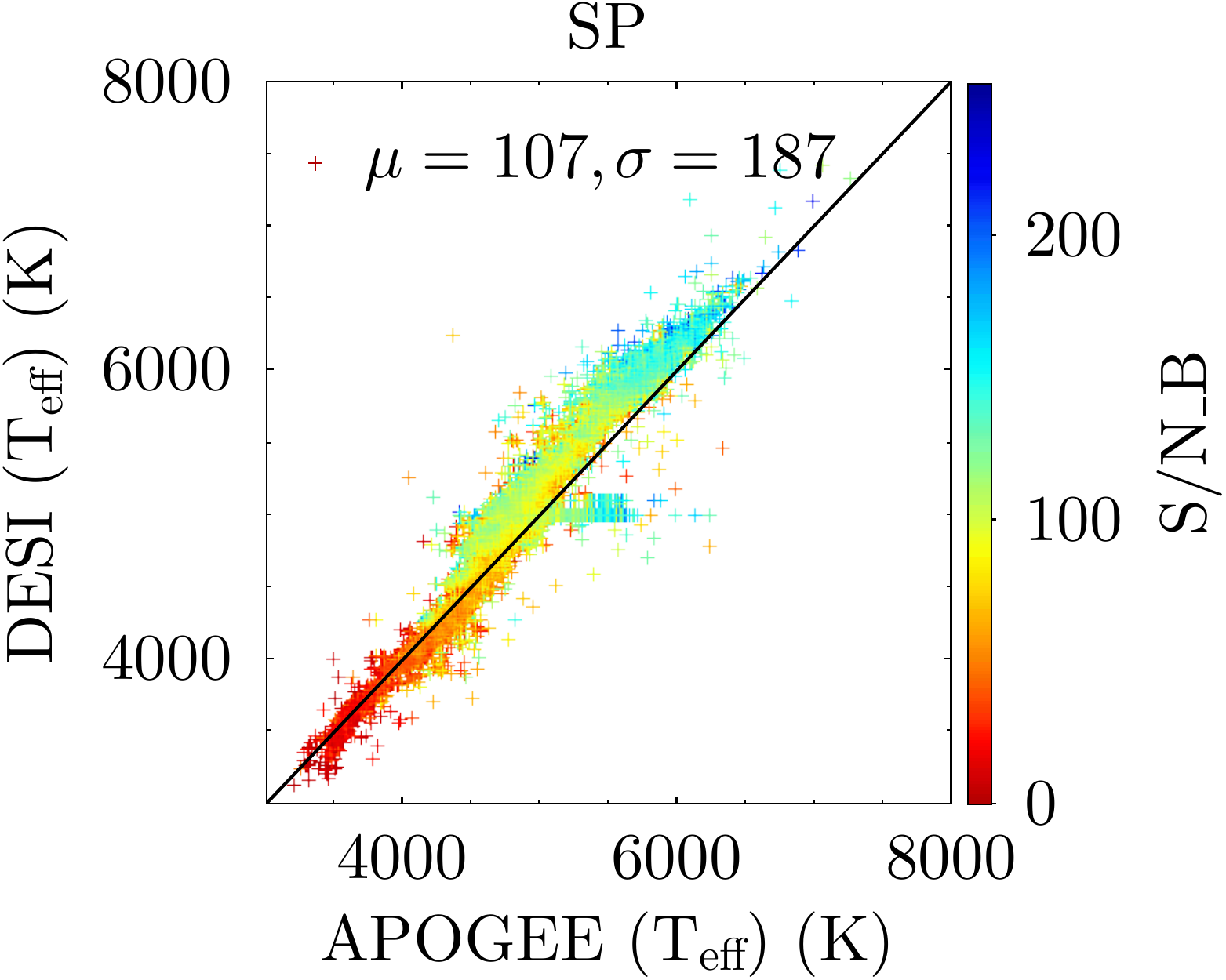}
		\includegraphics[width=0.24\linewidth]{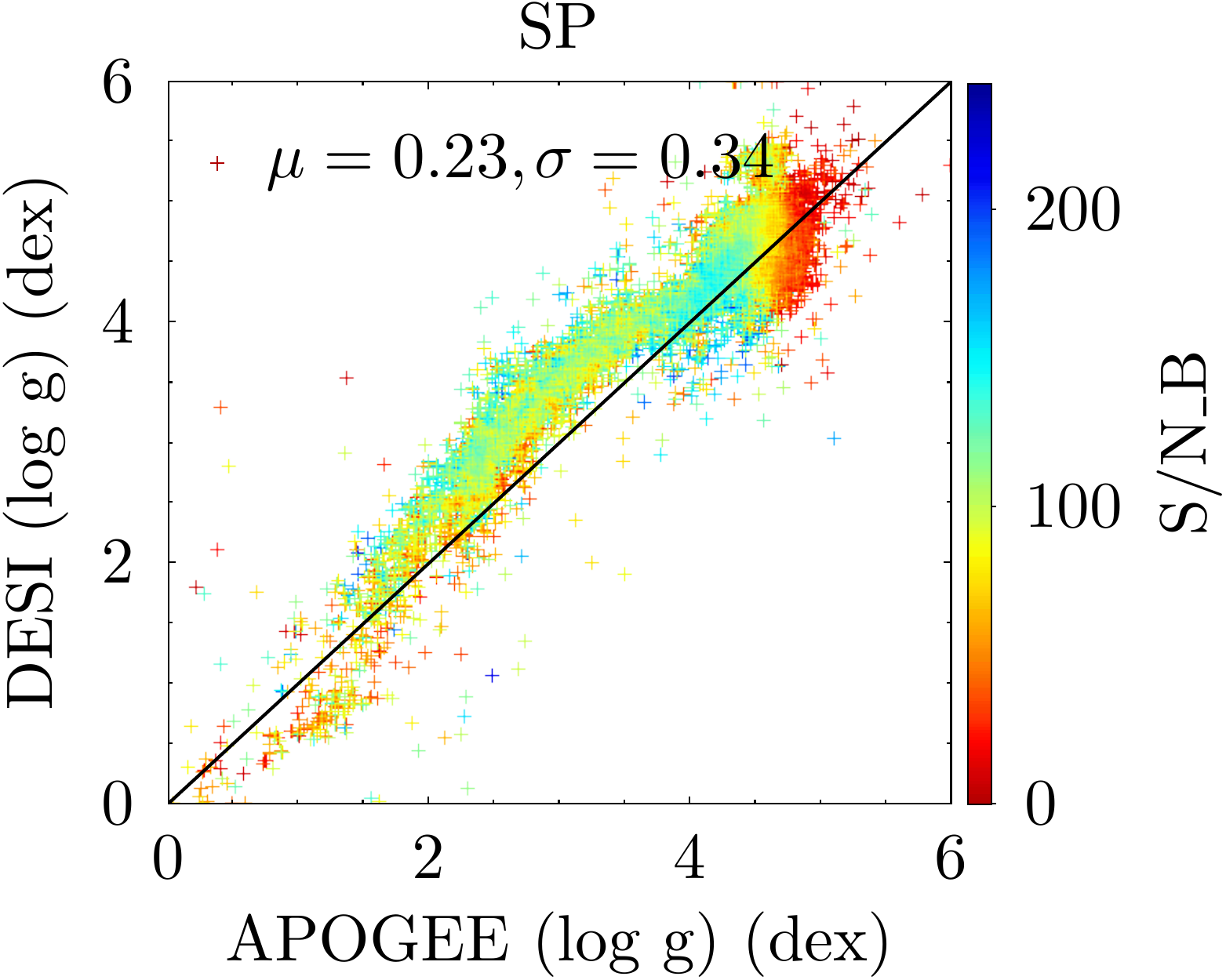}
		\includegraphics[width=0.24\linewidth]{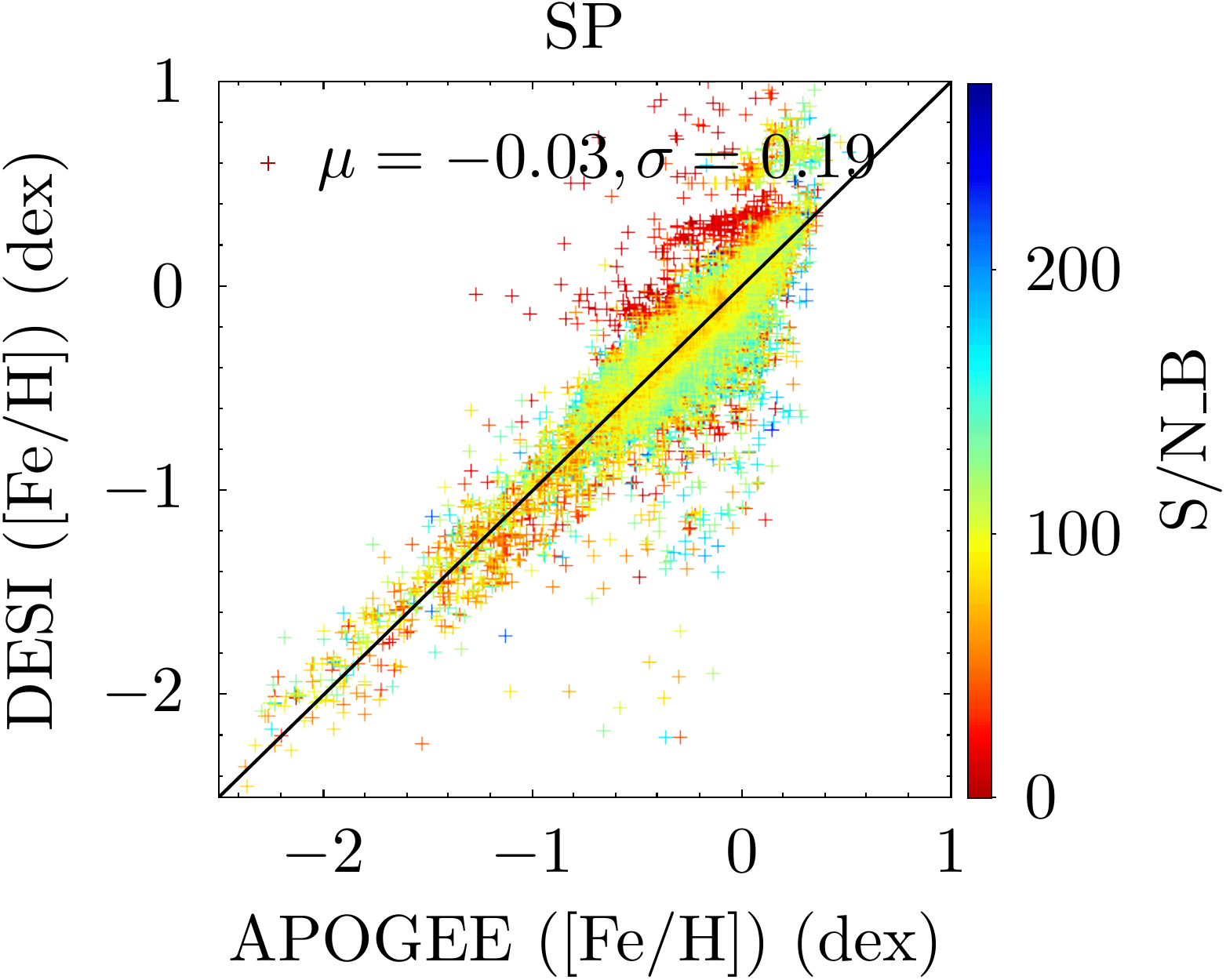} \\
		\includegraphics[width=0.24\linewidth]{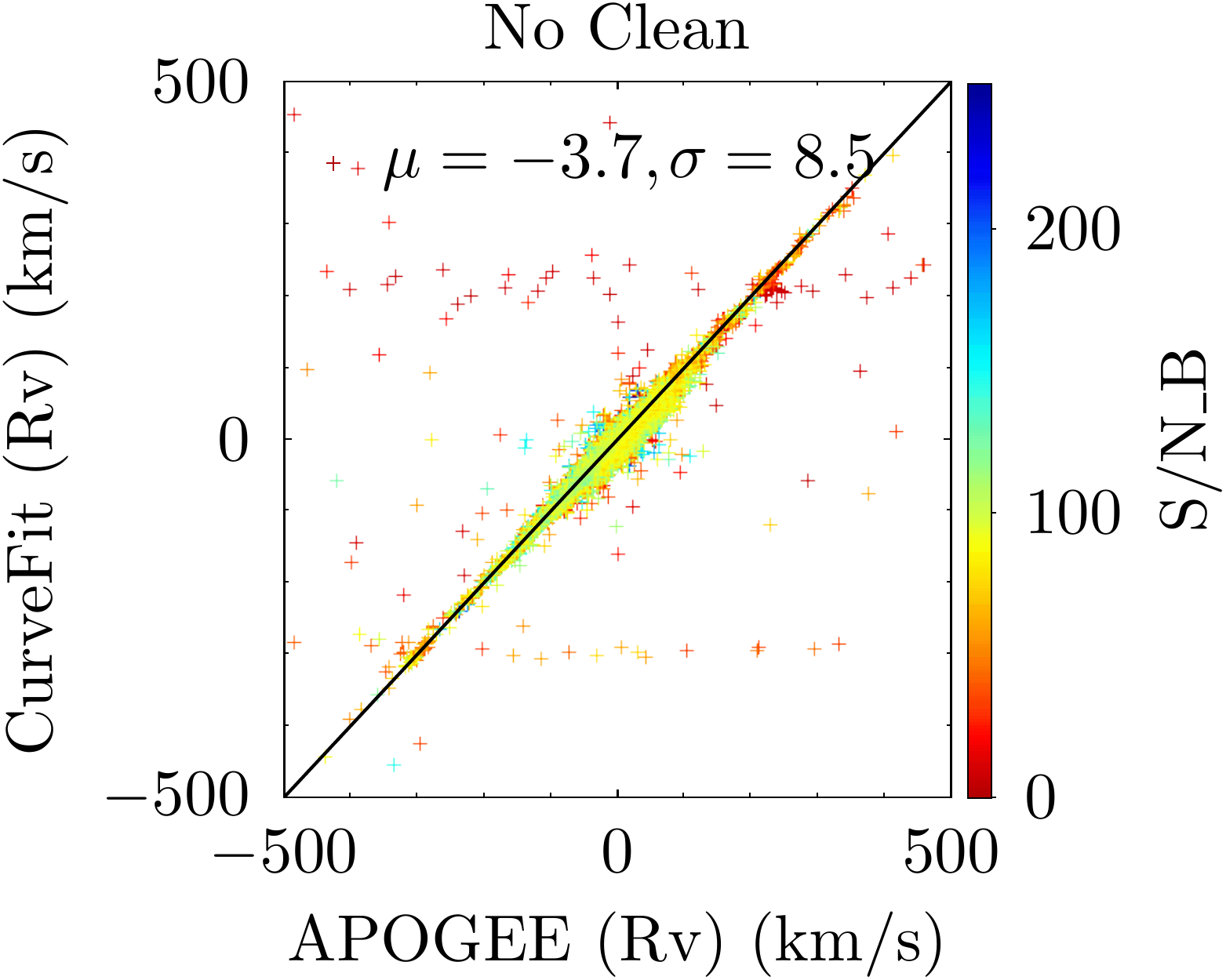}
		\includegraphics[width=0.24\linewidth]{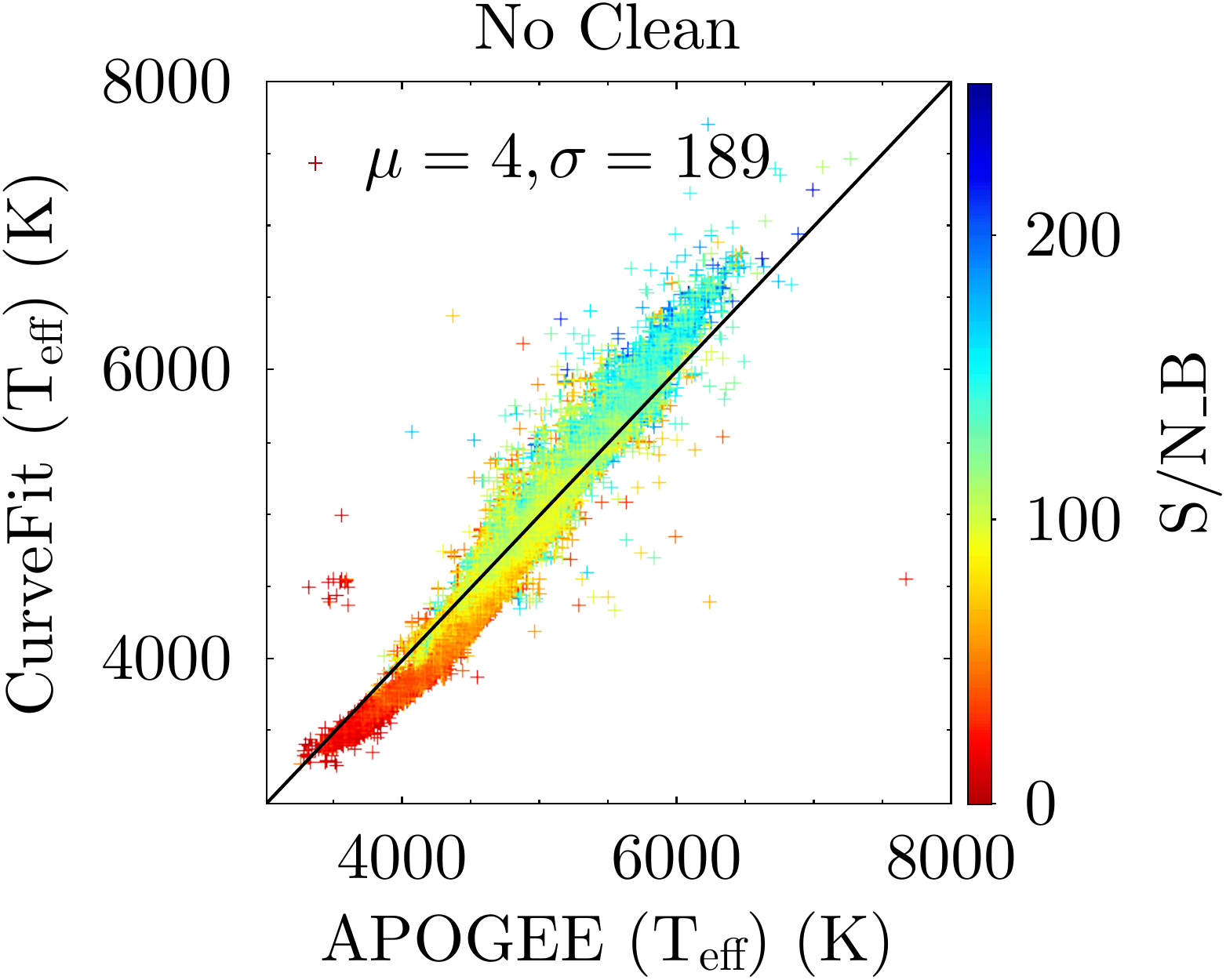}
		\includegraphics[width=0.24\linewidth]{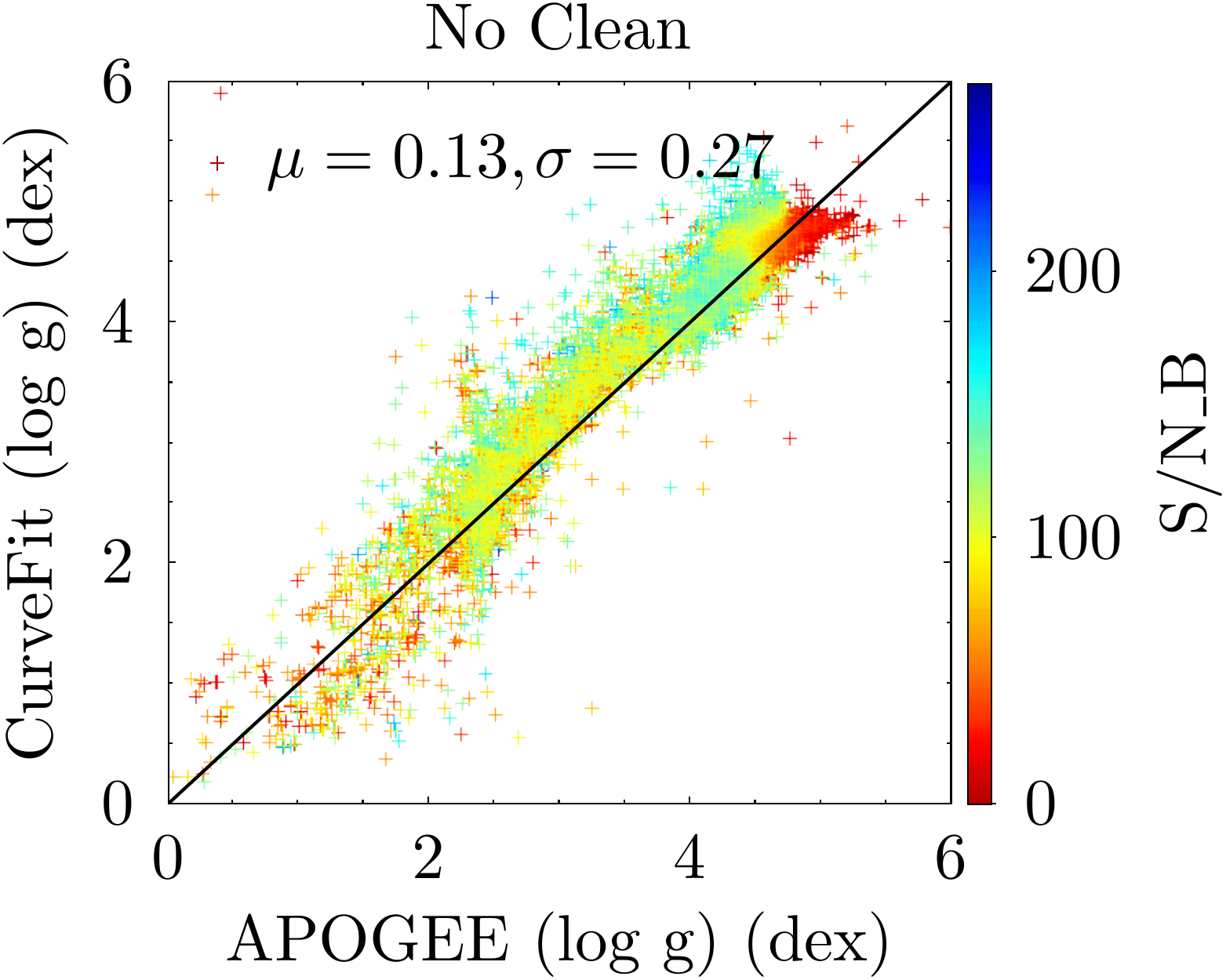}
		\includegraphics[width=0.24\linewidth]{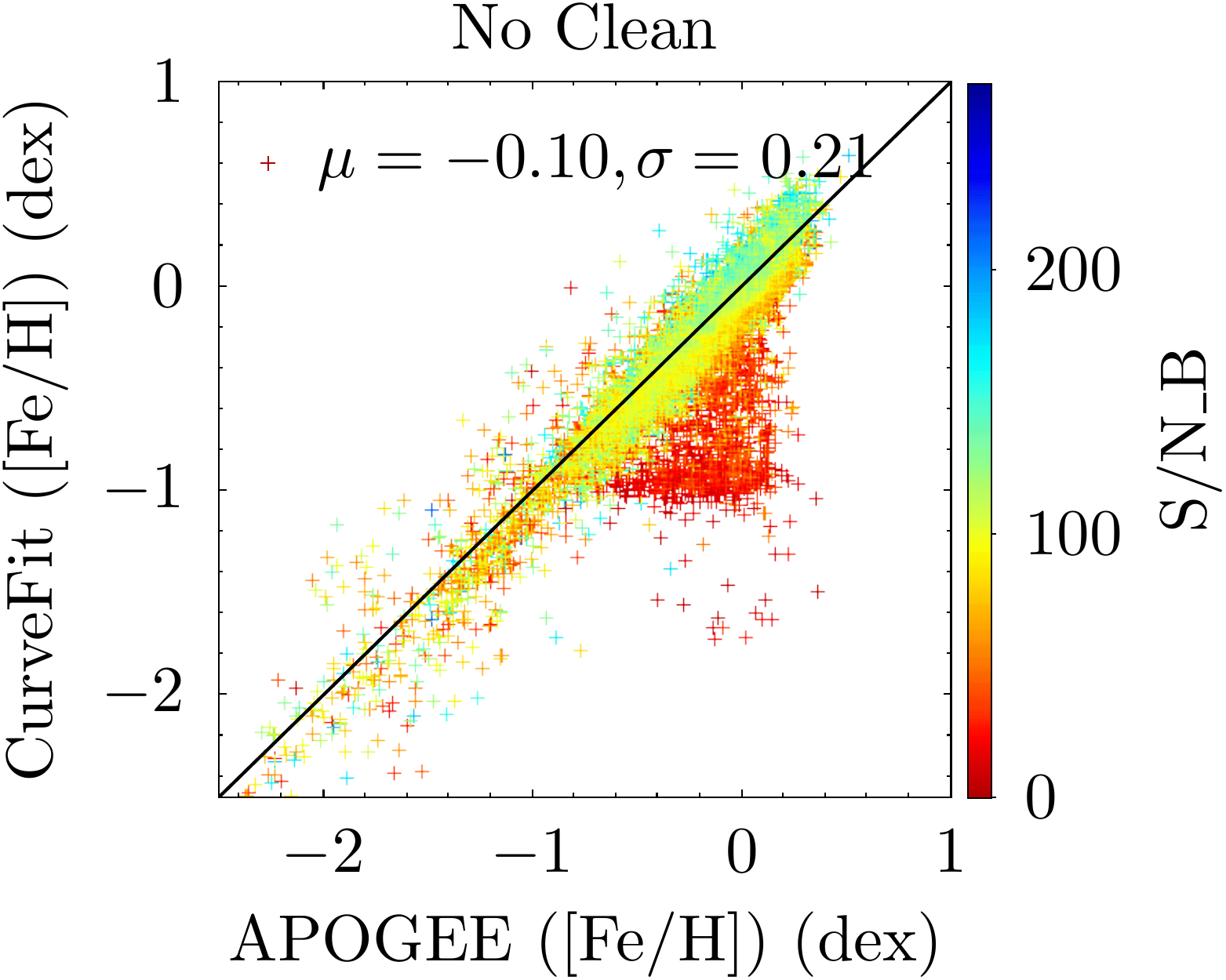} \\
		\includegraphics[width=0.24\linewidth]{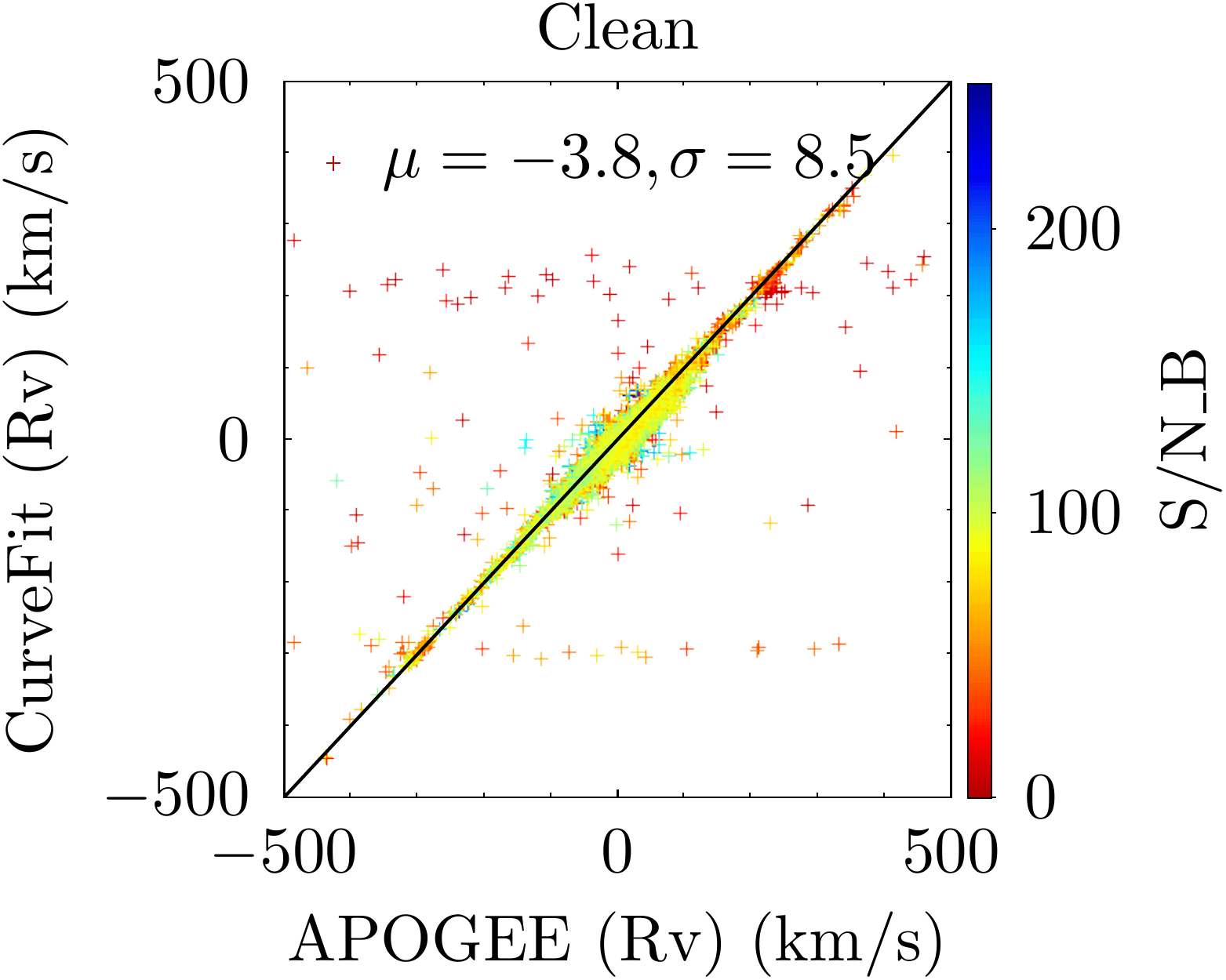}
		\includegraphics[width=0.24\linewidth]{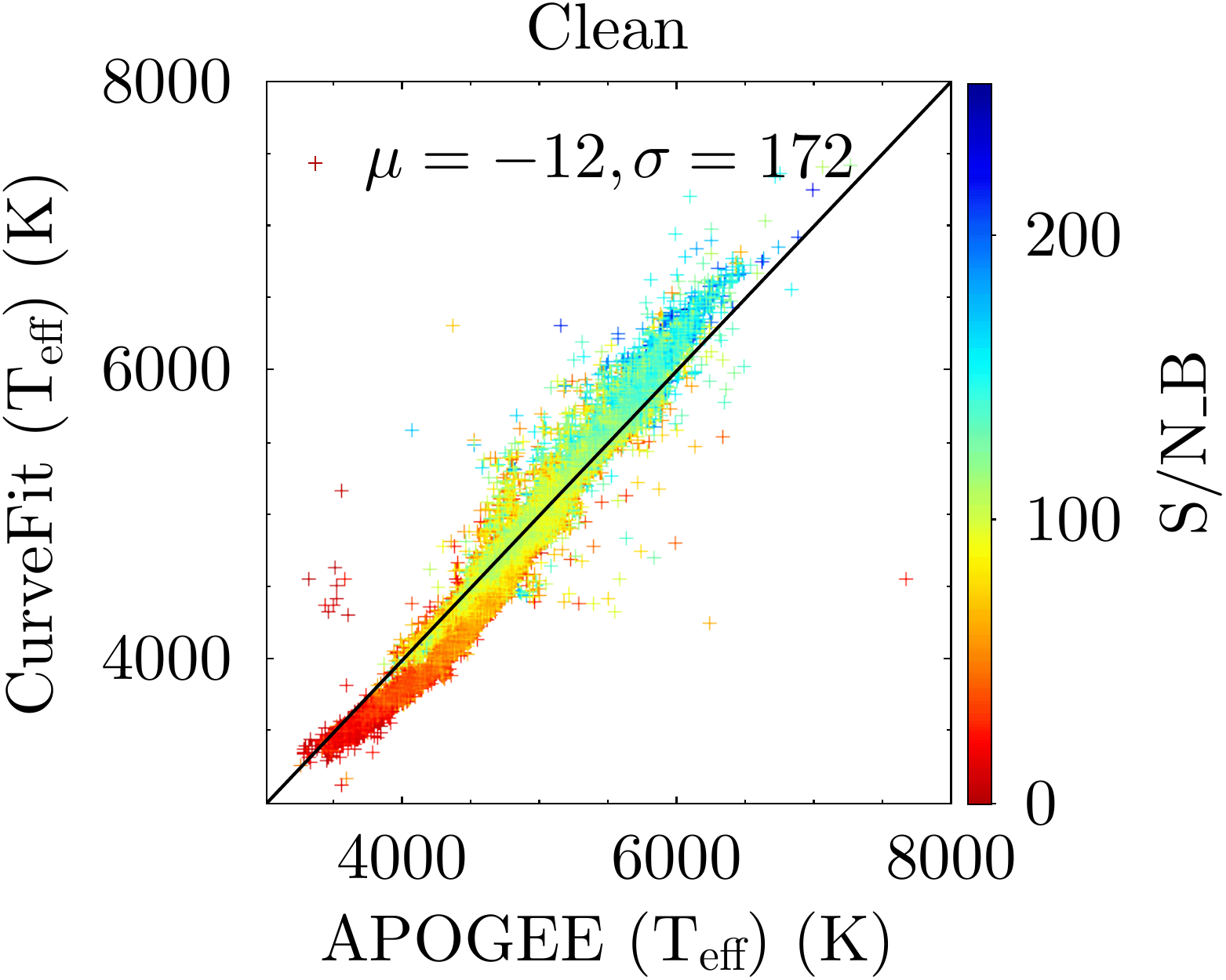}
		\includegraphics[width=0.24\linewidth]{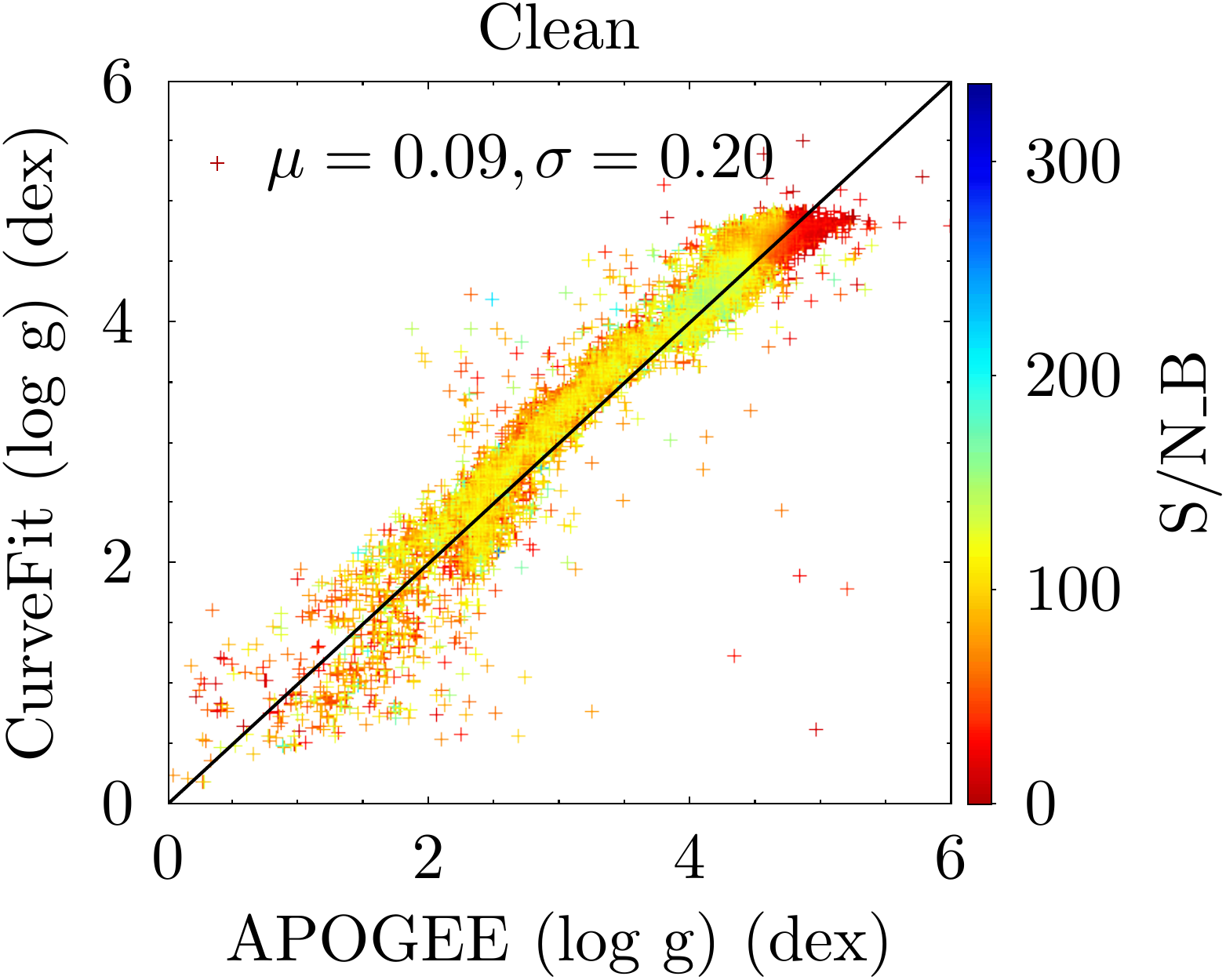}
		\includegraphics[width=0.24\linewidth]{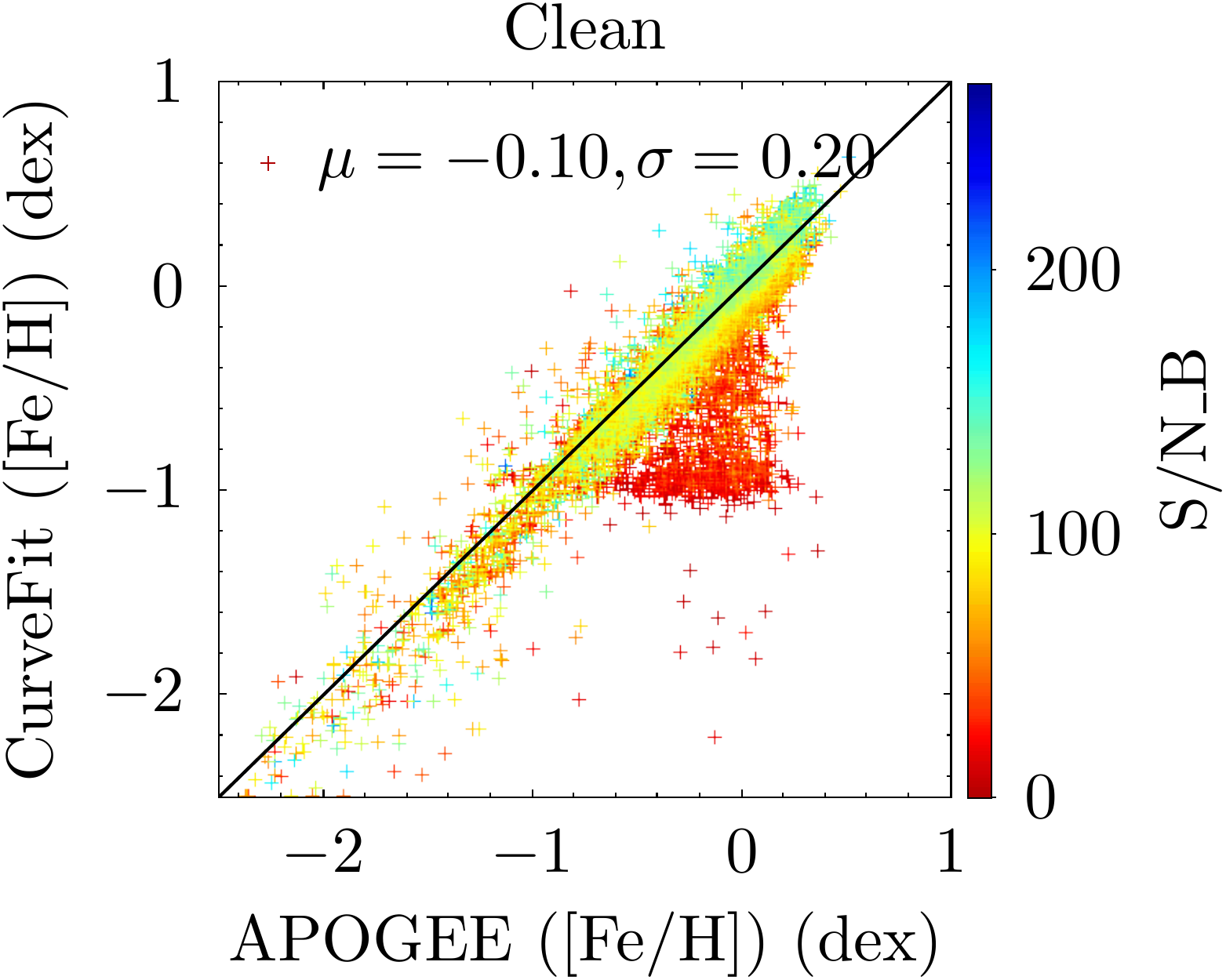}
		\includegraphics[width=0.24\linewidth]{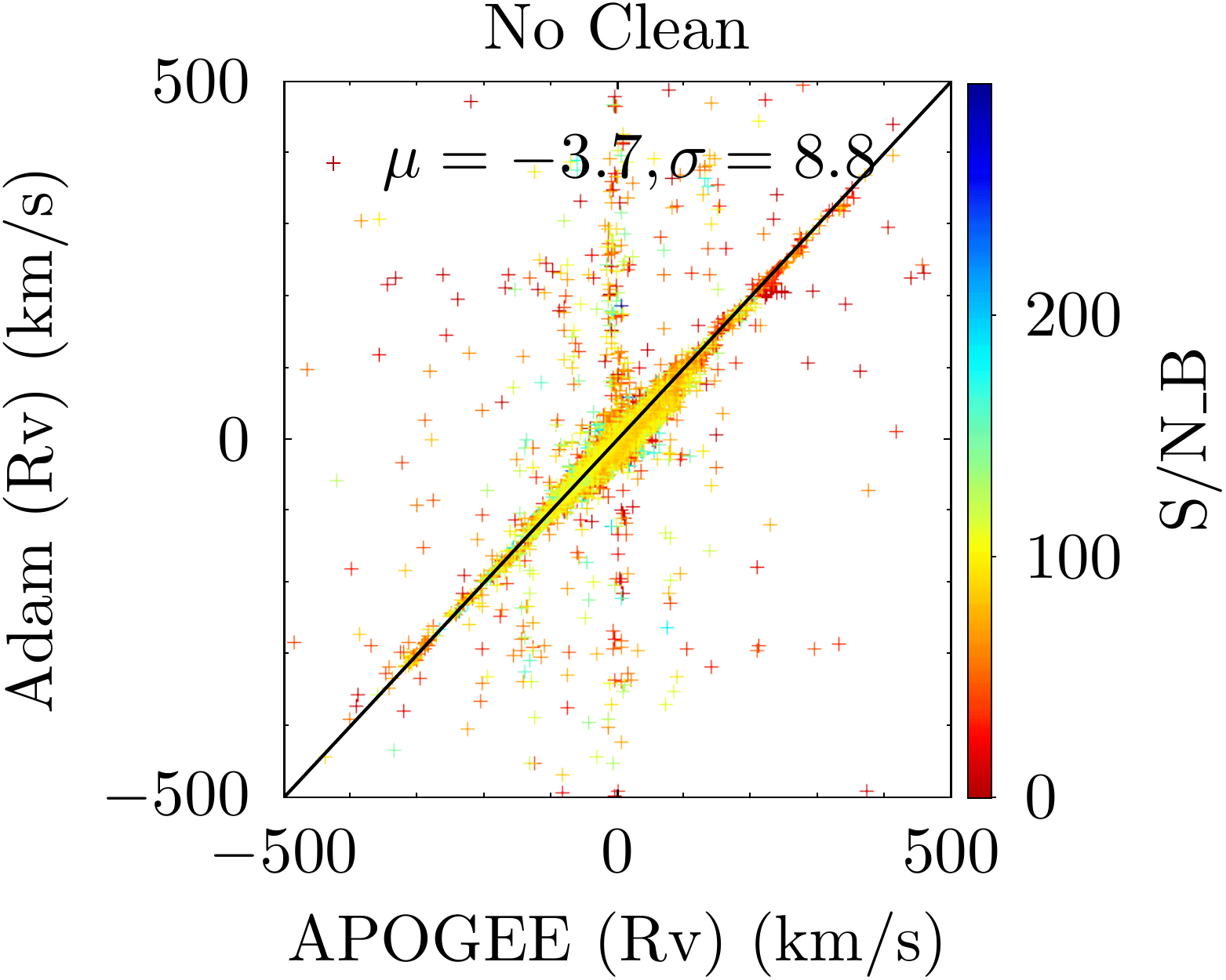}
		\includegraphics[width=0.24\linewidth]{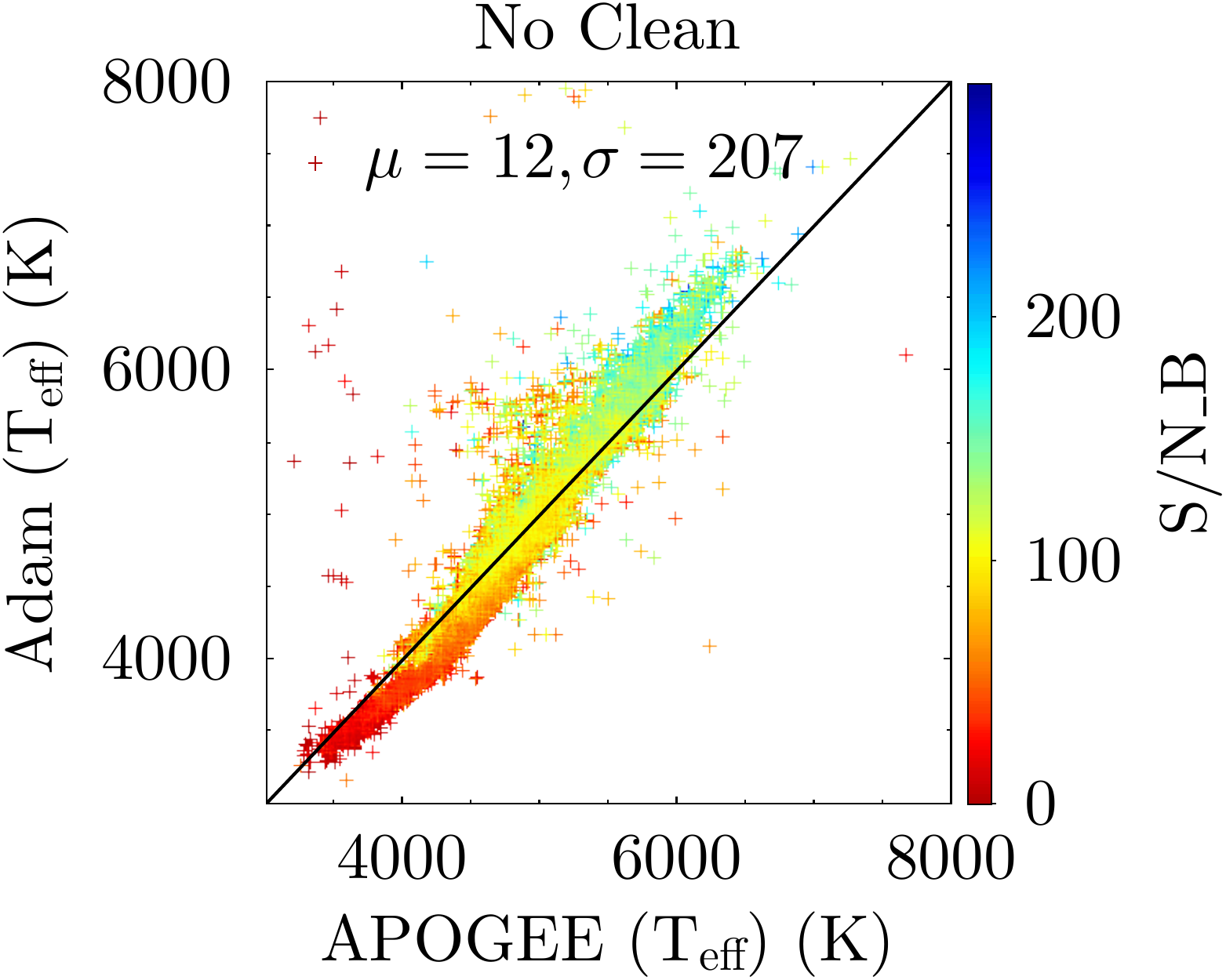}
		\includegraphics[width=0.24\linewidth]{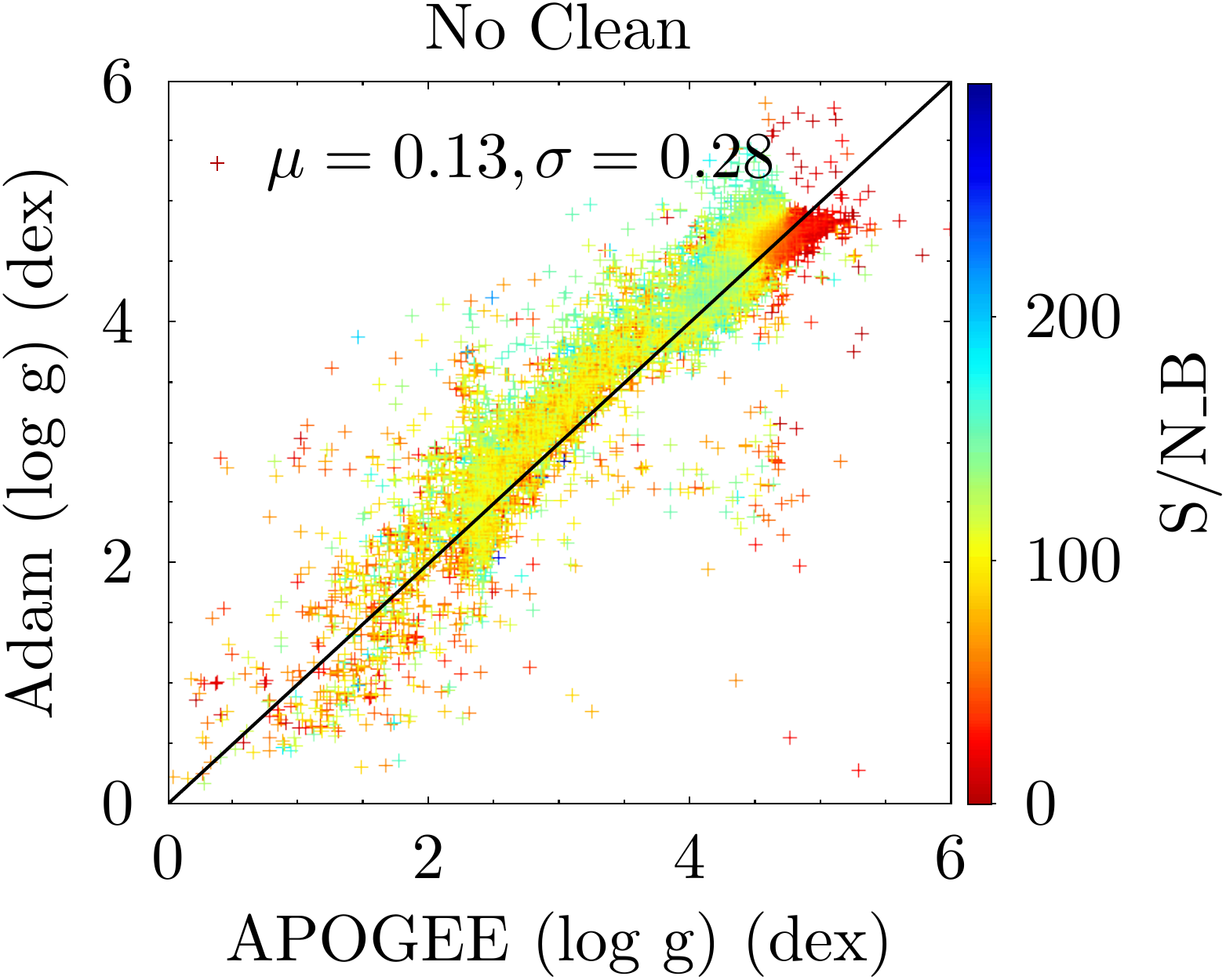}
		\includegraphics[width=0.24\linewidth]{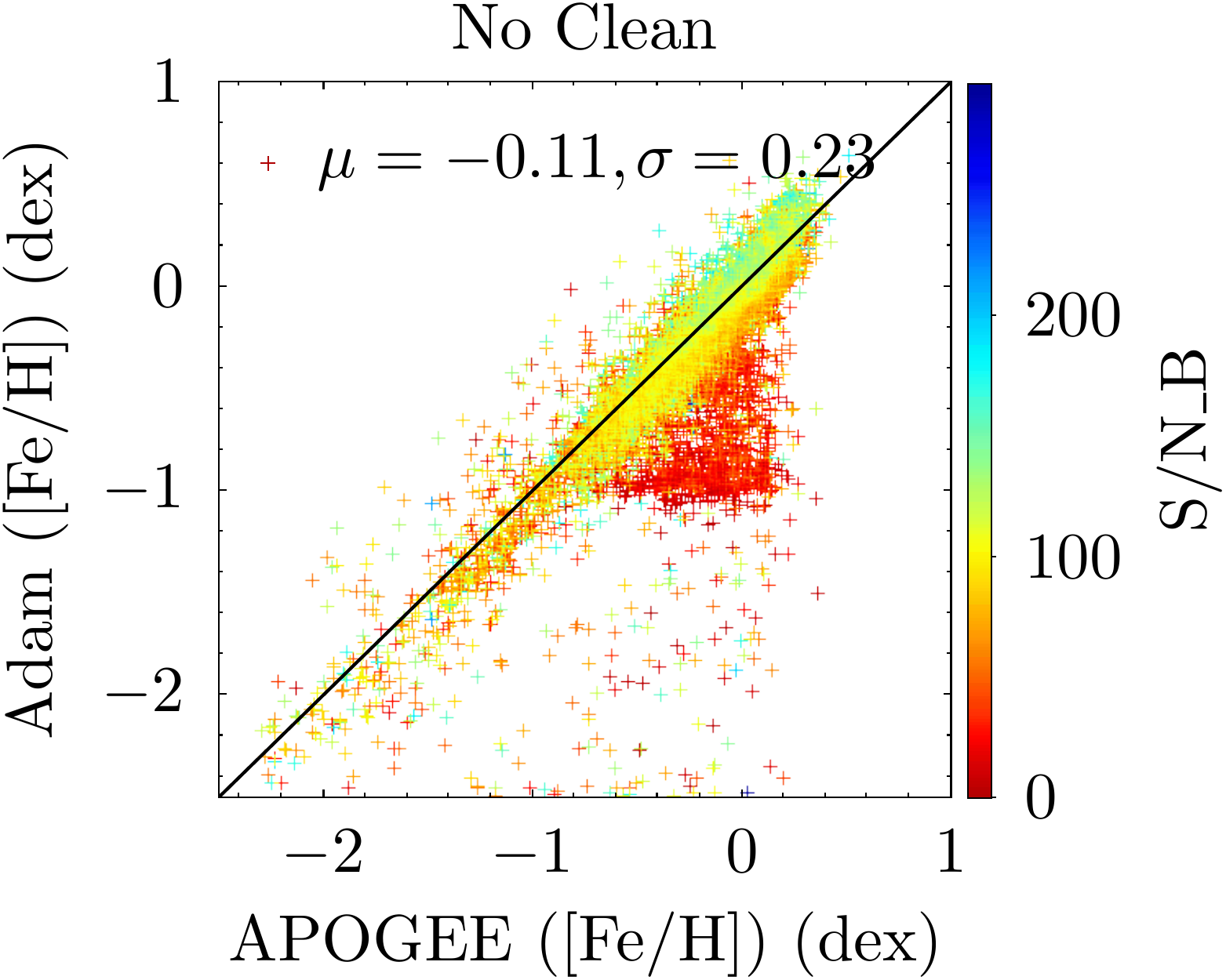} \\
		\includegraphics[width=0.24\linewidth]{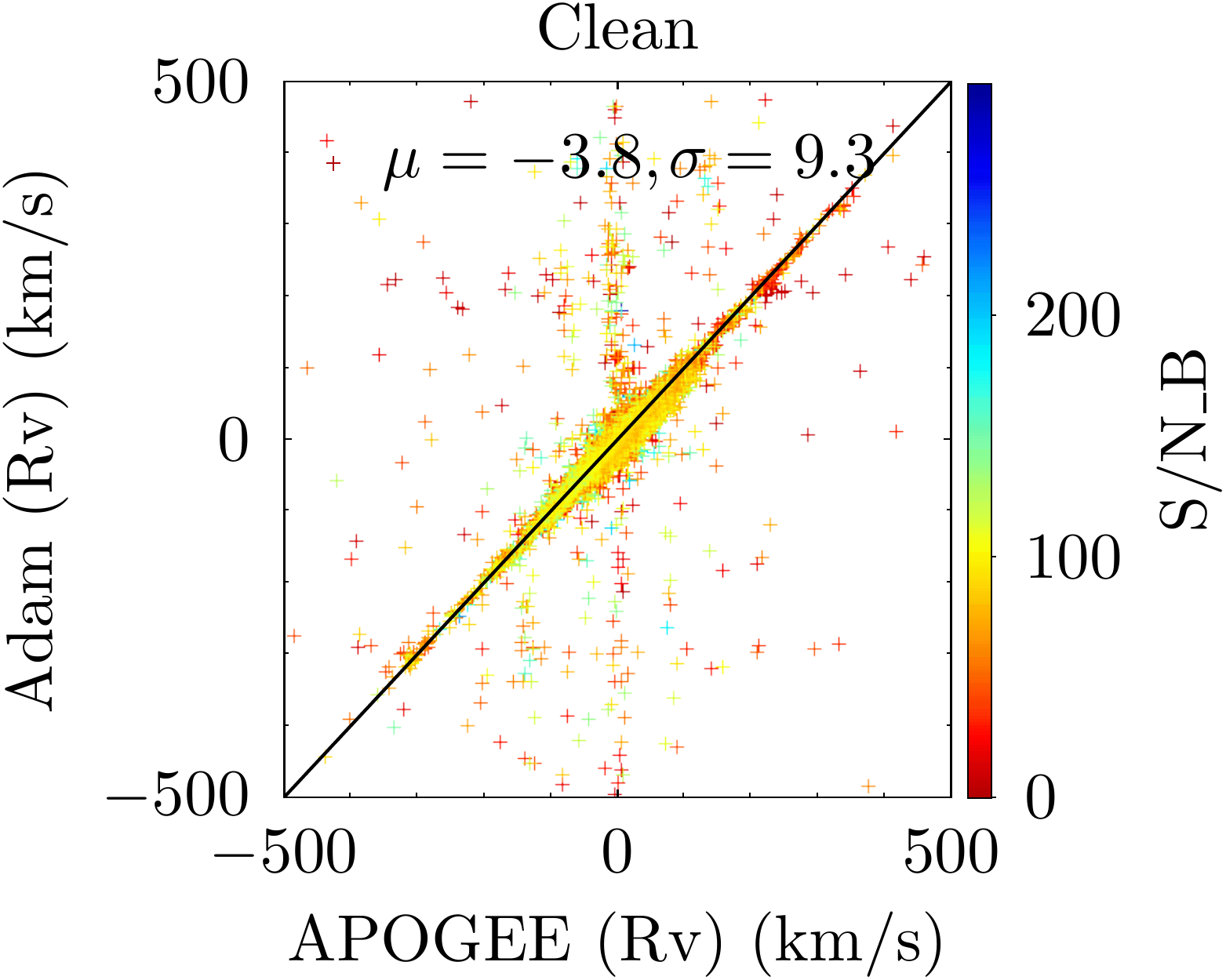}
		\includegraphics[width=0.24\linewidth]{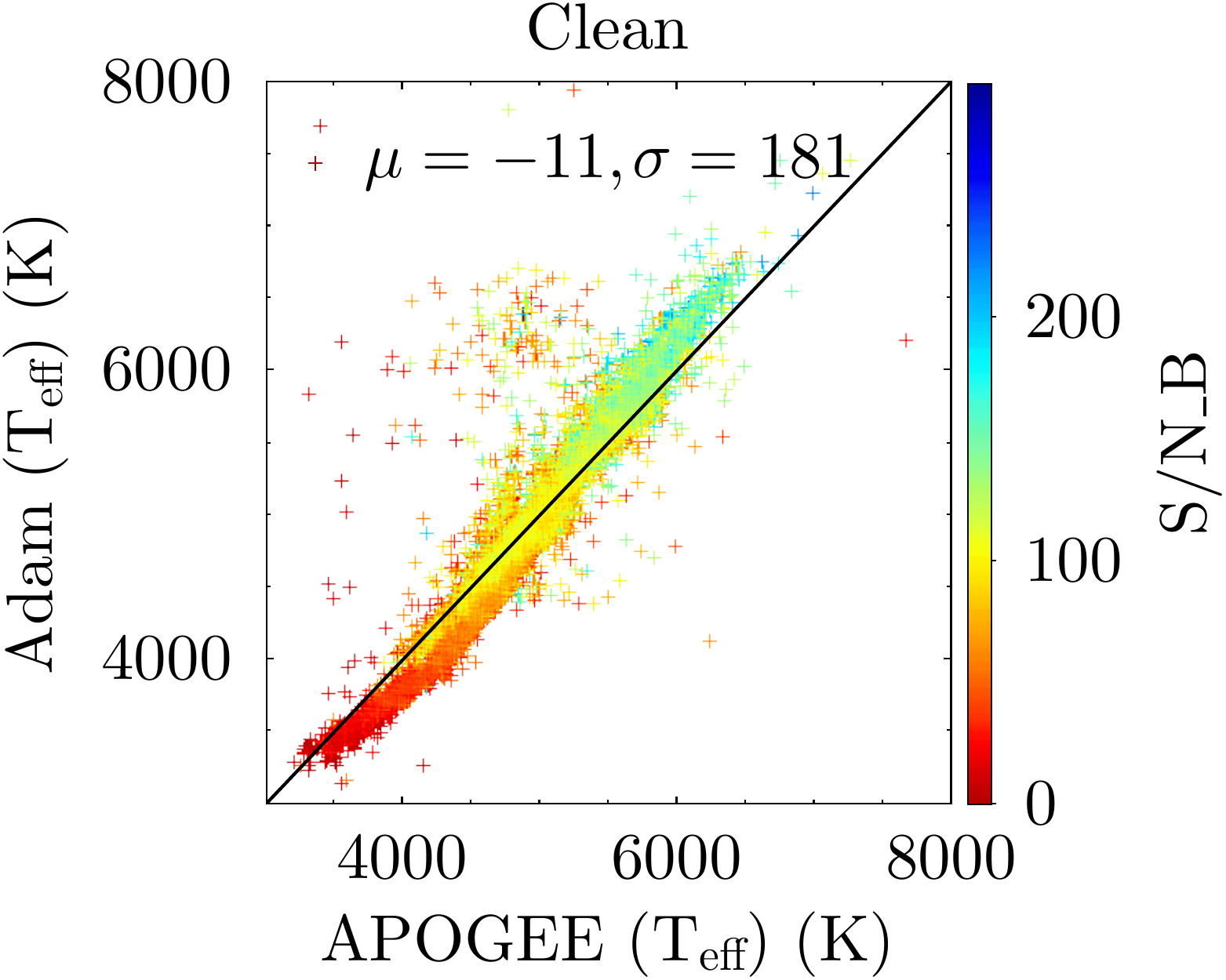}
		\includegraphics[width=0.24\linewidth]{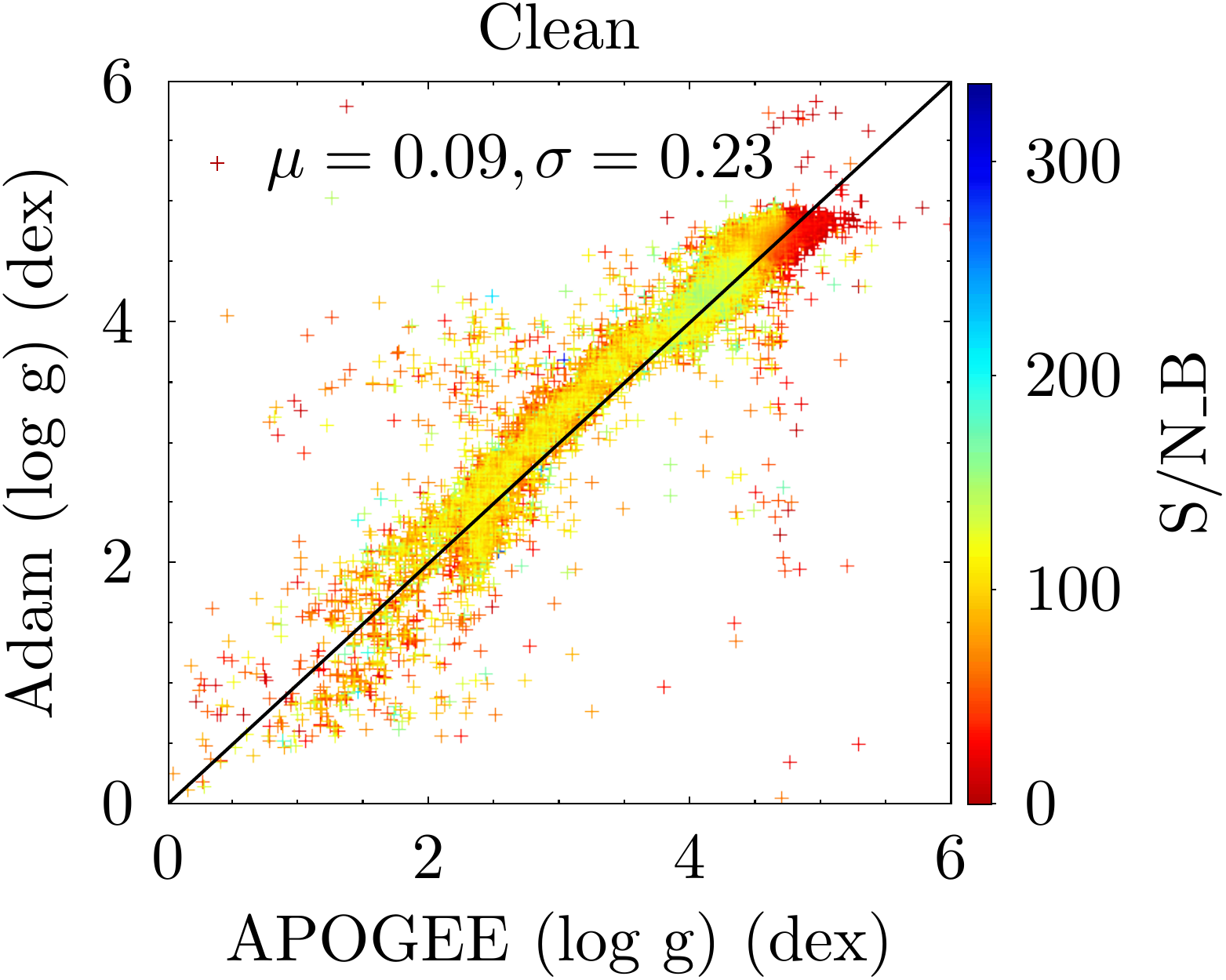}
		\includegraphics[width=0.24\linewidth]{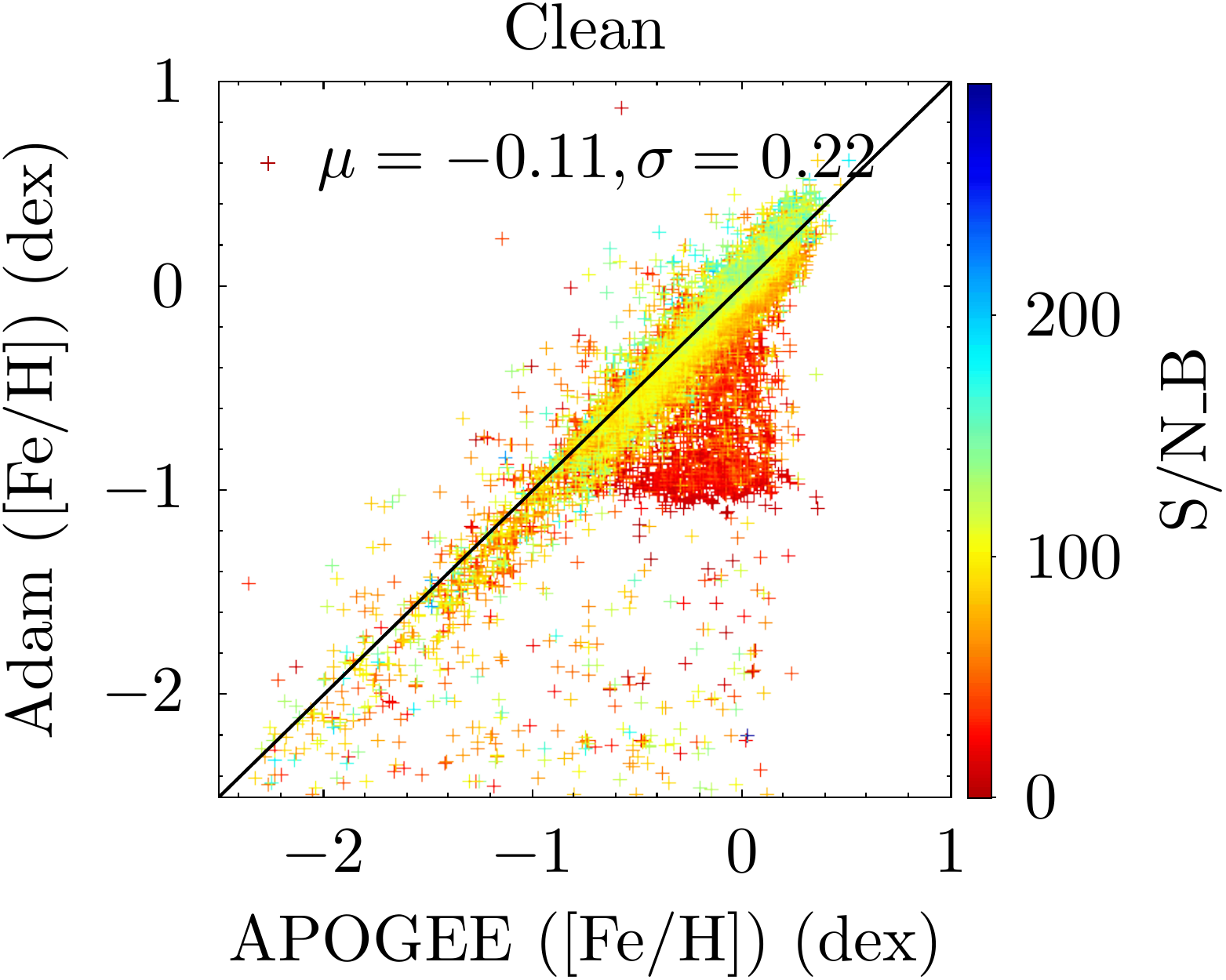}
		\caption{Comparison of inferred parameters with APOGEE DR$16$ labels. From top to bottom, the six rows correspond to comparisons between APOGEE DR16 and (1) the DESI RVS pipeline, (2) the DESI SP pipeline, (3) LASP-CurveFit with the No Clean strategy, (4) LASP-CurveFit with the Clean strategy, (5) LASP-Adam-GPU with the No Clean strategy, and (6) LASP-Adam-GPU with the Clean strategy. From left to right, the columns show $\mathrm{RV}$, $T_{\mathrm{eff}}$, $\log g$, and $\mathrm{[Fe/H]}$. The $\mu$ and $\sigma$ of the differences are computed using \texttt{astropy.stats.sigma\_clipped\_stats}, excluding values beyond $5\sigma$. For clarity, the \lq LASP\rq \ prefix is omitted in all figure labels.}
		\label{vsDESI}
	\end{figure*}
	
	\subsection{Error analysis} \label{Error}
	To evaluate the error estimation capability of PyLASP in stellar parameter inference, we examine the consistency between model errors estimated by the first-stage LASP-Adam-GPU and the corresponding repeat-observation errors of the same targets, and compare it with that between the official two-stage empirical errors from LAMOST and their respective repeat-observation errors. The model errors are derived from error propagation theory, which estimates how spectral noise propagates from pixel space to parameter space. In contrast, repeat-observation errors are computed from the dispersion among multiple independent observations of the same target, under the assumption of stable stellar properties and similar observing conditions. Since both types of errors reflect the influence of spectral noise, the repeat-observation errors serve as an empirical reference for evaluating the reliability of model errors. If the error model is reasonable, the two should be statistically consistent and exhibit similar trends with respect to S/N. For this analysis, we select $25{,}878$ spectra from a total of $177{,}848$ that have at least two repeat observations, S/N in the $g$ band (\texttt{S/N\_g}) greater than zero, and less than $10\%$ variation in \texttt{S/N\_g} across epochs, to ensure robust error assessment.
	
	As shown in Figure~\ref{LASPErr}, the model-based errors from LASP-Adam-GPU for $T_{\mathrm{eff}}$, $\log g$, and [Fe/H] are statistically consistent with the corresponding repeat-observation errors, suggesting that the model provides reasonable error estimates. At $\texttt{S/N\_g} {<} 20$, these model errors are generally larger than the official LAMOST empirical errors, while at $\texttt{S/N\_g} {>} 20$, they become smaller. This indicates that LASP-Adam-GPU provides an S/N-sensitive error response and better captures noise-driven variability than the official empirical formula. In contrast, the behavior of $\mathrm{RV}$ errors deviates notably. LASP-Adam-GPU consistently underestimates $\mathrm{RV}$ errors relative to repeat observations, which in turn lie below the official empirical values. A similar phenomenon is observed in the DESI RVS catalog. We speculate that this discrepancy mainly arises from the fact that the model does not include systematic effects such as wavelength calibration residuals and instrumental response variations in the error propagation process, and these factors have a more significant impact on $\mathrm{RV}$ measurements. Since repeat observations can partially reflect such systematic errors, while the model errors are solely derived from the propagation of spectral noise, the estimated results from the two approaches exhibit a systematic deviation of approximately $1.5\ \mathrm{km \ s^{-1}}$ in the $\mathrm{RV}$ dimension. These results suggest that, with the exception of $\mathrm{RV}$, the model errors estimated by LASP-Adam-GPU are generally consistent with the repeat-observation errors, and are more reasonable in the low-S/N regime.

	\subsection{Application to DESI} \label{DESI}
	DESI is a state-of-the-art, highly multiplexed spectroscopic facility mounted on the Mayall 4 m telescope at the Kitt Peak National Observatory \citep{Silber2023,Poppett2024,Miller2024}. In addition to its cosmological program, DESI conducts the Milky Way Survey (MWS) during bright-time conditions to obtain extensive spectroscopy of Galactic stellar populations. These stellar spectra span $3600{-}9800$\,\AA, with spectral resolution increasing from $R {\sim} 2000$ at $3600$\,\AA\ to $R {\sim} 5000$ at $9800$\,\AA. For the MWS program, these spectra are analyzed by two independent pipelines for stellar parameter inference \citep{Koposov2025}: the RVS pipeline uses RVSpecFit with neural network emulators trained on PHOENIX spectra to infer $\mathrm{RV}$, $T_{\rm eff}$, $\log g$, [Fe/H], and [$\alpha$/Fe], while the SP pipeline employs FERRE based on Kurucz models to infer $T_{\rm eff}$, $\log g$, [Fe/H], and the abundances of $10$ individual elements.
	
	To assess the applicability of PyLASP to other spectroscopic surveys, we apply it to the MWS in DESI DR$1$, comprising $6{,}369{,}991$\footnote{The full MWS catalog contains $6{,}372{,}607$ spectra; we exclude $2608$ labeled as BAD and eight as NON based on the \texttt{OBJTYPE} field.} spectra used to infer $\mathrm{RV}$, $T_{\mathrm{eff}}$, $\log g$, and [Fe/H]. The PyLASP inference is performed over the wavelength range $4200$–$5700$\,\AA, sampled in base-10 logarithmic wavelength space with a step size of $0.0001$. The initial values for atmospheric parameter optimization are set to $(7500\,\mathrm{K},\ 3\,\mathrm{dex},\ -0.5\,\mathrm{dex})$, and LASP-Adam-GPU is configured with a default learning rate of $0.1$ and a convergence threshold of $10^{-5}$. Under the No Clean strategy, LASP-CurveFit and LASP-Adam-GPU complete the inference in $35$ and $8$ hr, respectively; with the Clean strategy, the runtimes increase to $43$ and $14$ hr. The APOGEE stellar parameter catalog is adopted as the reference benchmark\footnote{APOGEE provides higher spectral resolution ($R {\sim} 22{,}500$ versus $R {\sim} 2{,}000$–$5{,}000$ for DESI), reduced interstellar extinction in the $H$ band, and well-established external calibrations of stellar parameters.}, and the parameters inferred by PyLASP and by DESI’s two official pipelines (RVS and SP) are independently compared against it.
	
	As shown in Figure~\ref{vsDESI}, the Clean strategy in PyLASP improves the consistency of $T_{\mathrm{eff}}$, $\log g$ and [Fe/H] over the No Clean strategy, indicating effective removal of bad pixels in DESI spectra. Compared to the RVS and SP pipelines, PyLASP shows better agreement with APOGEE for $T_{\mathrm{eff}}$ and $\log g$, while the DESI pipelines yield better consistency in $\mathrm{RV}$ and [Fe/H]. Notably, PyLASP exhibits a systematic underestimation of [Fe/H] in the low-S/N regime ($\texttt{S/N\_B} {<} 50$). This bias may arise from the limited wavelength coverage of the ELODIE spectra, which does not extend to the blue end of DESI spectra where many metal lines are located. In contrast, the broader wavelength range used in DESI's RVS and SP pipelines may provide greater robustness in this region. We further note that, beyond the systematic [Fe/H] bias, other outliers relative to APOGEE are mainly concentrated at low $\texttt{S/N\_B}$ for LASP-CurveFit, whereas a smaller subset of outliers is also exhibited at high $\texttt{S/N\_B}$ for LASP-Adam-GPU. In these high $\texttt{S/N\_B}$ cases, roughly one-third of the affected spectra contain negative or zero flux, while another portion exhibits pronounced absorption or emission features. These parameter anomalies in LASP-Adam-GPU results at high $\texttt{S/N\_B}$, relative to APOGEE, are mainly attributable to two factors. First, LASP-Adam-GPU, unlike LASP-CurveFit, lacks explicit failure detection mechanisms for parameter inference---for instance, it still treats results as valid even when the multiplicative correction factors $P(x)b_{i}$ are nonpositive, indicating that anomalous spectral features in the observed spectra cause an abnormal pseudocontinuum, which destabilizes the objective function (Equation~\ref{loss}) and results in unreliable parameter inferences. Second, LASP-Adam-GPU applies a fixed, intergroup Clean strategy rather than adapting the iterative process to individual spectra. As a result, anomalous spectral features within spectra of the same group may not be fully masked, potentially leading to unreliable parameter inferences. These limitations will be addressed in future versions of PyLASP. Additionally, implementation details such as the order of Legendre polynomials and the resampling step size used in PyLASP may also contribute to differences in $\mathrm{RV}$. A more detailed analysis will be provided in S. Li et al. (2025, in preparation).
		
	\subsection{Future improvements} \label{Future improvements}
	To further improve the efficiency, robustness, and reliability of parameter inference in PyLASP, we propose the following three directions for future development:
	\begin{itemize}
		\item \textbf{Enhancing curvature awareness in the optimizer.} 
		Compared to LASP-CurveFit, LASP-Adam-GPU generally requires more iterations to converge, primarily because the Adam optimizer relies solely on first-order gradients and lacks sensitivity to the curvature of the objective function. In contrast, LASP-CurveFit captures local curvature through an approximate Hessian, enabling faster convergence with fewer iterations. To reduce the number of iterations and improve optimization efficiency, we plan to incorporate curvature information into LASP-Adam-GPU using second-order central differences to approximate the Hessian. This approach can be designed to take advantage of parallel computation, balancing speed and accuracy. In addition, symbolic differentiation or learned derivative models may be employed to further reduce the cost of gradient evaluation.
		\item \textbf{Improving the adaptiveness of the Clean strategy.}
		Currently, LASP-Adam-GPU applies a fixed iteration schedule for outlier pixel rejection across all spectra in a group, without adjusting for individual spectral quality. This uniform scheme limits the ability to detect outliers in low-quality spectra and may contribute to discrepancies with LASP-MPFit under the Clean strategy. To enhance robustness, future versions will implement adaptive Clean control based on spectrum-specific properties, such as S/N or fitting residuals, allowing for dynamic adjustment within groups and improving consistency with the LASP-MPFit Clean implementation.
		\item \textbf{Refining parameter validation for catalog outputs.}
		Due to limited spectral quality, certain targets may exhibit abnormal emission lines or unflagged bad pixels not recorded in the official FITS files, which can compromise parameter reliability. Although the Clean strategy helps mitigate the effects of such artifacts, significant discrepancies between results from the Clean and No Clean modes often indicate unreliable parameters. We therefore propose the use of robust statistical metrics (e.g., sigma clipping) to implement a parameter consistency check, identify and exclude unreliable results, and prioritize Clean-mode parameters in the final catalog. This screening mechanism will improve the scientific utility and reliability of the recommended catalog.
	\end{itemize}
	
	\section{Conclusions} \label{Conclusions}
	Based on the original IDL version of LASP, we have developed a Python-based implementation tailored for large-scale stellar spectral analysis. The new framework incorporates two optimization strategies---LASP-CurveFit and LASP-Adam-GPU---within a modular architecture, and we systematically evaluate its performance in terms of inference efficiency, parameter consistency, error modeling, and cross-survey applicability. The PyLASP code and DESI-based catalog are available via \dataset[DOI: 10.12149/101679]{https://doi.org/10.12149/101679} and \dataset[DOI: 10.12149/101675]{https://doi.org/10.12149/101675}, respectively\footnote{Each of these DOIs represents version 1.0 of the specified repository. DOI: 10.12149/101678 and DOI: 10.12149/101674 will resolve to the latest version of each repository.}. The main conclusions are as follows:
	\begin{itemize}
		\item \textbf{Improved efficiency in parameter inference.} 
		On the same notebook platform, the runtime efficiency of LASP-Adam-GPU and LASP-CurveFit reaches $3.23$ and $1.75$ times that of LASP-MPFit, respectively. In addition, LASP-Adam-GPU enables fast parameter inference for tens of millions of stellar spectra on high-performance GPUs, completing $10$ million spectra in $8$ and $7$ hr on the RTX 3090 and A100, respectively. These results demonstrate the framework's strong parallel scalability and computational efficiency.
		\item \textbf{Consistent parameter estimates with the IDL version.}
		The Python and IDL versions yield highly consistent parameter estimates, with discrepancies observed in fewer than $0.35\%$ of cases. Further analysis indicates that abnormal flux patterns are the primary cause of outliers (Figure~\ref{5std}); applying the Clean strategy significantly reduces such inconsistencies. Comparison with APOGEE labels shows that the Python version achieves better agreement under the No Clean strategy, while LASP-Adam-GPU performs slightly worse under the Clean mode, possibly due to the use of a fixed number of Clean iterations without spectral-quality-aware adaptation.
		\item \textbf{Different sensitivity to initialization.}
		LASP-CurveFit is robust to initial values. In contrast, due to the piecewise structure of the spectral emulator in the $T_{\mathrm{eff}}$ dimension, LASP-Adam-GPU shows increased sensitivity to the initial temperature when $T_{\mathrm{eff}} > 8000$ K. To improve the inference accuracy of LASP-Adam-GPU in the high-temperature regime, we recommend using the initial values provided by the CFI when available. If CFI initialization is not accessible, we suggest adopting $(7500\ \mathrm{K},\ 3\ \mathrm{dex},\ -0.5\ \mathrm{dex})$ as the default initial guess.
		\item \textbf{Model-based errors are consistent with repeat observations.}
		The atmospheric parameter errors estimated by PyLASP are highly consistent with the random errors calculated from multiple observations of the same target, and outperform the empirical correction scheme based on fitted functions. The tendency of $\mathrm{RV}$ errors to be smaller than the corresponding repeat-observation errors is also observed in the DESI data, indicating this bias may be widespread and deserves further investigation.
		\item \textbf{Reliable performance across different surveys.}
		When applied to DESI DR$1$ data, PyLASP produces $T_{\mathrm{eff}}$ and $\log g$ values that agree well with APOGEE labels. For [Fe/H], performance is slightly worse than the DESI RVS and SP pipelines in low-S/N regions, likely due to the limited wavelength coverage of the ELODIE-based templates. $\mathrm{RV}$ precision is also slightly lower, possibly due to the impact of spectral sampling resolution.
		\item \textbf{Recommended use cases and application conditions.}
		PyLASP, built on the ELODIE library, targets optical spectra with $R{<}10{,}000$ and wavelength coverage $3900{-}6800$\,\AA{}, provided the targets’ atmospheric parameters lie within \(T_{\rm eff}{=}3100\text{--}59{,}000\,\mathrm{K}\), \(\log g{=}0\text{--}5\,\mathrm{dex}\), and \([\mathrm{Fe/H}]{=}-2.8\text{--}1\,\mathrm{dex}\). Within this domain, the framework adaptively matches observational resolution without retraining the spectral emulator. If the resolution, wavelength range, or parameter domain exceeds ELODIE’s coverage, extend the framework by replacing the model-spectrum generation module and adjusting the wavelength sampling. For applications, LASP-CurveFit with the Clean strategy is preferred for small-scale to medium-scale datasets (${\le} 100{,}000$) or low-quality spectra, owing to its robustness to initialization and effective handling of anomalous flux features; LASP-Adam-GPU is recommended for large-scale analyses (${>}100{,}000$) or joint optimization in high-dimensional parameter spaces, being better suited to multivariate optimization when the number of free parameters ranges from a few dozen to tens of thousands, where CurveFit may become inefficient.
	\end{itemize}
	
	In summary, PyLASP significantly improves the efficiency of large-scale spectral parameter inference and demonstrates reliable performance in terms of parameter accuracy, error modeling, and cross-survey applicability. These developments establish a robust computational foundation for extending LASP to high-dimensional label inference, chemical abundance modeling, and automated analysis pipelines in upcoming large-scale spectroscopic surveys.
	
	\section*{Acknowledgments}
	This work is supported by the National Natural Science Foundation of China (12273075, 12273078, and 12411530071) and the National Astronomical Observatories of the Chinese Academy of Sciences (No. E4ZR0516). We also acknowledge support from a Royal Society IEC\textbackslash NSFC\textbackslash233140 exchange grant.
	
	Guoshoujing Telescope (the Large Sky Area Multi-Object Fiber Spectroscopic Telescope LAMOST) is a National Major Scientific Project built by the Chinese Academy of Sciences. Funding for the project has been provided by the National Development and Reform Commission. LAMOST is operated and managed by the National Astronomical Observatories, Chinese Academy of Sciences.
	
	Funding for the Sloan Digital Sky 
	Survey IV has been provided by the 
	Alfred P. Sloan Foundation, the U.S. 
	Department of Energy Office of 
	Science, and the Participating 
	Institutions. 
	
	SDSS-IV acknowledges support and 
	resources from the Center for High 
	Performance Computing  at the 
	University of Utah. The SDSS 
	website is www.sdss4.org.
	
	SDSS-IV is managed by the 
	Astrophysical Research Consortium 
	for the Participating Institutions 
	of the SDSS Collaboration including 
	the Brazilian Participation Group, 
	the Carnegie Institution for Science, 
	Carnegie Mellon University, Center for 
	Astrophysics | Harvard \& 
	Smithsonian, the Chilean Participation 
	Group, the French Participation Group, 
	Instituto de Astrof\'isica de 
	Canarias, The Johns Hopkins 
	University, Kavli Institute for the 
	Physics and Mathematics of the 
	Universe (IPMU) / University of 
	Tokyo, the Korean Participation Group, 
	Lawrence Berkeley National Laboratory, 
	Leibniz Institut f\"ur Astrophysik 
	Potsdam (AIP),  Max-Planck-Institut 
	f\"ur Astronomie (MPIA Heidelberg), 
	Max-Planck-Institut f\"ur 
	Astrophysik (MPA Garching), 
	Max-Planck-Institut f\"ur 
	Extraterrestrische Physik (MPE), 
	National Astronomical Observatories of 
	China, New Mexico State University, 
	New York University, University of 
	Notre Dame, Observat\'ario 
	Nacional / MCTI, The Ohio State 
	University, Pennsylvania State 
	University, Shanghai 
	Astronomical Observatory, United 
	Kingdom Participation Group, 
	Universidad Nacional Aut\'onoma 
	de M\'exico, University of Arizona, 
	University of Colorado Boulder, 
	University of Oxford, University of 
	Portsmouth, University of Utah, 
	University of Virginia, University 
	of Washington, University of 
	Wisconsin, Vanderbilt University, 
	and Yale University.
	
	This research used data obtained with the Dark Energy Spectroscopic Instrument (DESI). DESI construction and operations is managed by the Lawrence Berkeley National Laboratory. This material is based upon work supported by the U.S. Department of Energy, Office of Science, Office of High-Energy Physics, under Contract No. DE–AC02–05CH11231, and by the National Energy Research Scientific Computing Center, a DOE Office of Science User Facility under the same contract. Additional support for DESI was provided by the U.S. National Science Foundation (NSF), Division of Astronomical Sciences under Contract No. AST-0950945 to the NSF’s National Optical-Infrared Astronomy Research Laboratory; the Science and Technology Facilities Council of the United Kingdom; the Gordon and Betty Moore Foundation; the Heising-Simons Foundation; the French Alternative Energies and Atomic Energy Commission (CEA); the National Council of Humanities, Science and Technology of Mexico (CONAHCYT); the Ministry of Science and Innovation of Spain (MICINN), and by the DESI Member Institutions: www.desi.lbl.gov/collaborating-institutions. The DESI collaboration is honored to be permitted to conduct scientific research on I’oligam Du’ag (Kitt Peak), a mountain with particular significance to the Tohono O’odham Nation. Any opinions, findings, and conclusions or recommendations expressed in this material are those of the author(s) and do not necessarily reflect the views of the U.S. National Science Foundation, the U.S. Department of Energy, or any of the listed funding agencies.
	
	\bibliographystyle{aasjournal}
	\footnotesize
	\bibliography{refs}
\end{document}